\documentclass[reqno,12pt,letterpaper]{amsart}
\usepackage{amsmath,amssymb,amsthm,graphicx,mathrsfs,url}
\usepackage[usenames,dvipsnames]{color}
\usepackage[colorlinks=true,linkcolor=Red,citecolor=Green]{hyperref}
\usepackage{amsxtra}
\usepackage{tikz}
\usepackage{comment}
\usepackage{amscd}
\usetikzlibrary{arrows.meta}
\usepackage[normalem]{ulem}
\usepackage{wasysym} 
\usepackage{graphicx}
\usepackage{subcaption}
\usepackage[normalem]{ulem}
\def\arXiv#1{\href{http://arxiv.org/abs/#1}{arXiv:#1}}
\usepackage{array}
\newcolumntype{P}[1]{>{\centering\arraybackslash}m{#1}}

\def\?[#1]{\textbf{[#1]}\marginpar{\Large{\textbf{??}}}}

\let\epsilon=\varepsilon 
\newcommand{\CC}{{\mathbb C}}

\newtheorem{theo}{Theorem}
\newtheorem{op}{Open Problem}

\numberwithin{equation}{section}

\DeclareMathOperator{\Spec}{Spec}

\DeclareMathOperator{\Hom}{Hom}

\let\Im=\Imag
\DeclareMathOperator{\loc}{loc}

\let\Re=\Real

\DeclareMathOperator{\tr}{tr}

\def\indic{\operatorname{1\hskip-2.75pt\relax l}}
\usepackage{scalerel}

\newcommand\reallywidehat[1]{\arraycolsep=0pt\relax%
\begin{array}{c}
\stretchto{
  \scaleto{
    \scalerel*[\widthof{\ensuremath{#1}}]{\kern-.5pt\bigwedge\kern-.5pt}
    {\rule[-\textheight/2]{1ex}{\textheight}} 
  }{\textheight} %
}{0.5ex}\\           
#1\\                 
\rule{-1ex}{0ex}
\end{array}
}

\def\blue#1{\textcolor{blue}{#1}}
\def\red#1{\textcolor{red}{#1}}

\begin{document}

\title[Mathematical results on the chiral model of TBG]{
Mathematical results on the chiral model of twisted bilayer graphene.}

\author{Maciej Zworski}
\email{zworski@math.berkeley.edu}
\address{Department of Mathematics, University of California,
Berkeley, CA 94720, USA}

\address{Department of Mathematics, University of California,
Berkeley, CA 94720, USA.}
\email{tzk320581@berkeley.edu}

\address{Department of Mathematics, University of California,
Berkeley, CA 94720, USA.}
\email{mxyang@math.berkeley.edu}

\maketitle 

\vspace{-0.2in}

\begin{center}
{\sc With an appendix by Mengxuan Yang and Zhongkai Tao}
\end{center}

\vspace{-0.1in}

\begin{abstract}
The study of twisted bilayer graphene (TBG) is a hot topic in condensed
matter physics with special focus on {\em magic angles} of twisting
at which TBG acquires unusual properties. Mathematically,  
topologically non-trivial flat bands appear at those special angles. The chiral model of TBG 
pioneered by Tarnopolsky--Kruchkov--Vishwanath \cite{magic} has
particularly nice mathematical properties and 
we survey, and in some cases, clarify, recent rigorous results which exploit them.
\end{abstract}

\tableofcontents

\section{Introduction}
\label{s:intr} 

\begin{figure}
\includegraphics[width=7.45cm]{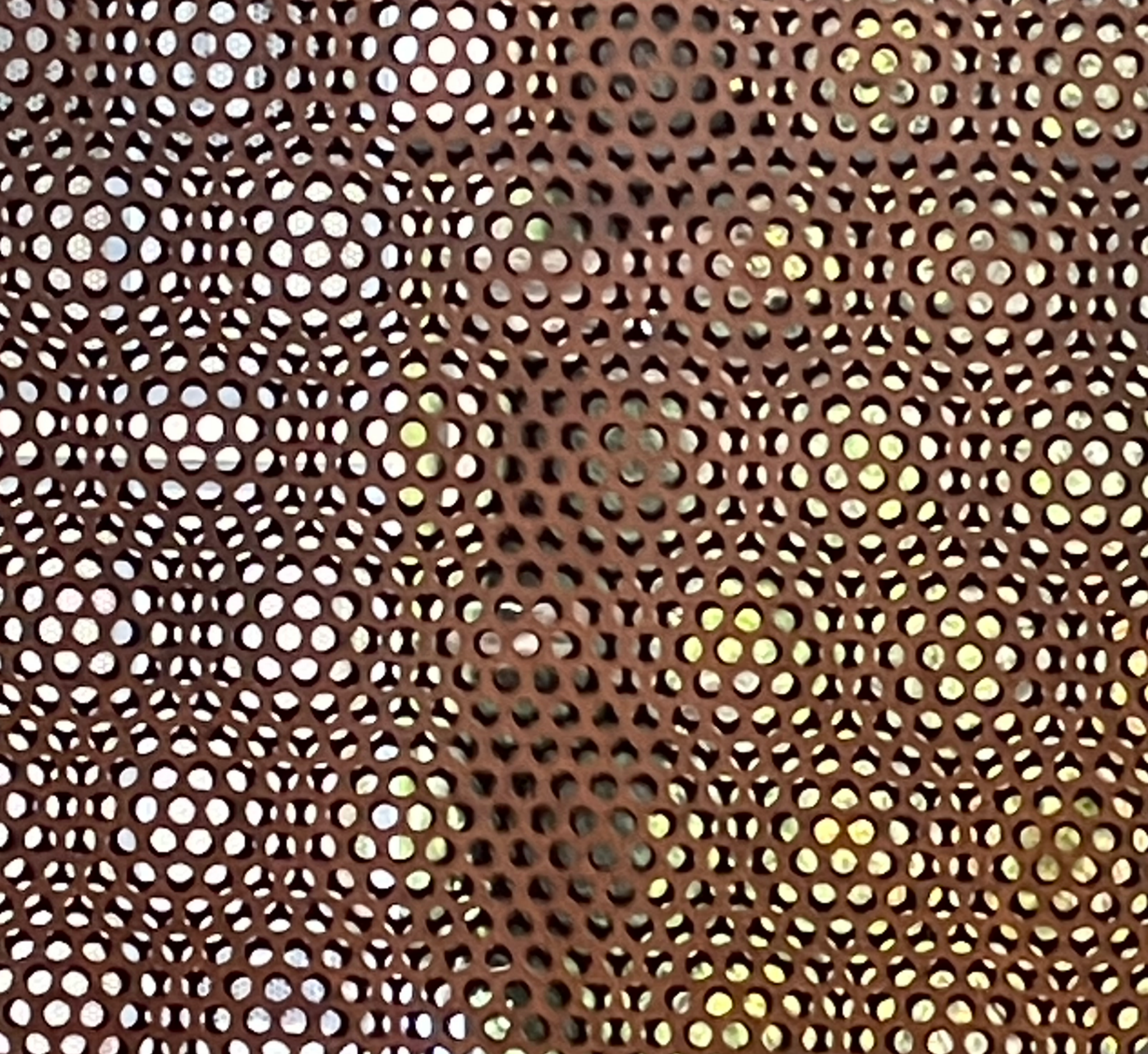}\includegraphics[width=8cm]{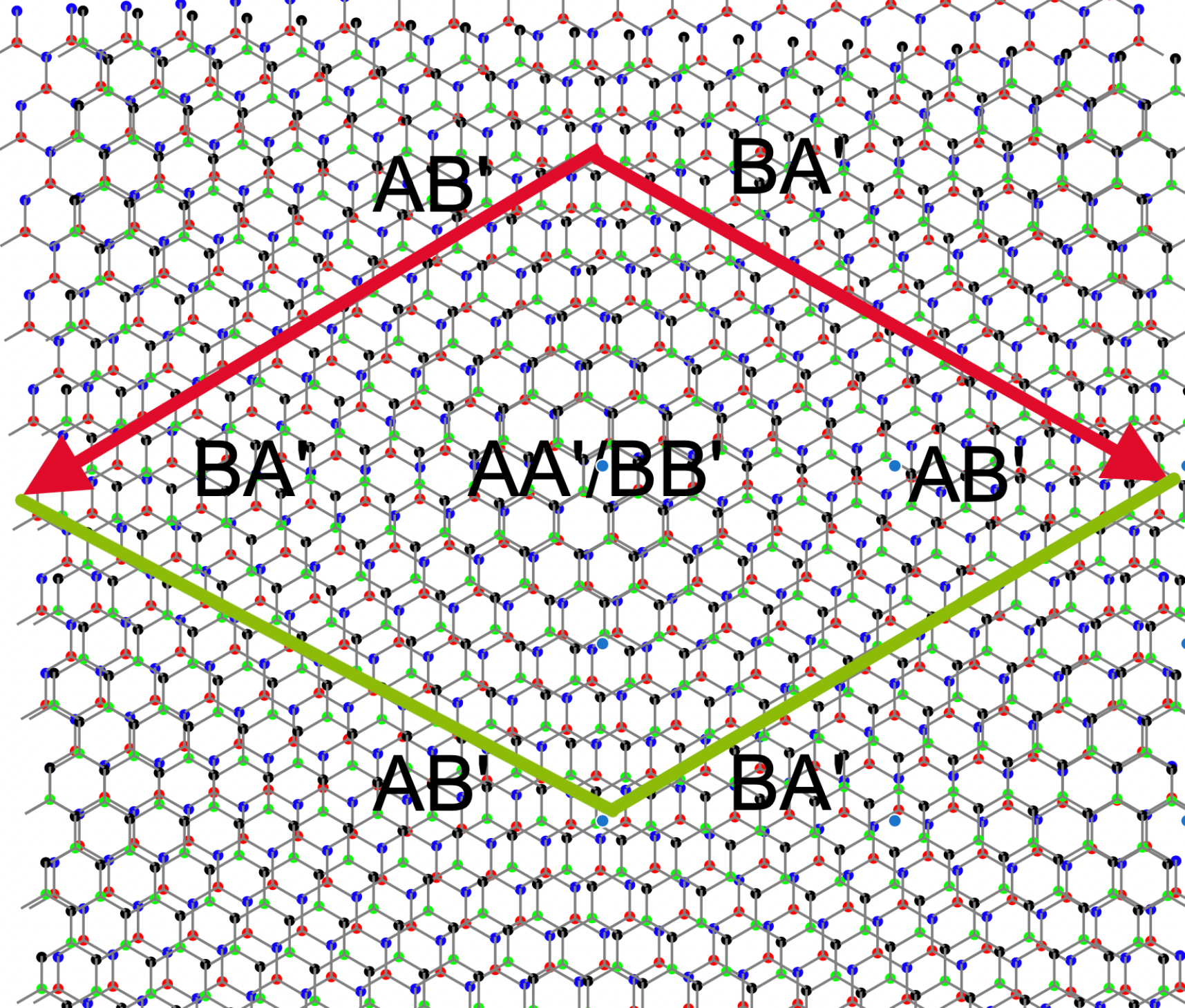}
\caption{\label{f:1} 
Left: a moir\'e pattern at CIRM in Luminy; right: a moir\'e fundamental cell with regions of different 
($ AA'$, $BB'$, $ AB' $... ) particle-type overlaps. Tunneling potential 
$\vert V(\mathbf r)\vert$ concentrates in $AA'/BB'$ regions and $\vert U(\mathbf r)\vert$  concentrates at $AB'$ regions.}.
\end{figure}

Investigation of physical properties of twisted bilayer graphene, and of similar 
structures, is a
hot topics in condensed matter physics. One feature which is present when
periodic structures are twisted is the emergence of {\em moir\'e patterns} -- 
see Figure \ref{f:1}. These patterns create new periodic (or quasi-periodic)
structures which now have much larger fundamental cells. That is very 
useful and, for instance, has led to experimental observation of the
{\em Hofstadter butterfly} \cite{Science} -- see \cite{avij} for the mathematical
derivation and history. 

The property on which we focus in 
this mathematical survey is existence of {\em flat bands} at certain angles
of twisting (see \S \ref{s:blfl} below for a review of the Bloch--Floquet theory
and for definition of band spectrum). Flat bands correspond to eigenvalues
of infinite multiplicity for the periodic Hamiltonian modeling the system. The
first thought would then suggest existence of highly localized eigenstates which
would prevent conductivity. If however the band topology is non-trivial 
(see \S \ref{s:top} below) the localization is weak and can lead to superconductivity,
in a somewhat mysterious mechanism, certainly not understood mathematically. 

The Bistritzer–MacDonald Hamiltonian (BMH) \cite{BM11} is widely considered to be a good model for the study of twisted bilayer graphene (TBG) and it achieved celebrity for an accurate prediction of the twisting angle at which superconductivity occurs \cite{Cao}.
The chiral limit of BMH is obtained by neglecting $ A A' /BB' $ tunneling (see Figure \ref{f:1} and
\S \ref{s:chili}). It has many advantageous properties and 
was studied with great success by Tarnopolsky--Kruchkov--Vishwanath
  \cite{magic}  and their collaborators, see for instance Ledwith et al \cite{led}. 
  One striking feature of the chiral limit, one which is not present in the BMH model, is the 
existence of {\em exact} flat bands.
  The Hamiltonian
 is of the form
 \[  H ( \alpha ) = \begin{pmatrix} 0 & D ( \alpha )^* \\
 D ( \alpha ) & 0 \end{pmatrix} , \ \ \  D ( \alpha )  : H^1 ( \mathbb C; \mathbb C^2 ) 
 \to L^2 (\mathbb C; \mathbb C^2 ), \]
 where $ D ( \alpha ) $ is a first order (non-self-adjoint) matrix valued operator and 
 $ \alpha $ is dimensionless constant (a much appreciated feature for mathematicians)
 with $ 1/\alpha $ corresponding to the angle of twisting. The bands are the eigenvalues
 of $ H_k (\alpha ) $ which is obtained by replacing $ D ( \alpha ) $ by $ D ( \alpha ) + k $
 in the definition of $ H (\alpha ) $ and by taking periodic boundary condition with respect to
 the lattice of periodicity of $ H ( \alpha ) $, $ \Gamma $. Hence, 
 \[  \text{ $ H (\alpha ) $ has a flat band at zero energy } 
 \ \Longleftrightarrow \   \Spec_{ L^2 ( \mathbb C/\Gamma ) } D ( \alpha ) = \mathbb C . \]
 It turns out (see \S\S \ref{s:chili},\ref{s:spec}) that the set of $ \alpha$'s for which this 
 happens is discrete -- at other $ \alpha $'s the spectrum is given by $ \Gamma^* $, 
 the reciprocal lattice of $ \Gamma $ (in the notation of \S \ref{s:stan}, $ \Gamma = 
 3 \Lambda $ and $ \Gamma^* = \frac13 \Lambda^*$). 

In this survey we discuss distribution of $ \alpha $ for which $ H ( \alpha ) $ has 
a flat band at zero energy and properties of the corresponding eigenfunctions. 
We  concentrate on presenting rigorous mathematical results familiar to the author with precise pointers to specific papers. In particular, we do not attempt to survey the vast physics literature on TBG. 
  The motivation comes from beautiful and mysterious properties of 
  the differential operator appearing in the chiral model (see Figure~\ref{f:magic} for an illustration).
  We also highlight some open mathematical problems. The most interesting are perhaps Problems \ref{1}
  and \ref{8} as they still attract attention in the physics literature.  Other problems concern finer aspects of the model and most are of purely mathematical interest -- I find Problems \ref{2},\ref{3},\ref{7},\ref{15},\ref{18}  and \ref{20} particularly 
  appealing. 

  Mathematical study of the chiral model of TBG started with the work of Watson--Luskin \cite{lawa} who showed existence of the  first magic angle, and of
  Becker--Embree--Wittsten--Zworski \cite{phys,beta} who 
  gave  a spectral characterization
  of magic angles and explained exponential squeezing of bands. 
 It has been developed in several directions by
  Becker--Humbert--Zworski \cite{bhz1,bhz2,bhz2} (trace formulas, existence of
  generalized magic angles, existence and properties of degenerate
  magic angles, topological properties), Becker--Humbert--Wittsten--Yang
  \cite{BHWY} (magic angles for trilayer graphene), 
  Becker--Oltman--Vogel \cite{bovo} (random perturbation of TBG), 
  Becker--Zworski \cite{bz1,bz2} (TBG in a magnetic field parallel to the graphene sheets, deformation 
  to the full Bistritzer--MacDonald model), Galkowski--Zworski \cite{gaz} (an abstract formulation of the spectral characterization, a scalar model for magic angles), 
  Hitrik--Zworski \cite{hiz}, Tao--Zworski \cite{tza} (classically forbidden regions
  for eigenstates), and Yang \cite{yang} (twisted multiple layer graphene). Many of these results are 
described in this survey.
  
 During the writing of this survey it became apparent that we did not have a reference to the fact
 that the chiral model of TBG exhibits Dirac cones away from $ \alpha $'s at which flat
 bands appear -- see Open Problem \ref{2}. Mengxuan Yang and Zhongkai Tao immediately provided
 an argument for that and it is included here as an appendix. 
  
\noindent
{\bf Notation}
In this paper we use the physics notation: for an
operator $ A $ on $ L^2 ( M, dm )  $, $ \langle u | A | v \rangle := \int_M Av \, \bar u \, dm $.
Also, $ | u \rangle $ denotes the operator $ \mathbb C \ni \mu \to \mu u \in L^2 $ and
$ \langle u | $, its adjoint $ L^2 \ni v \to \langle u | v \rangle \in \mathbb C$. 
For $ z, w \in \mathbb C \simeq \mathbb R^2 $, we use 
the real inner product, $ \langle z , w \rangle := \Re z \bar w$. If $ H $ is a function 
space (such as $ L^2 $, Sobolev space $ H^s $ or spaces with given 
periodicity conditions) then $ H ( M ; \mathbb C^n ) $ 
denotes functions in $ H $ on $ M $ with values in $ \mathbb C^n $. When the context is clear
we may drop $ M $ and $ \mathbb C^n $. 

\noindent
{\bf Acknowledgements.} I would like to thank Mike Zaletel for introducing me to TBG 
and pointing out the semiclassical nature of small angle asymptotics. 
I am grateful to my many collaborators on projects 
related to TBG, on whose work this survey is based, especially to Simon Becker who produced most of the figures (and movies) 
in our recent joint papers, some of which are re-used here. I would also like to thank
Patrick Ledwith, Lin Lin, Mitch Luskin, Allan MacDonald, Ashvin Vishwanath, and Alex Watson 
for valuable physics perspectives (many of which, alas, remain a mystery to this author). 
Simon Becker and Mengxuan Yang provided many insightful comments on the 
first drafts of this survey and I am very grateful for that great help. Thanks go also to Zhongkai Tao
for pointing out and clarifying a mistake in \S \ref{s:top}.
Partial support by the NSF grant DMS-1901462
and by the Simons Foundation under a ``Moir\'e Materials Magic" grant
is also most gratefully acknowledged.

\section{The Bistritzer--MacDonald Hamiltonian and  its
  chiral limit}
\label{s:ope}

In this section we consider the Bistritzer--MacDonald Hamiltonian (BMH) \cite{BM11} 
from the PDE point of view without addressing its physical 
motivation. It has been mathematically derived by Canc\`es--Garrigue--Gontier \cite{CGG} and
Watson--Kong--MacDonald--Luskin \cite{Wa22} and 
 we refer to these papers above and \cite{magic} for physics backgroup. As we will stress, its chiral limit exhibits beautiful and unusual mathematical properties which have been our main motivation.

The representation of BMH in the physics 
literature \cite{BM11}, \cite{magic} is given as follows: for two parameters $ \alpha $ and 
$ \lambda $ we define
 \begin{equation*}
 \begin{gathered} 
H_{\rm{BM} } (\alpha,\lambda)  = \begin{pmatrix} -i ( \sigma_1 \partial_{x_1} + \sigma_2 \partial_{x_2} )  & T(\alpha, \lambda)  \\  
T(\alpha, \lambda) ^* & -i ( \sigma_1 \partial_{x_1} + \sigma_2 \partial_{x_2}  ) \end{pmatrix}
: H^1 ( \mathbb R^2 ; \mathbb C^4 ) \to L^2 ( \mathbb R^2 ; \mathbb C^4 ) , 
\end{gathered} 
\end{equation*}
where we use Pauli matrices, 
\[  \sigma_1 := \begin{pmatrix} 0 & 1 \\ 1 & 0 \end{pmatrix} , \ \ 
 \sigma_2 := \begin{pmatrix} 0 & - i \\
i & \ 0 \end{pmatrix} , \ \ \  \mathbf r = ( x_1, x_2 ) \in \mathbb R^2 ,
\] 
and $ \bullet^* $ denotes the hermitian conjugate.

The interlayer tunnelling matrix is defined as follows
\[T(\alpha, \lambda ) = \begin{pmatrix}\lambda  V(\mathbf r) & \alpha \overline{U(- \mathbf r)} \\ \alpha U( \mathbf r)   & \lambda V( \mathbf r) \end{pmatrix}.\]
The non-equivalent pairs of atoms in a fundamental cell of the honeycomb lattice of graphene are labelled by $A,B$, with the labeling  $A',B'$ for the second sheet in TBG. 
In the matrix potential $T$, the  $U$ and $V$ model $AB'$ tunnelling and $AA'/BB'$ respectively, see Figure \ref{f:1}. 
The are defined as follows: with $ \omega := \exp ( 2 \pi i /3 ) $, 
\[ \begin{gathered}  U(\mathbf r) = \sum_{i=0}^2 \omega^\ell e^{-i q_\ell \cdot \mathbf r}   , \ \ \ 
V(\mathbf r) = \sum_{\ell=0}^2  e^{-i q_\ell\cdot \mathbf r}, \ \ \
q_\ell:= R^\ell(0,-1),  \ \ 
R := \tfrac{1}{2}\begin{pmatrix} -1 & -\sqrt{3} \\ \sqrt{3} & -1 \end{pmatrix}. \end{gathered} \]
(We note that $ R $ is the $2\pi/3$ rotation matrix.)
A useful equivalent representation of $ H_{\rm{BM}} $ is given as follows:
\[  A  H_{\rm{BM} } (\alpha, \lambda) A  =  \begin{pmatrix} \lambda C & D(\alpha)^* \\  D(\alpha) & \lambda C \end{pmatrix}, \ \ \ A := \begin{pmatrix} 1 & 0 & 0 \\
0 & \sigma_1 & 0 \\ 0 & 0 & 1 \end{pmatrix} : \mathbb C^4 \to \mathbb C^4 , \]
where (with $ D_{x_j } := (1/i) \partial_{x_j } $)
\[\begin{split} 
D(\alpha) &= \begin{pmatrix}D_{x_1} + i D_{x_2} & \alpha U(\mathbf r) \\ \alpha U(-\mathbf r) & D_{x_1}+i D_{x_2}\end{pmatrix} \text{ and }
C =\begin{pmatrix}0 & V(\mathbf r) \\ V(-\mathbf r) & 0 \end{pmatrix}. 
\end{split}\] 
In most of the figures we use the coordinates $ ( x_1, x_2 ) $ and corresponding
dual coordinates $ k $.

\subsection{Change to the standard lattice $ \mathbb Z + \omega \mathbb Z $}
\label{s:stan}
The potentials $ U $ and $ V $ are periodic with respect to the lattice
$ \Gamma = 4 \pi i ( \mathbb Z  + \omega \mathbb Z ) $ with finer 
twisted periodicity with respect to the moir\'e lattice $\tfrac{1}{3}\Gamma$. 
It is mathematically nicer, especially when dealing with theta functions, to use coordinates
in which the moir\'e lattice is given by $ \Lambda := \mathbb Z + \omega \mathbb Z $. This
corresponds to changing 
the physics coordinates $\mathbf r = (x_1,x_2) \in \mathbb R^2 $ to $ z \in \mathbb C \simeq \mathbb R^2  $
defined by 
\[ x_1 + i x_2=\tfrac{4}{3}\pi i z. \]
This leads to an equivalent Hamiltonian, 
\begin{equation}
\label{eq:defBM} H ( \alpha, \lambda ) := \begin{pmatrix} \lambda C & D ( \alpha)^* \\
D ( \alpha ) & \lambda C \end{pmatrix} : H^1 ( \mathbb C ; \mathbb C^4 ) \to 
L^2 ( \mathbb C ; \mathbb C^4 ),  \ \ \alpha \in \mathbb C, \ \ \lambda \in \mathbb R,  \end{equation}
where (with $ D_{\bar z }  = (1/i) \partial_{\bar z } = (1/i ) ( \partial_{x_1} + i \partial_{x_2} ) $)
\begin{equation}
\label{eq:Hamiltonian}
 D(\alpha) = \begin{pmatrix} 2D_{\bar z }& \alpha U (z ) \\ \alpha U(- z) & 2 D_{\bar z } \end{pmatrix}, \ \ \ \
  C := \begin{pmatrix} 0 & V ( z )  \\
V  ( -z ) & 0 \end{pmatrix} , 
\end{equation}
where the parameter $\alpha$ is proportional to the inverse relative twisting angle. 
With $\omega = e^{2\pi i /3}$ and  $ K := \frac 4 3 \pi $,  we assume that 
\begin{equation}
\label{eq:defU}
U(z  + \gamma ) = e^{ i \langle \gamma, K \rangle } U  (z ), \ 
\gamma \in \Lambda ,  \quad U  (\omega z ) = \omega U  (z ), \ \ 
 \overline{U( \bar z ) } = - U ( - z ), \end{equation}
$ \Lambda := \mathbb Z \oplus \omega \mathbb Z $, and 
 \begin{equation}
 \label{eq:defV}  V (  z ) = V ( \bar z ) = \overline{ V ( - z ) } ,  \ \ V ( \omega z ) = V ( z ) , 
 \ \ V ( z + \gamma ) = e^{ i \langle \gamma, K \rangle } V ( z) . 
 \end{equation}

The specific potentials in $ H_{\rm{BM}} $ are, with $ K = \tfrac43 \pi $,
\begin{equation}
\label{eq:defUV}  U ( z ) =  U_{\rm{BM} } ( z) :=  - \tfrac{4} 3 \pi i \sum_{ \ell = 0 }^2 \omega^\ell e^{ i \langle z , \omega^\ell K \rangle },
\ \ V( z) = V_{\rm{BM}} (z ) := \sum_{ \ell = 0 }^2  e^{ i \langle z , \omega^\ell K \rangle }, 
  \end{equation} 
and these are the potentials used in (most) numerical experiments in the papers cited in the abstract.

\subsection{The chiral limit}
\label{s:chili} 
When we put $ \lambda = 0 $ in \eqref{eq:defBM} (or equivalently in $ H_{\rm {BM}} $) we obtain an operator built from $ D ( \alpha ) $ only and satisfying a chiral symmetry:
\begin{equation}
\label{eq:chiH}  \begin{gathered}
H ( \alpha ) := H ( \alpha, 0 ) = \begin{pmatrix} 0 & D ( \alpha)^* \\
D ( \alpha ) & 0 \end{pmatrix} ,  \\ \begin{pmatrix} -1 & 0 \\
\ 0 & 1 \end{pmatrix}  H( \alpha ) \begin{pmatrix} -1 & 0 \\
\ 0 & 1 \end{pmatrix}   = - H ( \alpha ) . \end{gathered} \end{equation}
In particular, the spectrum of $ H ( \alpha ) $ is symmetric with respect to $ 0$.
The great advantage comes from reducing some properties of $ H ( \alpha )$ to those
of the operator $ D ( \alpha ) $. We will see in \S \ref{s:sym} that $ H ( \alpha ) $ has
a perfect \emph{flat band} at energy zero if and only if
\begin{equation}
\label{eq:exotic}   \Spec_{ L^2 ( \mathbb C/3 \Lambda ; \mathbb C^2 ) } D ( \alpha ) = \mathbb C , \end{equation}
\begin{figure}
\includegraphics[width=15.5cm]{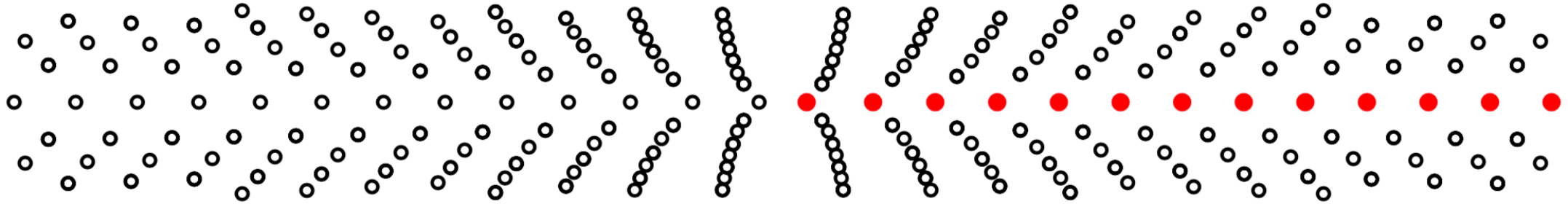}
\caption{The set of $ \alpha$'s for which \eqref{eq:exotic} holds (with the potential 
given by \eqref{eq:defUV}), that is for which the  chiral Hamiltonian has a perfectly flat band
at $ 0$ energy. The regular distribution becomes less apparent when the potential is 
relaxed while all the properties \eqref{eq:defU} are maintained. \label{f:magic}
}
\end{figure}
and that the set of $ \alpha $'s for which this happens is discrete. Outside of 
that discrete set the spectrum on $ L^2 ( \mathbb C/3 \Lambda )$ is given by 
$ \frac13 \Lambda^* $. The domain of $ D ( \alpha ) $ is given by $ H^1 ( \mathbb C/3 \Lambda) $
and it is a Fredholm operator of index $ 0 $. (In \S \ref{s:sym} we will consider a finer space $ L^2_0 ( \mathbb C ; 
\mathbb C^2 )$ which is more suitable for Floquet theory and the study of flat bands; 
the reason for  $ \mathbb C/3 \Lambda $ is periodicity of potentials with respect to 
the lattice $ 3 \Lambda $.) The set,  $ \mathcal A $, of $ \alpha $'s for which 
\eqref{eq:exotic} satisfies the following symmetries (see 
\cite{beta}, \cite[\S 2.3]{bhz2}):
\begin{equation}
\label{eq:symmA}  \mathcal A = - \mathcal A = \bar {\mathcal A} . \end{equation}

Another advantage of the operator $ D ( \alpha ) $ is that scalar valued holomorphic functions
act as scalars:
\[  D ( \alpha ) ( f u ) = f D ( \alpha ) u , \ \ u \in  H^1_{\rm{loc}}  ( \mathbb C ; \mathbb C^2 ), \ \ 
f \in \mathcal O ( \mathbb C ; \mathbb C ).  \]
This was emphasized in \cite{magic} and was a basis of the argument recalled in \S \ref{s:theta} below.

A crucial feature of $ D ( \alpha ) $ is its non normality, $ [ D( \alpha) , D ( \alpha )^* ] \neq 0 $. 
This allows for exotic phenomena such as \eqref{eq:exotic}, which in turn produce exactly flat 
bands appreciated by physicists. As indicated in \cite{beta} it also results in less desirable
features such as exponential squeezing of bands (see \S \ref{s:squeeze}) and spectral 
instability (see Figure \ref{f:inst}). Those effects are exploited in \cite{bovo} where 
small random perturbations produce dramatic changes in spectral behaviour, suggesting
high instability of all but the first magic angle.

The set of (complex) $ \alpha$'s for which \eqref{eq:exotic} holds for the potential \eqref{eq:defUV}
is shown in Figure \ref{f:magic}. Its structure remains a mystery. One striking observation
made in \cite{magic} is the even spacing of real $ \alpha$'s (shown in \red{red} and 
labeled $ 0 <  \alpha_1 < \alpha_2 < \cdots $) roughly given by
\begin{equation}
\label{eq:threeh}
\alpha_{ j+1 } - \alpha_j \simeq \tfrac32. 
\end{equation}
(A more accurate computation based on the spectral characterization \cite{beta} -- 
see Theorem \ref{t:spec} -- 
suggests the spacing $ \simeq 1.515 $). 

\begin{op}
\label{1} For $ U $ given in \eqref{eq:defUV} establish an asymptotic
quantization rule \eqref{eq:threeh}. At the moment, there are no convincing arguments.
A more general question is obtaining asymptotics of real $ \alpha$'s for more general
potentials satisfying \eqref{eq:defU}. In that case, a simple law similar to \eqref{eq:threeh}
is harder to observe -- see the movie linked to Figure \ref{fig:degeneracy}. See 
also \S \ref{s:scal} and \S \ref{s:semi} for discussions of related issues.
\end{op}

\section{Basic symmetries and band theory of TBG}
\label{s:sym}

The translation symmetry of BMH are given as follows:  for $ u \in L^2_{\rm{loc}} ( \mathbb C ; \mathbb C^2 ) $ we define
\begin{equation}
\label{eq:defLag}  
L_{\gamma } u (z )  :=  
\begin{pmatrix} e^{ i \langle \gamma, K \rangle }   & 0  \\
0 & e^{-  i \langle \gamma , K \rangle } 
\end{pmatrix}  u ( z + \gamma )   ,   \ \ \ 
\gamma \in \Lambda,  \ \  K = \tfrac43 \pi .  \end{equation}
We extend this action diagonally for $ w \in L^2_{\rm{loc}} ( \mathbb C ; \mathbb C^4 ) $:
\[  \mathscr L_\gamma w = \begin{pmatrix} L_\gamma w_1 \\L_\gamma w_2 \end{pmatrix}, \ \ \ 
w = \begin{pmatrix} w_1 \\ w_2 \end{pmatrix}, \ \ w_j \in L^2_{\rm{loc} } ( \mathbb C; \mathbb C^2 ). \]
We then have, in the notation of \eqref{eq:defBM}, \eqref{eq:Hamiltonian} with $ U $, $ V$
satisfying \eqref{eq:defU} and \eqref{eq:defV}, 
\begin{equation}
\label{eq:transl}  L_\gamma D ( \alpha ) = D ( \alpha ) L_\gamma , \ \ \ 
 \mathscr L_\gamma H ( \alpha, \lambda  ) = H ( \alpha , \lambda ) \mathscr L_\gamma . \end{equation}

We also define the pull back of the rotation by $2 \pi /3$:
\begin{equation}
\label{eq:OC} 
\begin{gathered} 
 \Omega : L^2_{\rm{loc}} ( \mathbb C; \mathbb C^2) \to L^2_{\rm{loc}} ( \mathbb C; \mathbb C^2) , \ \ \ 
 \mathscr C :  L^2_{\rm{loc}} ( \mathbb C; \mathbb C^4) \to L^2_{\rm{loc}} ( \mathbb C; \mathbb C^4) ,\\
 \Omega u ( z ) := u ( \omega z ) , \ \  \mathscr C \begin{pmatrix} w_1 \\ w_2 \end{pmatrix} :=
 \begin{pmatrix} \Omega w_1 \\ \bar \omega \Omega w_2 \end{pmatrix}. \end{gathered} \end{equation} 
This gives 
\begin{equation}
\label{eq:rota}
 \Omega D( \alpha ) = \omega D ( \alpha ) , \ \ \  \mathscr C H( \alpha, \lambda ) = 
 H ( \alpha, \lambda ) \mathscr C . \end{equation}
 
 The natural subspaces of $ L^2_{\rm{loc}} ( \mathbb C ; \mathbb C^p ) $, $ p = 2, 4 $, are given by
 \begin{equation}
 \label{eq:Lk} 
 \begin{gathered}
 L^2_k ( \mathbb C ; \mathbb C^2 ) := \{ u \in L^2_{\rm{loc} } ( \mathbb C, \mathbb C^ 2) : 
 L_\gamma u = e^{ i \langle k , \gamma\rangle } u  \}, \ \  \| u \|_{ L^2_k } = \int_{\mathbb C/\Lambda }
 |u ( z )|^2 dm ( z ) , 
  \end{gathered}
 \end{equation}
 and similarly for $ p = 4 $ with $ L_\gamma$ replaced by $ \mathscr L_\gamma $.
  We also define Sobolev spaces $ H^s_{k} := L^2_k \cap H_{\rm{loc}}^s $. With $ s = 1 $
 they can be used as domains of our operators. 

These spaces depend only on the congruence class of $ k $ in $ \mathbb C/\Lambda^* $, 
\begin{equation}
\label{eq:defz} \Lambda^* := \frac{ 4 \pi i }{\sqrt 3 } \Lambda , \ \   k \mapsto z ( k ) :=\frac{ \sqrt{3}k }{ 4 \pi i } ,
\  \ \Lambda^* \to \Lambda, \ \  \langle p , \gamma \rangle \in 2 \pi \mathbb Z,  \ \ p\in \Lambda^*, \ 
\gamma \in \Lambda . \end{equation}

The points of high symmetry, $ \mathcal K $, are defined by demanding that
\[  p \in \mathcal K  \ \Longrightarrow \ \omega p \equiv p \! \mod \! \Lambda^* .\]
They are given by 
\begin{equation}
\label{eq:defK}  \mathcal K = \{ K, - K , 0 \} + \Lambda^* , \ \ \  K = \tfrac 4 3 \pi . 
\end{equation}
Mathematically, these are the fixed points of the action of $ z \mapsto \omega z $ on 
$ \mathbb C/\Lambda^* $. Physically, $ \pm K $ are called the K-points at which 
Dirac points are present (see \S \ref{s:theta}) and $ 0 $ is called a $ \Gamma$-point --
see Figure~\ref{f:2}.
(A different choice of $ L_\gamma $ in \eqref{eq:defLag}  can result in different 
sets of $ K$-points -- see \cite[\S 2]{bz1}.)

For $ k \in \mathcal K/\Lambda^* $ and $ p \in \mathbb Z_3 $ we also define
\begin{equation}
\label{eq:Lkp} 
L_{k,p}^2 ( \mathbb C ; \mathbb C^4 ) := \{ 
u \in L^2_{k} ( \mathbb C; \mathbb C^4 ) :  \mathscr C^p u = \bar \omega^p u \} , 
\end{equation}
with the definition of $ L_{k,p}^2 ( \mathbb C ; \mathbb C^2 )$ obtained by replacing $\mathscr C $
by $ \Omega $. We have orthogonal decompositions $ L^2_k = \bigoplus_{ p \in \mathbb Z_3}
L_{k,p} $, $ k \in \mathcal K/\Lambda^*$. Also, the actions of
$ \mathscr L_\gamma $ and $ \mathscr C $ on $ L^2_{p,k} $ commute. In general, 
 $ \mathscr L_\gamma \mathscr C = \mathscr C \mathscr L_{\omega \gamma}$
 and the group generated by the action $ \mathscr L_\gamma $ and $ \mathscr C $
(or the actions of $ L_\gamma $ and $ \mathscr C $) is a discrete Heisenberg group -- 
see \S \cite[\S 2.1]{beta}. These spaces play an important role in the study of protected states, 
multiplicities and trace formulas for magic angles.

\subsection{Bloch--Floquet theory}
\label{s:blfl} 
The ``twisted" translations $ \mathscr L_\gamma $ can be used to define 
a Bloch transform
\[    \mathcal  B u ( k,z ) := | \mathbb C / \Lambda^* |^{-\frac12} \sum_{ \gamma \in \Lambda } e^{ -  i \langle  z + \gamma, k \rangle}   \mathscr L_\gamma  u ( z ) ,  \ \ \ u \in \mathscr S ( \mathbb C ) .
\]
We then easily check that 
\[ \begin{gathered}  \mathcal B u ( k + p ,  z ) =  
e^{- i \langle z , p \rangle}\mathcal B u  ( k , z )  , \ \ p \in \Lambda^*, \\ 
\mathscr L_{\alpha  } \mathcal  B u ( k , \bullet) = | \mathbb C / \Lambda^* |^{-\frac12} 
\sum_{\gamma} e^{ -  i \langle z + \alpha + \gamma , k \rangle } \mathscr L_{\alpha + \gamma } 
u ( z ) =  \mathcal B u (k ,  \bullet ) , \ \ \alpha  \in \Lambda
\end{gathered} \] 
We can check that for $ u \in \mathscr S ( \mathbb C ) $, 
\[  \begin{split} \int_{ \mathbb C/\Lambda } \int_{\mathbb C/\Lambda^* } 
| \mathcal B u ( k, z ) |^2 dm ( z ) d m ( k ) & = 
\int_{ \mathbb C } | u ( z ) |^2 dm ( z ) , 
\end{split} 
\]
and that
\[ \mathcal C \mathcal B u (z ) = u ( z ) , \ \ \ 
\mathcal C u ( z )   := | \mathbb C / \Lambda^* |^{-\frac12} \int_{ \mathbb C/ \Lambda^* } v ( z, k ) e^{  i \langle z, k \rangle} dm ( k ). \]
This shows that $ \mathcal B $ extends to a unitary map $ \mathcal B : L^2 ( \mathbb C ; \mathbb C^4 ) \to \mathscr H$, where 
\[  \mathscr H := 
\{ v ( k , z ) \in L^2_{\rm{loc}} ( \mathbb C ; L^2_0 ( \mathbb C ; \mathbb C^4 ) ), \ \ 
v ( k + p , z ) = e^{ - i \langle z , p \rangle } v ( k , z ), \ p \in \Lambda^* \} . \]
We then define 
\begin{equation} \begin{gathered} 
\label{eq:defHk} 
H_k ( \alpha , \lambda ) : \mathscr D \to \mathscr H, \ \ \mathscr D := \mathscr H \cap L^2_{\rm{loc}} ( \mathbb C_k ; H^1_0 ( \mathbb C, \mathbb C^4 ) ),  \\
H_k ( \alpha , \lambda ) := e^{ -i \langle z, k \rangle} H ( \alpha, \lambda ) e^{ i \langle z, k \rangle } = 
\begin{pmatrix} \lambda C & D(\alpha) ^* + \bar k \\
D( \alpha) +k & \lambda C \end{pmatrix},  \\
[ H_k ( \alpha , \lambda ) \mathcal B u ] ( k , z ) = [ \mathcal B H ( \alpha , \lambda ) u ] ( k , z ) .
\end{gathered} \end{equation}
We see that 
$ \Spec_{ L^2 _0 } ( H_k ( \alpha, \lambda ) ) $ (with the domain given by $H^1 _0 $) 
is discrete and 
\[ \Spec_{ L^2 ( \mathbb C ; \mathbb C^{4} ) } ( H ( \alpha , \lambda ) ) = \bigcup_{ k \in \mathbb C/\Lambda^* } 
\Spec_{ L^2 _0 } H_k ( \alpha , \lambda ) . \]

 The Hamiltonian \eqref{eq:defBM} possesses other important symmetries called the \emph{parity-inversion/time-reversal} 
symmetry, the \emph{particle-hole} symmetry and the \emph{mirror} symmetry -- see \cite[\S 2.2]{bz2} for
a concise review. One consequence of the symmetries is the existence  and properties 
of protected states:

\begin{theo}[\cite{phys,beta}]
\label{t:prot}
For the Hamiltonian \eqref{eq:defHk} with $ U $ and $ V $ satisfying \eqref{eq:defU},\eqref{eq:defV}
and $ \alpha, \lambda \in \mathbb R $, 
\begin{equation}
\label{eq:prot}
\dim \ker_{ H^1_0 } H_{\pm K } ( \alpha, \lambda ) \geq 2 , \ \  K = \tfrac43 \pi , 
\end{equation}
In addition, for $ \alpha \in \mathbb C $
\begin{equation}
\label{eq:prot1}
\dim \ker_{ H^1_0 } ( D ( \alpha) \pm K ) \geq 1 . 
\end{equation}
Moreover we can find a holomorphic function  
\[ \mathbb C \ni \alpha \to u_{\pm K } ( \alpha ) 
\in (C^\infty \cap L^2_0) ( \mathbb C; \mathbb C^2 ) \setminus \{ 0 \}  , \]
such that
\begin{equation}
\label{eq:uKK}
\begin{gathered}
( D (\alpha ) \pm K  ) u_{ \pm K } ( \alpha ) = 0 , \ \ \ \ 
u_{-K } ( \alpha ) = \tau ( K ) \mathscr E \tau (K ) u_K ( \alpha ) , 
\\ \tau( K ) u_K ( 0 ) =  \begin{pmatrix}
1 \\ 0 \end{pmatrix} , \ \ \ \  \tau ( \pm K ) u_{\pm K } ( \alpha) \in L^2_{\pm K , 0 } ,  \\ 
\mathscr E \begin{pmatrix} u_1 ( z ) \\ u_2 ( z ) \end{pmatrix} := \begin{pmatrix}  \ u_2 ( - z ) \\
- u_1 ( -z ) \end{pmatrix},  \ \ \  \ \tau (k ) u ( z ) := e^{ i \langle z, k \rangle } u ( z).
\end{gathered}
\end{equation}
\end{theo}
This was essentially established in \cite{phys,beta} but for a streamlined proof of \eqref{eq:prot} see \cite[Proposition 2]{bz2}, and 
for the proofs of \eqref{eq:prot1} and \eqref{eq:uKK},  \cite[Propositions 2.2, 2.3]{bhz2}, respectively.
An alternative proof of \eqref{eq:prot1} which does not involve $ H ( \alpha, 0 ) $ is presented 
in \cite{BHWY}. 

\begin{op}
\label{2}  
Do upper and lower bands for the Bistritzer--MacDonald Hamiltonian 
have conic singularities at $ \pm K $ for all real values of $ \alpha $ and $ \lambda $? 
That would mean that $ \pm K $ are Dirac points:
\begin{center}
\includegraphics[width=8cm]{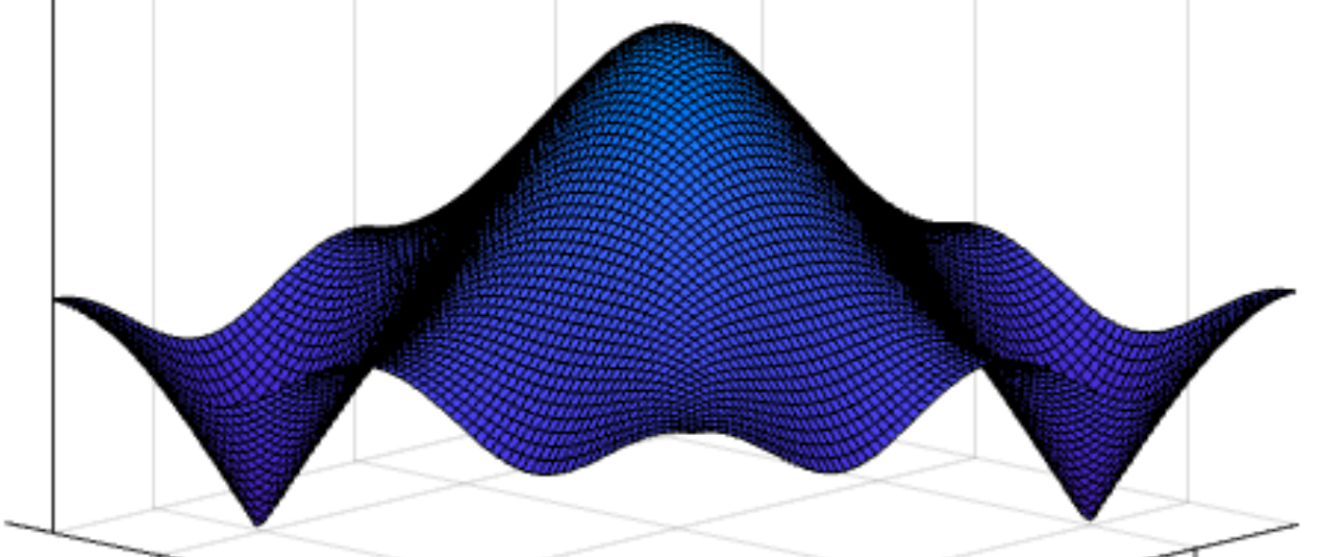}
\end{center}
\end{op}

\subsection{Flat bands in the chiral limit} 
\label{s:flat}
The first advantage of the chiral model \eqref{eq:chiH}
is that the spectrum of $ H_k ( \alpha) := H_k ( \alpha, 0 ) $ is symmetric with respect to $ 0 $
(that is not true in the case of BMH -- see \S \ref{s:chi2BMH}). In view of \eqref{eq:prot1} 
we know that that two bands always touch at $ 0 $. Hence it is natural to label the 
spectrum of $ H_k ( \alpha ) $ as follows:
\begin{equation}
\label{eq:specHk} 
\begin{gathered} \Spec_{L^2_0}  H_k ( \alpha ) = \{ E_\ell (\alpha, k ) \}_{ \ell \in \mathbb Z \setminus 0 }, 
\ \ \ E_{ \ell + 1 } ( \alpha, k ) \geq E_\ell ( \alpha, k ) , 
\\ 
E_\ell ( \alpha, k ) = - E_{-\ell } ( \alpha, k ) , \ \ \ \  E_{\pm 1} ( \alpha, \pm K ) = 0 , \text{ for all $ \alpha \in \mathbb C$.} \end{gathered}
\end{equation}
We note that $ E_\ell ( \alpha, k ) $, $ \ell \geq 1 $, are the ordered sequence of
the  singular values of the non-self-adjoint operator $ D ( \alpha ) + k $. 

A flat band at zero energy occurs at a given value of the parameter $ \alpha $ if 
$  E_1 ( \alpha, k ) = 0 $ for all $ k \in \mathbb C $.
We recall that in the BMH, $ 1/\alpha $ is proportional to the angle of twisting of the two sheets of
graphene. 
For a specific potential $ U $ satisfying \eqref{eq:defU} the magic $ \alpha $ (that is {\em magic angles})
and their multiplicities were defined as follows in \cite{bhz3}:

\noindent
{\bf Definition} (Magic angles and their multiplicities). {\em A value of $ \alpha $ in \eqref{eq:Hamiltonian} is called magical if 
$ H ( \alpha ) $ has a flat band at zero
\begin{equation} 
\label{eq:flat} E_1 ( \alpha, k ) \equiv 0 , \ \ k \in \mathbb C . 
\end{equation}
The set of magic $ \alpha$'s is denoted by $ \mathcal A $ or $ \mathcal A (U ) $
if we specify the dependence on the potential. 
The multiplicity of a magic $ \alpha $ is defined as 
\begin{equation} 
\label{eq:defm} m ( \alpha ) = m_U ( \alpha ) := \min \{ j > 0 : \max_k  E_{j+1} ( \alpha, k ) > 0  \}. 
\end{equation}
Magic angles are (up to physical constants) reciprocals of $ \alpha \in \mathcal A $.}

\begin{figure}
\includegraphics[width=7.65cm]{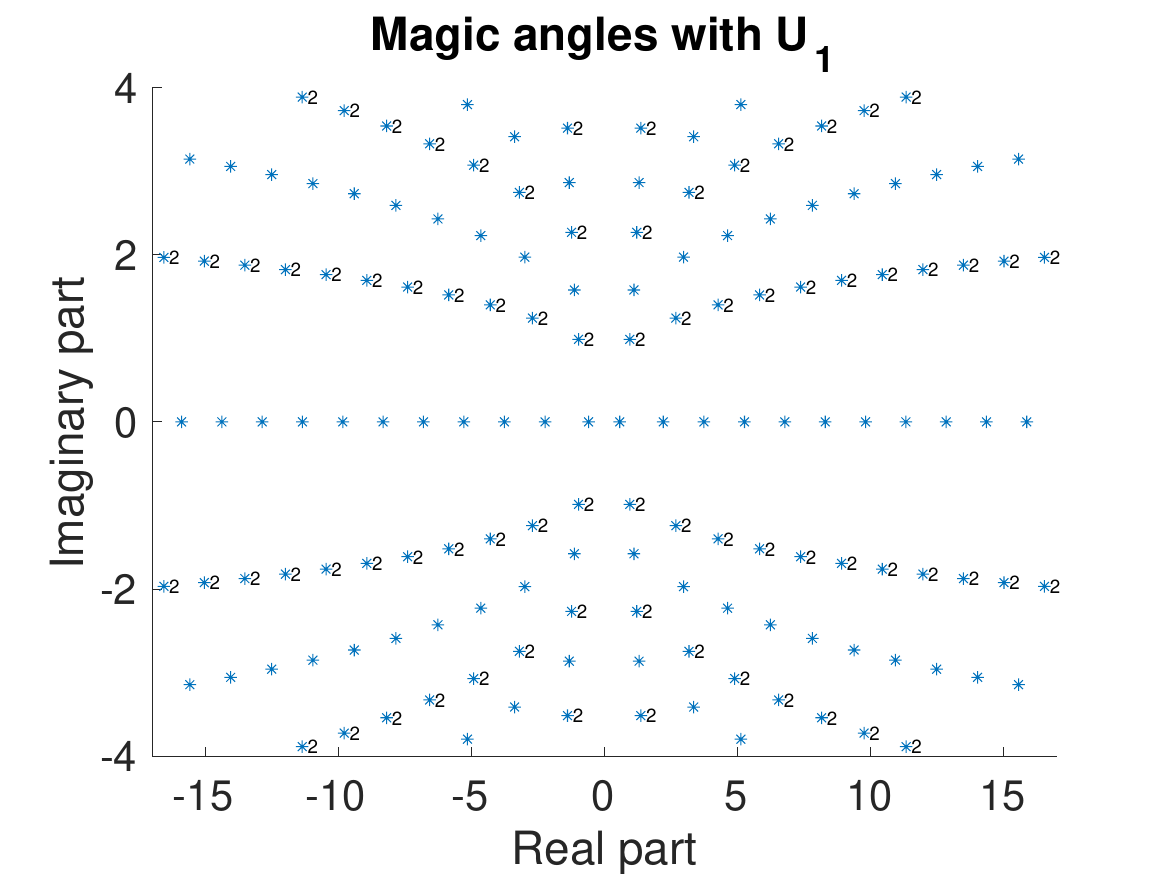}
\includegraphics[width=7.65cm]{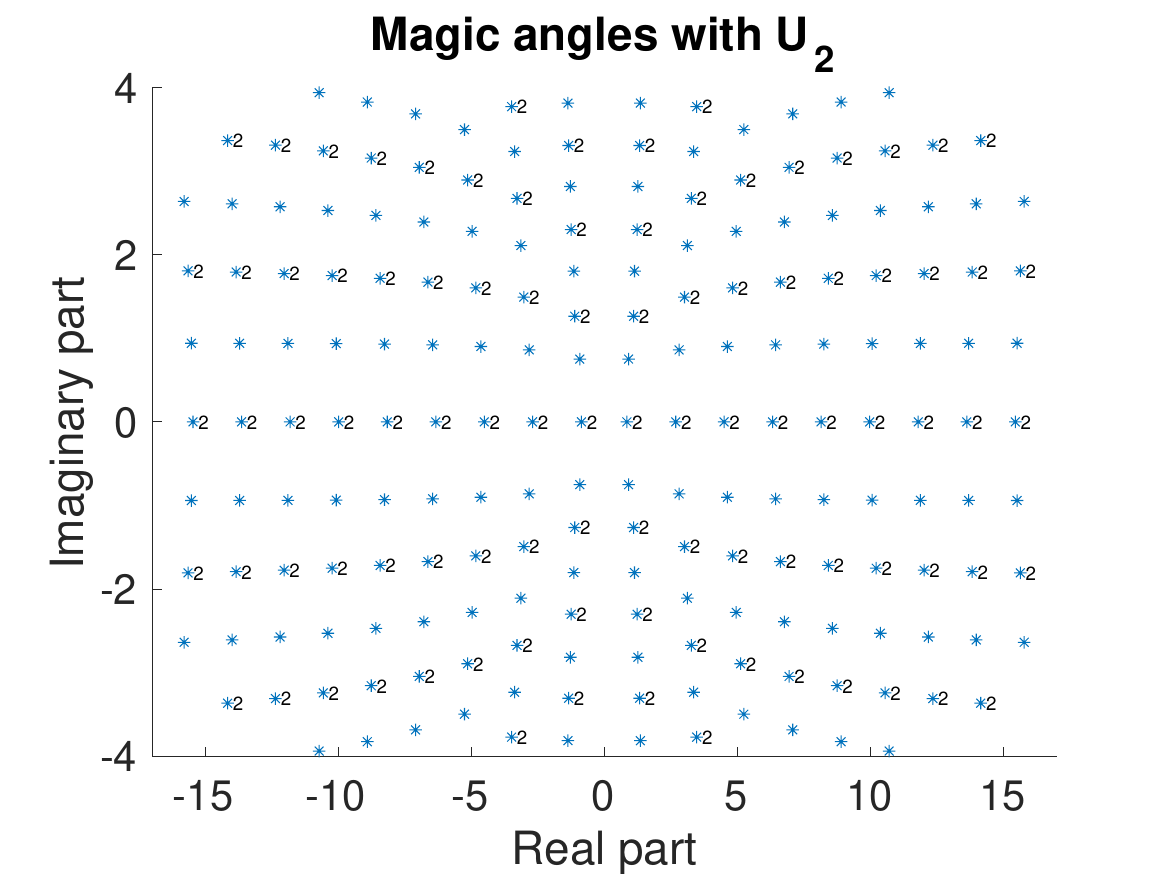}
\caption{Magic angles $\alpha$ for $U_1 ( z ) = U_{\rm{BM}}( z ) $ given in \eqref{eq:defUV} (left) 
 and $U_2 ( z )  = ( U_{\rm{BM}} ( z)  - U_{\rm{BM}} ( -2 z ) )/\sqrt 2 $ (right).
Multiplicity of the flat bands (no number $\rightarrow$ simple magic angle, 2 $\rightarrow$ two-fold degenerate magic angle) in the figure. The movie
\url{https://math.berkeley.edu/~zworski/Interpolation.mp4} shows 
the magic angles for interpolation between these potentials:
$ U ( z ) = ( \cos \theta - \sin \theta )  U_1 ( z ) + \sin \theta U_2  $;
multiplicity one magic angles are coded by $ \red{*} $ and multiplicity two by $ \blue{*} $. }
\label{fig:degeneracy}
\end{figure}

A numerical illustration of the sets $ \mathcal A $ for different potentials satisfying 
\eqref{eq:defU} is shown in Figure~\ref{fig:degeneracy}. Multiplicities are indicated 
there and in the linked animation. The computation was done based on the spectral 
characterization described in the next section. The protected nature of multiplicities
one and two will be reviewed in \S \ref{s:exi}. 

Although the proof relies on the material presented in \S \ref{s:spec} we recall here
a result stating that if $ E_1 ( \alpha, k ) $ touches $ 0 $ at some $ k $ away from the K-points
then the band has to be perfectly flat:

\begin{theo}[\cite{beta}, \cite{bhz2}]
\label{t:bands} 
For any $ U $ satisfying \eqref{eq:defU} and $ \alpha \in \mathbb C $, 
\begin{equation}  
\label{eq:alphac1}  
\exists \, \ k \notin \{- K, K  \} + \Lambda^*   \ \ \  E_1 ( \alpha, k ) =  0 
\ \Longrightarrow \ \forall \, \ k \in \mathbb C  \ \ \  E_1 ( \alpha, k ) =  0 
. \end{equation}
\end{theo}

For the Bistritzer--MacDonald Hamiltonian \eqref{eq:defBM} perfectly flat bands are not expected.
That the antichiral model $ H ( 0 , \lambda ) $ cannot have flat bands was shown in \cite{phys}. 

A perfectly flat band at $ 0 $ energy for a periodic Hamiltonian corresponds to an eigenvalue of infinite 
multiplicity at $ 0 $ for the Hamiltonian acting on $ L^2 $ (in our case $ L^2 ( \mathbb C; \mathbb C^4 ) $
with the domain given by $ H^1 ( \mathbb C; \mathbb C^4 ) $). Physical properties, such as 
superconductivity, are then related to the decay of the corresponding eigenfunctions.
That in turn is related to the {\em topology of the flat band} -- see \cite[\S 8.5]{notes} and references
given there. Trivial topology gives exponential decay while nontrivial topology forces the blow
up of moments of the probability distribution of the Wannier functions \cite[Theorem 9]{notes}.
We will discuss the topology of flat bands for TBG in \S \ref{s:top}.

\begin{op}
\label{3} Show that the Hamiltonian \eqref{eq:defBM}, $ H ( \alpha, \lambda ) $,
with $ U $ and $ V \not \equiv 0 $ satisfying \eqref{eq:defU},\eqref{eq:defV},  
cannot have flat bands when $ \lambda \neq 0 $. (Or give a counterexample to this claim.)
\end{op}

\begin{op}
\label{4}  Numerics indicate (see \cite[Figure 2]{bhz2}) that 
for $ U = U_{\rm{ B M}} $, 
$ k \mapsto E_1 ( \alpha, k) / [ \max_{ p \in \mathbb C } E_1 ( \alpha, p )] $ 
does not vary much with $ \alpha $, in particular in neighbourhoods of $ \alpha \in \mathcal A $,
and  its graph is close to that of $ k \mapsto | U_{\rm{BM}} ( z ( k ) ) | $, where $ z  : \Lambda^* \to 
\Lambda$, see \eqref{eq:defz}. What is the explanation of this phenomenon? For an animation 
of rescaled bands see {\rm{\url{https://math.berkeley.edu/~zworski/KKmovie.mp4}.}}
\end{op}

\section{BMH as a perturbation of the
  chiral model} 
  \label{s:chi2BMH}

The Bistritzer--MacDonald Hamiltonian (BMH) \eqref{eq:defBM} could, for 
small values of the coupling constant $ \lambda $, be considered as a perturbation of
the chiral model. The actual physical value of $ \lambda $ (see \cite{BM11,magic})
is approximately given by $ \lambda = 0.7 \alpha $. 

The simplest case to consider is of 
$ \alpha \in \mathcal A $ which is positive and simple (which, in the case
of the potential in \eqref{eq:defUV}  we know 
rigorously for the smallest magic $ \alpha $ and numerically for other real $ \alpha$'s -- see 
\S \ref{s:exi}).
Then, in the notation of \eqref{eq:specHk} , 
\[    E_{-2} ( \alpha , k ) < E_{-1} ( \alpha, k ) = 0 = E_1 ( \alpha, k ) <   E_2 ( \alpha, k )  , 
\ \text{ for all $ k $.} \]
This means that for  $ | \lambda | \ll 1 $ in \eqref{eq:specHk}, the bands 
$ E_{\pm 1} ( \alpha, \lambda, k ) $ are well defined. 

A standard application of perturbation theory (see \S \ref{s:dynip}), the symmetries of 
$ D ( \alpha) $ and $ H ( \alpha, \lambda ) $, and of some basic properties
of theta functions (see \eqref{eq:defFk},\eqref{eq:ukz},\eqref{eq:u*kz} below) gives the
following simple, but to us, surprising result:
\begin{theo}[\cite{bz2}]
  \label{t:pert} Suppose that $ \alpha \in \mathcal A \cap \mathbb R $ is simple and 
that $ k \mapsto E_{\pm 1 } ( \alpha, \lambda , k ) $ are the two lowest bands (in 
absolute value) of BMH in 
\eqref{eq:defBM}. Then there
exist  $e(\alpha,\bullet),f(\alpha,\bullet)\in C^{\infty}(\mathbb C/\Lambda^*)$ such that
\begin{equation}
\label{eq:Epm1}  E_{\pm 1 } ( \alpha, \lambda , k ) = e ( \alpha, k ) \lambda 
\pm |f ( \alpha, k )| \lambda^2 + 
 \mathcal O ( \lambda^3 ) , \ \ 
\lambda \to 0 , \end{equation}
$ f ( \pm K ) = 0 $, ($ \omega K \equiv K \!\! \! \mod \! \Lambda^*$, $ K \neq 0 $), and 
\begin{equation}
\label{eq:prope}   e (\alpha,   k ) = -e ( \alpha,  - k ) = - e( \alpha, \bar k ) = e(\alpha,  \omega k ) , \ \ \omega = e^{ { 2 \pi i } /3} .  
\end{equation}
\end{theo}

\begin{figure}
\includegraphics[width=16cm]{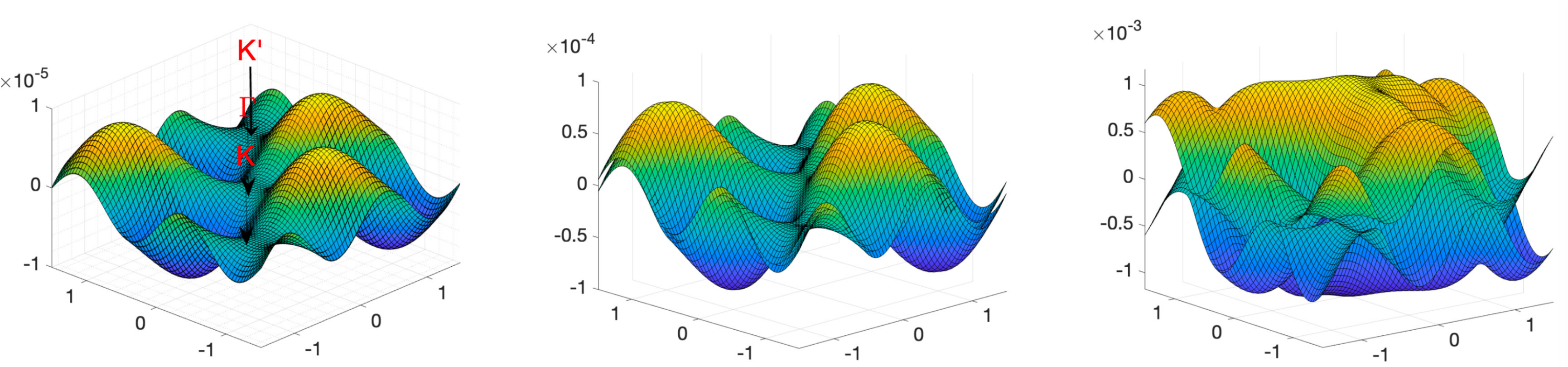}
\caption{\label{f:2} Plots of $ k \mapsto E_{\pm 1} ( \alpha, \lambda, k ) $ for 
$ \alpha $ the first real magic element of $ \mathcal A $ and 
$ \lambda =  10^{-3}, 10^{-2}, 10^{-1 } $. We see that for very small coupling the flat bands
``move together" and split only when the coupling gets larger; the quadratic term controls
the splitting of the bands, see Figure~\ref{f:1}. For an animated version see 
\url{https://math.berkeley.edu/~zworski/Chiral2BM.mp4}.
}
\end{figure}

The surprising fact is that the leading linear term (for very small $ \lambda $) does not depend
on the band: when $ \lambda $ is switched on the two bands initially move together -- see Figure~\ref{f:2}. However, $ | e ( \alpha , k ) | \ll | f ( \alpha, k ) | $ 
(except at the crossing points $ k = \pm K $) and hence the quadratic term quickly dominates
and is responsible for the splitting of the bands -- see \cite[Figure 2]{bz2}. For the first magic
$ \alpha $ (and the potential in \eqref{eq:defUV}), the quadratic approximation provides an 
accurate description of the bands when $ \lambda = 0.7 \alpha $ (the physical $ \lambda $). 
For a discussion of the splitting of bands in the case of double $ \alpha $'s see \cite[\S 5]{bz2}.

\begin{op}
\label{9}  Show that $ |f ( \alpha,  \pm K + \zeta )| \sim |\zeta|$
which is equivalent to showing that the Jacobian does not vanish: $ | \partial_k f ( \alpha, \pm K ) |^2 - | \partial_{\bar k } f ( \alpha, \pm K ) |^2  \neq 0 $.
This is a simpler (infinitesimal) version of Problem 2 at a magic angle.
\end{op}

\section{Spectral characterization of magic angles}
\label{s:spec}

In \S \ref{s:sym} we gave the definition of $ \mathcal A \subset \mathbb C $, the set of magic parameters $ \alpha $ (corresponding to the reciprocals of magic angles). The purpose of this section is
to give a general argument \cite{gaz} for the discreteness of $ \mathcal A $ which relies only on 
holomorphy of $ \alpha \mapsto D ( \alpha ) $, Fredholm properties, and existence of protected
states. In the case of operators appearing in \cite{magic,beta,BHWY,yang} it also characterizes
magic angles as eigenvalues of a compact operator, which in turn allows their accurate
numerical computation (see Figure \ref{fig:degeneracy}). 

We replace the operator $ D( \alpha ) + k $ by a family of operators acting between Banach spaces
$ X $ and $ Y $. We let $ \Omega \subset \mathbb C $ be an open set and assume that 
for $ ( \alpha, k ) \in \Omega \times \mathbb C $
\begin{equation}
\label{eq:defQ}  
\begin{gathered} Q ( \alpha, k ) : X \to Y,  \  \text{ is a holomorphic family of
Fredholm operators of index $ 0 $,} \\
\tau_Y ( p ) Q ( \alpha, k ) \tau_X ( p )^{-1} = Q ( \alpha, k + p ) , \ \ k \in \mathbb C , \ \ p \in \Lambda^* , 
\end{gathered}
\end{equation}
where 
the maps $ \tau_\bullet ( p ) : \bullet  \to \bullet $, $ \bullet = X, Y $, are invertible bounded linear maps, 
and $ \Lambda^* $ is a lattice in $ \mathbb C $. (The last condition can be significantly weakened but
we leave in the form relevant to period problems.) 

We have the following for dichotomy: for a fixed $ \alpha \in \Omega $
\begin{gather}
\label{eq:mer} k \mapsto Q ( \alpha , k )^{-1} \text{ is a meromorphic for $ k \in \mathbb C $ with poles of finite rank} \\
\nonumber \text{ or } \\
\label{eq:ker} \ker_X Q  ( \alpha, k ) \neq \{ 0 \} \text{ for all $ k \in \mathbb C $. } 
\end{gather}
(See \cite{gaz} and also \cite[Appendix C]{res} for a brief introduction to Fredholm theory and families of 
meromorphic operators.)

We now define multiplicity 
as follows:  if \eqref{eq:mer} holds then 
 \begin{equation}
\label{eq:mult}
m (\alpha, k ) := \frac{1}{ 2 \pi i } \tr \oint_{\partial D }   Q( \alpha, \zeta ) ^{-1} 
\partial_\zeta Q ( \alpha , \zeta ) d \zeta , 
\end{equation} 
where the integral is over the positively oriented boundary of a disc 
$ D $ which contains $ k$ as the
only possible pole of $ \zeta \mapsto Q ( \alpha, \zeta ) $. 
Otherwise, that is when \eqref{eq:ker} holds. we  put $ m ( \alpha, k ) =  
\infty $ for all $k\in \mathbb{C}$. 

Although seemingly very general and abstract, this definition is necessary in natural 
examples as will be indicated in \S\S \ref{s:schar},\ref{s:scal}. 

\begin{theo}[\cite{gaz}]
\label{t:gaz}  Suppose that \eqref{eq:defQ} holds and that 
for some  $ \alpha_0 \in \Omega  $ and every $ k \in \mathbb C $, we
have,  
\begin{equation}
\label{eq:hyp} 
\begin{gathered}
m ( \alpha, k  ) \geq m ( \alpha_0, k ) \neq \infty .
\end{gathered} 
\end{equation}
Then there exists a discrete set $ \mathcal A\subset \Omega  $ such that for all $ k \in \mathbb C $
\begin{equation}
\label{eq:magic}   \begin{gathered}
m ( \alpha, k ) = \left\{ \begin{array}{ll} 
\ \ \infty &  \alpha \in \mathcal A, \\
m( \alpha_0, k ) &  \alpha \notin \mathcal A . \end{array} \right. 
\end{gathered}
\end{equation}
\end{theo} 

We illustrate the theorem with some simple examples.

\noindent
{\bf Examples.} 1. Consider 
\[  Q ( \alpha , k ) = e^{ i x } D_x + ( \alpha - \tfrac12 )  e^{ix } + k , \ \  x \in \mathbb R/ 2 \pi \mathbb Z ,  \ \ \
D_x := (1/i) \partial_x.  \]
Then, in the notation of  Theorem \ref{t:gaz}, 
$ X = L^2 (  \mathbb R/ 2 \pi \mathbb Z )$, $ Y =  H^1( \mathbb R/ 2 \pi \mathbb Z )$
and 
\begin{equation}
\label{eq:see}  m ( k , 0 ) \equiv 0 ,  \ \  \Lambda^* = 2 \pi \mathbb Z , \ \ \mathcal A = \mathbb Z + \tfrac12.  \end{equation} 
In this case we do not have the second condition in \eqref{eq:defQ} but the proof in \cite{gaz}
still applies as $ m ( k , 0 ) \equiv 0 $. A direct elementary verification is of course much simpler.
This is a special case of the class of one dimensional examples constructed by Seeley \cite{see} 
to show pathological properties of non-normal operators.

\noindent 2.  We can consider $ Q ( \alpha, k  ) = D ( \alpha ) + k  $ given in \eqref{eq:Hamiltonian} with 
$ U $ satisfying \eqref{eq:defU}. In \cite{beta} we took 
\[  X= L^2 ( \mathbb C/ 3 \Lambda ; \mathbb C^2 ) , \ \ \ Y = H^1 ( \mathbb
C / 3\Lambda ; \mathbb C^2 ) . \]
In that case the assumptions were satisfied by 
\[   m ( 0 , k ) = 2 \indic_{\frac13 \Lambda^* } (k ) ,  \ \ \tau(p) u ( z ) := e^{ i \langle p, z \rangle}
u ( z ) . \]

\noindent 3. In \cite{bhz2} we took the point of view closer to the physics literature and had $ D ( \alpha ) $ act 
on
\[X  = L^2_0 ( \mathbb C; \mathbb C^2 ) , \ \ \ Y = H^1_0 ( \mathbb C;
\mathbb C^2 ) , \]
where the spaces were defined in \eqref{eq:Lk}. 
so that
\[ m ( 0, k ) = \indic_{\mathcal K_0} ( k), \ \ \mathcal K_0 := \{ K, - K \} + \Lambda^* , \ \ \tau(p) u ( z ) := e^{ i \langle p, z \rangle}
u ( z ) . \]
(The protected states were reviewed in Theorem \ref{t:prot}.) 
The sets $ \mathcal A $ are the same in both cases. However, there are multiplicity issues
illustrated in \cite[Figure 4]{bhz2}. 

More interesting examples, in which $ m ( \alpha_0 , k ) > \dim \ker Q ( \alpha_0, k ) $, will be given the next two sections. 

\subsection{Spectral characterization}
\label{s:schar}

For operators appearing in TBG (see the examples above) but also in the study of multilayer graphene -- see 
\cite{BHWY},\cite{yang} and references given there -- the structure of operators $ Q ( \alpha, k ) $
in Theorem \ref{t:gaz} is more special. 

A natural generalization of $ D ( \alpha ) $ in \eqref{eq:Hamiltonian} is given as follows
 \begin{equation}
 \label{eq:defD} \begin{gathered} D ( \alpha ) := 2 D_{\bar z} \otimes I_{\mathbb C^n}  + 
 W ( z )  + \alpha V ( z ) : H^1_{\rm{loc} } ( \mathbb C ; \mathbb C^n ) 
\to L^2_{\rm{loc}} ( \mathbb C ; \mathbb C^n ) , \\
H( \alpha ) := \begin{pmatrix} \ \ 0 & D ( \alpha)^* \\
D ( \alpha ) & 0 \end{pmatrix} ,  \end{gathered} \end{equation}
where $ V ( z ), W ( z ) \in  C^\infty ( \mathbb C ; \mathbb C^n \otimes \mathbb C^n ) $. Here
$ 2 D_{\bar z } := \partial_{x_1} + i \partial_{x_2} $, $ z = x_1 + i x_2 $,  and we 
will write $ 2 D_{\bar z } $ for the diagonal action on $ \mathbb C^n $-valued functions.

In \eqref{eq:Hamiltonian} we had $ n = 2 $ and $ W = 0 $ but the presence of $ W $ is needed for other models. Mathematically,
having that term seems essential when $ n > 3 $ is considered as it helps in controlling the 
number of protected states, see \eqref{eq:Prho} below. We could consider an even  more general 
case of $ W ( z ) + \alpha V (z ) $ replaced by $ V ( \alpha, z ) $. 

Let 
\[ \Lambda = c_\Lambda ( \mathbb Z + \omega \mathbb Z ) , \ \ \  c_\Lambda \in 
\mathbb C^* ,  \ \ \   \omega = e^{ 2 \pi i/3}  . \]
 One nice choice is $ c_\Lambda = 1 $  (used in \cite{bz1} and later papers and in \S \ref{s:stan} above) but the lattices in the physics
literature have different $ c_\Lambda $. Let $ \Lambda^* := c_\Lambda^{-1} ({4 \pi i}/  {\sqrt 3})  \Lambda$, 
be the dual (reciprocal) lattice.

The class of very general periodicity conditions is given as follows:
\begin{equation} 
\label{eq:defV}  \begin{gathered} 
 V ( z + \gamma ) = \rho( \gamma)^{-1} V ( z ) \rho ( \gamma ) ,    \ \   W ( z + \gamma ) = \rho( \gamma)^{-1} W ( z ) \rho ( \gamma ), \\
  \rho ( \gamma ) := {\rm{diag}} \, \left[ ( \exp ( i \langle \gamma , k_j \rangle )   )_{j=1}^n \right] , \ \ \
k_j \in \mathbb C/ \Lambda^* . \end{gathered} 
\end{equation}
We remark that $ \rho ( \gamma ) $ is, up to a change of coordinates on $ \mathbb C^n $ 
a general unitary representation of the group $ \Lambda$ on $ \mathbb C^n$. 

We then have  
\[  L_\gamma D ( \alpha ) = L_\gamma D ( \alpha ) , \ \ \ 
L_\gamma u ( z ) := \rho ( \gamma ) u ( z + \gamma ), \]
and Bloch--Floquet theory follows the same path as in \S \ref{s:blfl} by considering the spectrum of 
\[  \begin{gathered}
H_k ( \alpha ) := \begin{pmatrix} \ \ 0 & D ( \alpha)^* + \bar k \\
D ( \alpha ) + k & 0 \end{pmatrix} : H^1_\rho \to L^2_\rho , \\
L_\rho^2 := \{ u \in L^2_{\loc} ( \mathbb C; \mathbb C^n ) , \ \ L_\gamma u = u \}, \ \ \
H^1_\rho := H^1_{\loc} \cap L^2_\rho . \end{gathered} \]
Equivalently we can consider
\begin{equation}
\label{eq:Prho} \begin{gathered}
 D_\rho ( \alpha ) : =  \rho ( z )  D ( \alpha ) \rho ( z )^{-1}  = 
  {\rm{diag}} \, \left[ ( 2D_{\bar z } - k_j )_{j=1}^n \right] + 
W_\rho ( z ) + \alpha  V_\rho ( z ) , \\
   \bullet_\rho ( z + \gamma ) = \bullet_\rho ( z ) , \ \ \  \bullet_\rho ( z ) :=  \rho ( z ) \bullet( z ) \rho ( z )^{-1} ,
   \ \ \ \bullet = V, W , 
\end{gathered}
\end{equation}
which is a periodic operator with respect to $ \Lambda $ and look at the corresponding
$ H_{\rho, k } ( \alpha ) $ on $ \Lambda $-periodic functions. 

By putting 
\[ Q ( \alpha , k ) := D ( \alpha ) + k , \ \ \ X = L^2_\rho , \ \ \ Y = H^1_\rho , \]
we can apply Theorem \ref{t:gaz} to this case provided that 
$ D ( 0 ) $ (corresponding to $ \alpha_0 = 0 $) has discrete spectrum. If
the eigenvalues of $ D ( 0 ) $ are {\em semisimple} then
\begin{equation}
\label{eq:semis}  m ( k, 0 ) = \dim \ker_{ H^1_{\rho } } ( 2 D_{\bar z} + W ( z )  + k ) . \end{equation}
This happens when  $ W (z ) \equiv 0 $ in which case 
\[ m ( k , 0 ) = | \{  j \in [ 1, \cdots , n ] : k \equiv k_j \!\! \! \! \mod \! \Lambda^* \}| .\]

The advantage of the special form of $ D ( \alpha )$ is that for
$ k \notin \Spec_{ H_\rho } D ( 0 ) $,  $ ( D ( 0 ) + k )^{-1} : L^2_\rho \to L^2_\rho $
is a compact operator. Combined with Theorem \ref{t:gaz} this gives
\begin{theo}[\cite{beta},\cite{bhz2},\cite{gaz}]
\label{t:spec}
Suppose that  $ Q ( \alpha, k  ) := D ( \alpha ) + k  $ where 
$ D ( \alpha ) $ is given in \eqref{eq:defD} and that $ D ( 0 ) $ has discrete 
spectrum. If for all $ k $ (see definition \eqref{eq:mult}) 
\begin{equation}
\label{eq:prot2} m ( \alpha, k ) \geq m ( 0, k ) , 
\end{equation}
then the 
 {\em Birman--Schwinger} operator, 
  \begin{equation}
 \label{eq:defBS} T_z :=( D ( 0 ) - z )^{-1} W ( z ) : 
 L^2_\rho \to H^1_\rho \hookrightarrow L^2_\rho 
 , \ \ \   z \notin \Spec ( P ( 0 ) )  ,  \end{equation} 
 has discrete spectrum independent of $ z $ and, in the notation of
 Theorem \ref{t:gaz}, 
 \begin{equation}
 \label{eq:a2T}    m ( k , \alpha ) = \left\{ \begin{array}{ll} \ \ \  \infty, &  1/\alpha \in \Spec  ( T_z ), 
 \\ m ( k, 0 ), & \ \  \text{otherwise.} \end{array} \right.
  \end{equation}
  In particular, $ H ( \alpha ) $ in \eqref{eq:defD} has a flat band at $0 $ if and only
  if $ 1/\alpha \in \Spec ( T_z) $.
  
  Conversely, if the spectrum of $ T_z $ is independent of $ z \notin \Spec D ( 0 ) $, 
  then \eqref{eq:prot2} and \eqref{eq:a2T} hold.
 \end{theo}

\begin{figure}
\hspace*{-10pt}
\includegraphics[width=8cm]{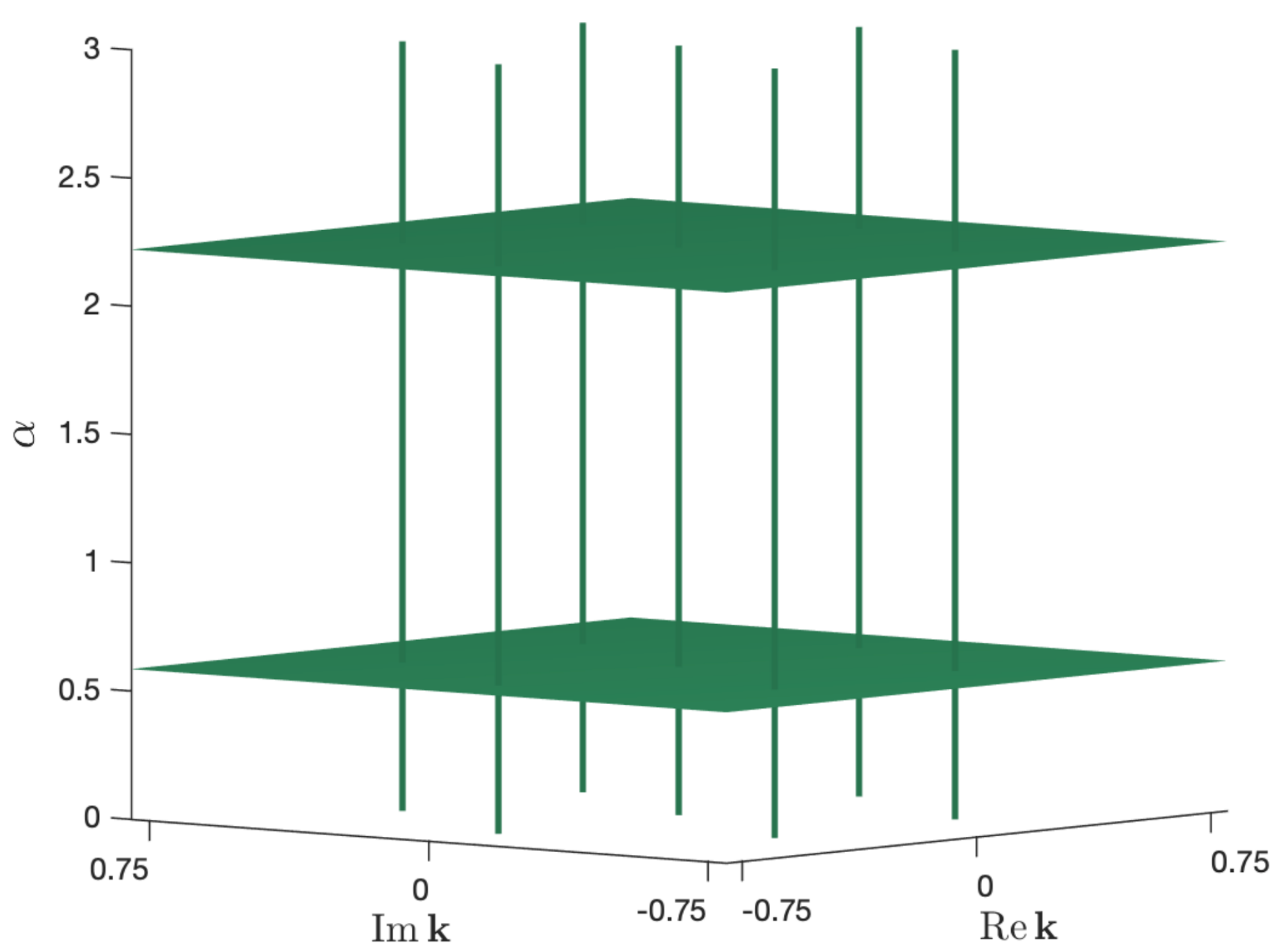}
\hspace*{-25pt}
\includegraphics[width=8cm]{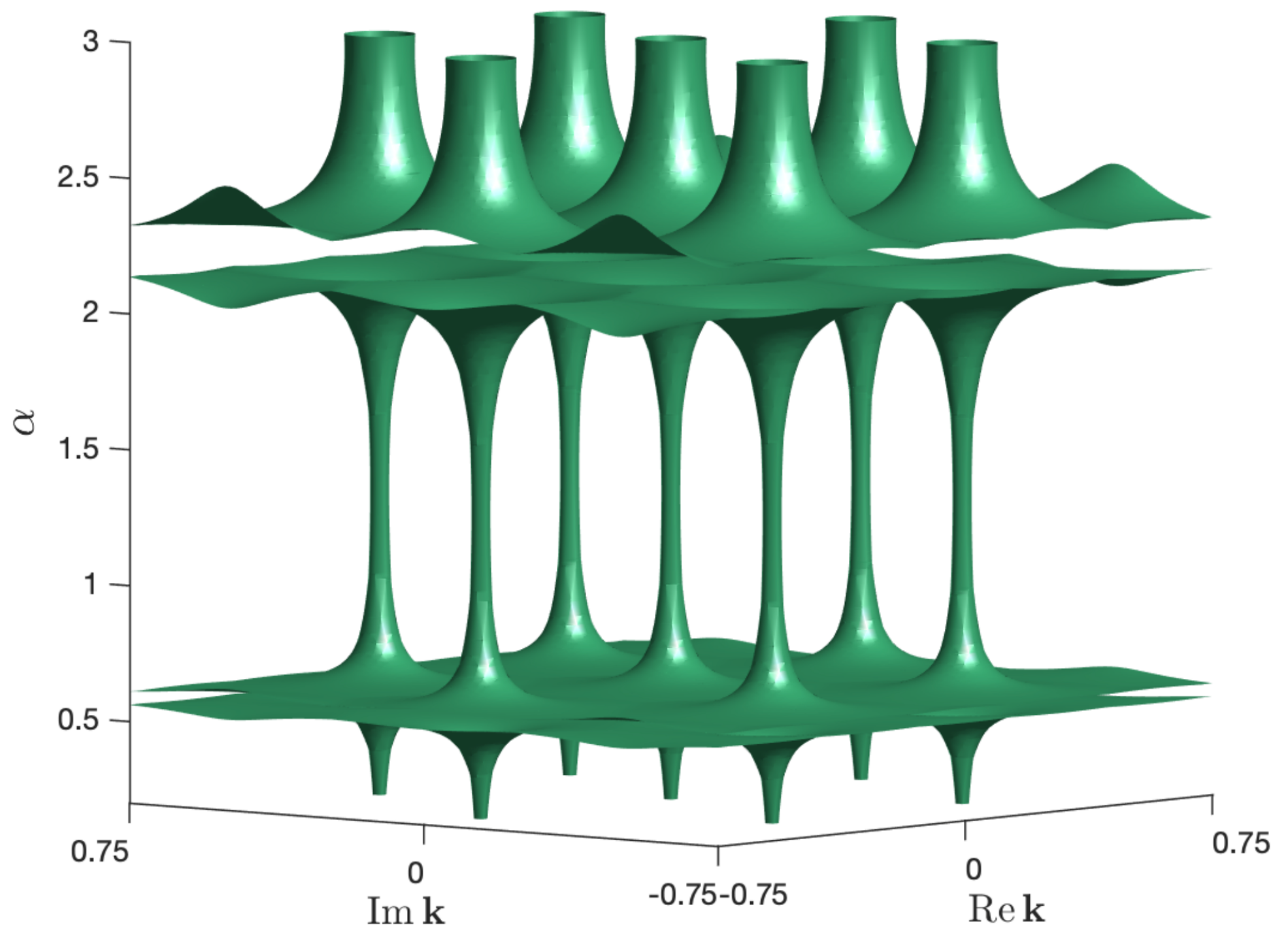} 
\caption{\label{f:inst}: Left: the spectrum of $ D (\alpha) $ 
(in the $ k $ plane) as $\alpha $ varies (vertical axis). Flat surfaces indicate that $1/\alpha$ is a magic angle. 
Righ: level surface of $ \| ( D ( \alpha  ) -  k )^{-1} \| = 10^{2}$ as a function of
$ k $ and
 $\alpha$:
the norm blows up at magic angles for all $ k $ ($\alpha$ near the magic values $ 0.586$ and  $2.221$). The thickening of the ``trunks'' reflects the exponential squeezing
of the bands. This figure comes from \cite{beta}.}
\end{figure}

As pointed out above, this spectral characterization, with magic angles as the
spectrum of a compact operator, has been very useful in computing
elements of $ \mathcal A $. Since $ T_z $ is non-selfadjoint, 
pseudospectral issues (see \cite{dsz} and references given there), that is 
the large size of the norm of the resolvent of $ T_z $, enter for large values of $ \alpha $.
An explanation of this is provided in \S \ref{s:squeeze} but a striking numerical illustration
is given in Figure \ref{f:inst}.

An example of an operator with $ n = 3 $ can be found in \cite{BHWY} where
trilayer graphene was studied (and \eqref{eq:semis} holds). A more interesting case
is given by twisted $ m $-sheets of graphene studied mathematically 
in \cite{yang}:

\noindent
{\bf Example.} 
Let us rename the operator $ D ( \alpha ) $ in \eqref{eq:Hamiltonian} as $ D_1 ( \alpha ) $. Following
\cite{yang} and the physics papers cited there, we put, for $ N > 1 $, 
\begin{equation}
\label{eq:Yang}  
D ( \alpha ) = D_N ( \alpha , \mathbf t ) 
 := \left(\begin{matrix} D_1 (\alpha) & t_1T_+ & \\ t_1T_- & D_1(0) & t_2T_+ & \\ & t_2T_- & D_1(0) & \ddots \\ && \ddots && \\ &&&& t_{N-1}T_+ \\ &&&t_{N-1}T_- & D_1 (0)\end{matrix}\right),
\end{equation}
with $\mathbf{t} =(t_1, t_2, \cdots, t_{N-1})$ and 
\[     T_+ = \left(\begin{matrix}1&0\\0&0\end{matrix}\right), \ \ \ \ 
    T_- = \left(\begin{matrix}0&0\\0&1\end{matrix}\right). \]
    To find a suitable $ \rho$ in \eqref{eq:defV} we first choose $ k_1 $ and $ k_2 $ which 
work for $ D_1 ( \alpha ) $ (for instance, as in
\eqref{eq:defLag}) and then check that $ k_j $ for $ 3 \leq j \leq 2N $ can be chosen 
consistently so that \eqref{eq:defV} holds. Then $ D ( \alpha ) $ is an example of an operator
to which Theorem \ref{t:spec} applies with $ n = 2 N $.  In this case 
$ m ( 0 , k ) = N \indic_{ \mathcal K } ( k ) > \dim \ker ( D_N ( 0 ) + k ) =   \indic_{ \mathcal K } ( k ) $
($ \mathcal K = \{ K, - K \} + \Lambda $ in the case of \eqref{eq:defLag}).
A direct argument in \cite[\S 4.2]{yang} showed
that set of $ \alpha $'s for which the spectrum of $ D ( \alpha ) $ is a discrete set and that implies that the 
spectrum $ T_z $ in
\eqref{eq:defBS} is independent of $ z \notin \mathcal K $. Hence Theorem \ref{t:spec} implies that
\eqref{eq:prot2} holds but it would be interesting to have a direct argument for that.

\subsection{A scalar model} 
\label{s:scal}

One of the difficulties of dealing with the operator $ D ( \alpha ) $ given in \eqref{eq:Hamiltonian} is
that it acts on vector valued functions -- some of that will be highlighted in \S \ref{s:semi}. 
By increasing the order of the operator a scalar model non-equivalent to $ D ( \alpha ) $ but 
exhibiting flat bands was proposed in \cite{gaz}. 

We first observe that $ D ( -\alpha ) $ is the co-adjoint matrix of $ D( \alpha ) $ and hence
\begin{equation}
\label{eq:scalar} 
\begin{gathered}  D ( - \alpha ) D ( \alpha ) = Q ( \alpha ) \otimes I_{\mathbb C^2 } + 
\begin{pmatrix}  \ \ 0 &  \alpha 2 D_{\bar z } U ( z ) \\
- \alpha [ 2 D_{\bar z} U ] ( -z ) & \ \ 0 \end{pmatrix}, \\
Q ( \alpha ) := ( 2 D_{\bar z } )^2 - \alpha^2 U ( z ) U ( -z )  . 
\end{gathered} 
\end{equation}
From the semiclassical point (as $ \alpha \to \infty $) of view the non-scalar term in \eqref{eq:scalar} is of 
lower order (see \S \ref{s:semi}) and is natural to consider the operator $ Q ( \alpha ) $ on its own. 

We can then consider a self-adjoint Hamiltonian, on $ L^2 ( \mathbb C ; \mathbb C^2) $ with the domain given by 
$ H^2 ( \mathbb C ; \mathbb C^2 ) $ (note that $ D_{\bar z }^2 $ is an elliptic operator),
\begin{equation}
\label{eq:Q2H1}  H ( \alpha ) := \begin{pmatrix} 0 & Q ( \alpha )^* \\
Q ( \alpha ) & 0 \end{pmatrix} . \end{equation}
This is a periodic operator with respect to the lattice $ \Lambda $ (note that for 
$ U $ satisfying \eqref{eq:defU}, $ U ( z ) U ( -z ) $ is $ \Lambda$-periodic). And Floquet 
theory (as reviewed in \S \ref{s:blfl})  corresponds to studying the spectra of 
\[ H( \alpha, k ) = \begin{pmatrix} 0 & Q ( \alpha, k )^* \\
Q ( \alpha, k ) & 0 \end{pmatrix} , \ \ Q ( \alpha, k ) := ( 2 D_{\bar z } + k )^2 - \alpha^2 U ( z ) U ( -z ), \]
on $ L^2 ( \mathbb C/\Lambda ; \mathbb C^2) $ and with the domain 
$ H^2 ( \mathbb C/\Lambda; \mathbb C^2 ) $.  

\begin{figure}
 \includegraphics[width=15cm]{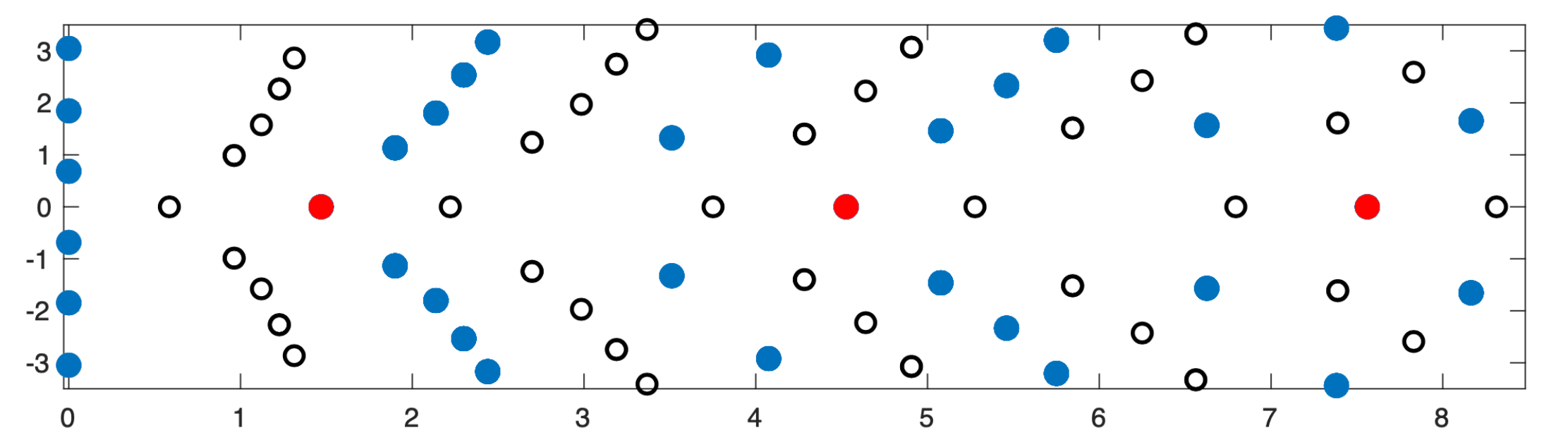}
\caption{\label{f:A2B} 
Comparison of the set of magic $ \alpha $'s, $ \mathcal A $ for the potential $ U = U_{\rm{BM}} $
given in \eqref{eq:defUV} (shown as $ \circ $) and $ \mathcal A_{\rm{sc}} $ the set for which \eqref{eq:flatQ} holds (with the same $ U $; shown as \blue{$\bullet$}).
The real elements of $ \mathcal A_{\rm{sc}} $ are shown as \red{$\bullet$}. They appear to have multiplicity 
two.  When we interpolate between the chiral model and the 
scalar model, the multiplicity two real $ \alpha$'s  split and travel in opposite directions to become
magic $ \alpha $'s for the chiral model: see \url{https://math.berkeley.edu/~zworski/Spec.mp4}.  }
\end{figure}

A flat band of $ H ( \alpha ) $ given in \eqref{eq:Q2H1} corresponds to 
\begin{equation}
\label{eq:flatQ} 
\forall \, k \in \mathbb C \ \ 0 \in \Spec H_k ( \alpha ) \ \Longleftrightarrow \ 
\forall \, k \in \mathbb C \ \  \ker_{ H^1 ( \mathbb C/\Lambda) } Q ( \alpha, k ) \neq \{ 0 \} . 
\end{equation}
To apply Theorem \ref{t:gaz} we need to verify the first inequality in  (using the definition \eqref{eq:mult})
\[  m ( \alpha , k ) \geq m ( 0 , k ) = 2 \indic_{\Lambda^*} ( k ) > \dim \ker Q ( 0 , k ) = \indic_{\Lambda^* } ( k ) , \]
see \cite[\S 3]{gaz}. (Just as in the case of \eqref{eq:Yang} it is important to consider the generalized
multiplicities.) 
It then follows that there exists a discrete set $ \mathcal A_{\rm{sc}} $ such that
\eqref{eq:flatQ} holds if and only if $ \alpha \in \mathcal A_{\rm{sc}} $ -- see 
 Figure \ref{f:A2B}.

The next two problems are probably the most doable on the list.

\begin{op}
\label{5}  Adapt the theta function argument recalled in \S \ref{s:theta} 
to the scalar model and show that the multiplicity of the flat bands is at least $ 2 $. 
\end{op}

\begin{op}
\label{6} Adapt the trace argument is \S \ref{s:exi} to show that for the
potential $ U = U_{\rm{BM}} $ in \eqref{eq:defU}, $ | \mathcal A_{\rm{sc}} | = \infty $. 
\end{op}

The situation is less clear for the next 

\begin{op}
\label{7}  Is the spectrum of $ Q ( \alpha , k ) $ discrete for all $ \alpha $ 
and $k $?  Characterize the set for which $ \Spec Q ( \alpha, k ) = \emptyset $. (We should
stress that this is a mathematical curiosity: only the fact that $ 0 \in \Spec Q ( \alpha, k ) $
is relevant to the question of band theory, bands being given by characteristic values of 
$ Q ( \alpha, k ) $ as $ k $ varies.
\end{op}

The next problem is the analogue of Open Problem \ref{1}. One would like to hope that the
scalar nature of the operator could be of some help in the semiclassical analysis (see
\S \ref{s:semi}). 

\begin{op}
\label{8}  If $ \beta_1 < \beta_2 < \cdots $ is the ordered sequence
of elements of $ \mathcal A_{\rm{sc}} \cap [ 0, \infty ) $ then, for the potential \eqref{eq:defUV}, 
\[   \beta_{j+1}  - \beta_j = 2 \gamma + o ( 1 ) , \ \ j \to \infty  \]
where $ \gamma \simeq \frac32 $ is the asymptotic spacing between the elements
of $ \mathcal A \cap [ 0 , \infty ) $ (see \eqref{eq:threeh}). What happens for more 
general potential satisfying \eqref{eq:defU}?
\end{op}

\section{Theta function argument for magic angles}
  \label{s:theta} 
  
Magic angles for the chiral model were described in \cite{magic} using a different 
approach than that recalled in \S \ref{s:spec} and coming from \cite{beta}. It was based on an idea which appeared earlier, in a different but related context, in the
work of Dubrovin--Novikov \cite{dun}. It was revisited in \cite{bhz2} and here we present a slightly different
variant.

The operator $ 2 D_{\bar z } := \partial_{x_1} + i \partial_{x_2} $, $ z = x_1 + i x_2 $,  acting on $ L^2 ( \mathbb C/\Lambda;\mathbb C ) $ with the
domain given by $ H^1 ( \mathbb C/\Lambda ; \mathbb C ) $ is a
normal operator (a sum of two commuting sel-adjoint operators). It spectrum is given by 
$ \Lambda^* $ with simple eigenvalues and normalized eigenfunctions given by $ v(p) =  \tau (p ) v( 0 ) $, 
$ v( 0 ) := |\mathbb C/\Lambda|^{-\frac12}$, $[ \tau ( p ) u] ( z ) = e^{ i \langle z , p \rangle } u ( z ) $, $ p \in \Lambda^* $. Hence, its resolvent 
\[ ( 2 D_{\bar z } + k )^{-1} :  L^2 ( \mathbb C/\Lambda ; \mathbb C  ) \to H^1( \mathbb C/\Lambda;
\mathbb C ), 
\]
is a {\em meromorphic family} of operators with simple poles at $ p \in \Lambda^* $ 
and residues $ | v( p ) \rangle \langle v ( p ) | $. Since $ 2 D_{\bar z } $ is translation invariant,
we have 
\[ ( 2 D_{\bar z } + k )^{-1} f ( z ) = \int_{ \mathbb C/\Lambda } G_k ( z - \zeta ) f ( \zeta ) d m ( \zeta) , \ \
( 2 D_{\bar z } + k ) G_k ( z ) = \delta_0 (z ), \ \  k \notin \Lambda^* . 
\]
If $ a ( k ) $ is {\em any} entire function with the zero set given by simple zeros at 
$ \Lambda^* $, then 
\begin{equation}
\label{eq:propFk} \begin{gathered}    k \mapsto F_k ( z ) := a(k) G_k ( z ) , \ \ \text{ is a holomorphic family of distributions,} \\ 
( 2 D_{\bar z } + k ) F_k ( z) = a(k) \delta_0 ( z ) . \end{gathered}
\end{equation} 

If $ u_{K } $ is the protected state described in Theorem \ref{t:prot} and 
$ z_0 \in \mathbb C/\Lambda $, then \eqref{eq:propFk} gives 
(note that $ u_K $ is valued in $ \mathbb C^2$ and $ F_k $ is scalar valued)
\begin{equation}
\label{eq:green}  ( 2 D_{\bar z } + k ) ( F_{k-K} ( z - z_0  ) u_K ( \alpha, z ) ) = a ( k - K ) u_K ( \alpha,  z_0 ) \delta (z -z_0 ) . \end{equation}
Hence,  
\begin{equation}
\label{eq:z2u}  \exists \, z_0  \ \ u_K ( \alpha, z_0 ) = 0 \ \Longrightarrow \ \forall k \, \exists \, u(k) \in H^1_0 \ \ 
( D ( \alpha ) + k ) u ( k) = 0 , \ \ \| u ( k ) \|_{L^2_0} = 1 . \end{equation}
The required vanishing condition is strong: we are looking for simultaneous vanishing of
two complex valued functions of the complex variable (components of $ u_K $).

Following \cite{magic} we observe 
\[   \tau ( K ) u_K ( \alpha, z ) = \begin{pmatrix} \psi ( z ) \\ \varphi ( z ) \end{pmatrix}  \ \Longrightarrow 
\ \forall \, \alpha \ \ \varphi ( z ( K ) ) = 0 .\]
(Here we use the notation of \eqref{eq:defFk}  and recall from \eqref{eq:Lkp} and \eqref{eq:uKK} that $  L_\gamma \tau ( K ) u_K ( \alpha)  = e^{ i \langle  \gamma, K \rangle }  \tau ( K ) u_K ( \alpha ) $ and that 
$ \tau ( K ) u_K ( \alpha, \omega z ) = \tau ( K ) u_K ( \alpha, z ) $ which then implies, following
the definitions, that $ \varphi ( z ( K ) ) = \bar \omega \varphi ( z ( K ) ) $.)  Using \eqref{eq:uKK} we have
\[ \tau ( - K ) u_K ( \alpha, z ) = \begin{pmatrix} \ \varphi ( - z ) \\ - \psi ( - z ) \end{pmatrix} . \]
Since  $ D ( \alpha ) ( \tau ( \pm K ) u_{\pm K } ( \alpha ) ) = 0 $, the Wronskian of $ 
\tau ( \pm K ) u_{\pm K } $ is a holomorphic $ \Lambda$-periodic function. Hence it is 
a constant depending only on $ \alpha $:
\begin{equation}
\label{eq:defv}  v_F ( \alpha ) := \frac{ \psi( z ) \psi( -z ) + \varphi ( z ) \varphi ( - z ) } {\| u_K ( \alpha ) \|^2} = 
\frac{ \psi( z( K ) ) \psi ( - z ( K ) ) } {\| u_K( \alpha)  \|^2}  .  \end{equation}
(For an interesting physical interpretation of $  v_F (\alpha ) $ as the Fermi velocity 
 see \cite[(8),(21),(22)]{magic}. We lose holomorphy in $ \alpha $ because of the normalization.) We conclude that
\begin{equation} 
\label{eq:propv}  \exists \, z_0 \ \ u_K ( \alpha, z_0 ) = 0 \ \Longleftrightarrow \ v_F ( \alpha ) = 0 \
\Longleftrightarrow \ \exists \, \varepsilon \in \{ + , - \} \ \  u_K ( \alpha, \varepsilon z ( K ) ) = 0 . 
\end{equation}
This argument, essentially from \cite{magic}, establishes one implication in the first statement of
\begin{theo}[\cite{magic},\cite{beta},\cite{bhz2}]
\label{t:v}
For any potential $ U $ satisfying \eqref{eq:defU},  $ \mathcal A $ defined in \S \ref{s:flat} 
and $ v_F ( \alpha ) $ defined in \eqref{eq:defv}, we have
\begin{equation}
\label{eq:tv1} 
v_F ( \alpha ) = 0 \ \Longleftrightarrow \ \alpha \in \mathcal A . 
\end{equation}
Moreover, if $ \alpha \in \mathcal A $ is {\em simple} then 
\begin{equation}
\label{eq:zS} 
  u_K ( \alpha ,  z_0) = 0 \ \Longrightarrow \  z_0 = z ( K ) , 
\end{equation}
and the zero is simple: $ u_K ( \alpha, z ) = ( z - z(K ) ) w(z )$, $ w \in C^\infty $, 
$w ( z( K ) ) \neq 0 $. 
\end{theo}
The implication $ v_F ( \alpha ) \neq 0 \Rightarrow \alpha \notin \mathcal A $ follows easily 
from building a formula for $ ( D ( \alpha ) + k )^{-1} $ using $ u_{\pm K } ( \alpha ) $ -- see 
\cite[Proposition 3.3]{beta}. The implication \eqref{eq:zS} is a special case of
\cite[Theorem 3]{bhz2}. The point $ z ( K ) =: - z_S $ is called a stacking point -- see
Figure~\ref{f:hiz}. The proof of \eqref{eq:zS} was simplified in \cite{BHWY} 
in a way which allowed an adaptation to the trilayer case.
For an animation showing the behaviour of $ u_K (\alpha ) $ as 
$ \alpha $ increases along the real axis (for the potential \eqref{eq:defUV}), 
see \url{https://math.berkeley.edu/~zworski/magic.mp4}. 

We recall another characterization of simple $ \alpha \in \mathcal A $:

\begin{theo}[\cite{bhz2}]
\label{t:simple}
We have the following equivalence (using definition \eqref{eq:defm} and denoting
$ \mathcal K_0 := \{ K , - K \} + \Lambda^*$)
\begin{equation}
\label{eq:tsim} 
\begin{split} 
m ( \alpha ) = 1   & \ \Longleftrightarrow \ \forall \,  k \in \mathbb C 
\ \ 
\dim \ker_{L^2_{0} ( \mathbb C/\Lambda) } ( D ( \alpha ) + k ) = 1 \\
& \ \Longleftrightarrow \ \exists \,  p \notin \mathcal  K_0  \ \ 
\dim \ker_{L^2_{0 } ( \mathbb C/\Lambda) }(  D ( \alpha ) +p ) = 1. \end{split} 
\end{equation}
\end{theo}

Returning to \eqref{eq:green} and \eqref{eq:z2u} we see that for $ \alpha \in \mathcal 
A$, simple, we can take (see \cite[(3.43)]{bz1})
\begin{equation}
\label{eq:ukz}  u ( k , z) = c ( k ) F_k ( z ) u_0 ( z ) , \ \ \  \ker_{H^1_0 } D ( \alpha) = \mathbb C u_0, \ \ 
\ker_{ H^1_0} ( D ( \alpha ) + k ) =  \mathbb C u(k) 
 , \end{equation} 
where $ c ( k ) $ is the normalizing constant so that $ \| u ( k ) \|_{L^2_0} = 1$.
(We know that in this case $ u_0 $ has a simple zero at $ 0 $ -- see 
\cite[Proposition 3.6]{bz1}. Please note that $ u_0 \in L^2_0 $ exists only for $ \alpha
\in \mathcal A $, unlike $ \tau (\pm K ) u_{\pm K } \in L^2_{\pm K } $, $ D ( \alpha ) \tau (\pm K ) u_{\pm K } = 0 $  which exist 
for all $ \alpha $.) Using symmetries of  $ D ( \alpha ) $ we can also describe 
the kernel of  $ ( D ( \alpha ) + k )^* $ and that can be done in different ways.
Following \cite[(3.44)]{bz1} we can take (with the advantage that it works also for 
more general potentials \eqref{eq:Hamiltonian1})
\begin{equation}
\label{eq:u*kz}  u^* ( k , z) = c ( k ) \overline{ F_{-k} ( z )} \begin{pmatrix} \ \  \overline{ \varphi_0 ( z) } \\
- \overline{ \psi_0 ( z ) } \end{pmatrix}
 , \ \   u_0 =:  \begin{pmatrix} \psi_0 \\ \varphi_0 \end{pmatrix},  \ \ 
\ker_{ H^1_0} ( D ( \alpha ) + k )^* =  \mathbb C u^*(k) , \end{equation} 
and $   \| u^* ( k ) \|_{L^2_0} = 1$. (For other choices of $ u^* ( k ) $ when $ D ( \alpha ) $ is
given by \eqref{eq:Hamiltonian} see \cite[(2.9)]{bz2}.)

There are many choices for $ F_k $ (that is, choices of entire functions with simple zeros precisely 
at $ \Lambda^* $) and we can for instance follow \cite{bz1} and take
\begin{equation}
\label{eq:defFk}  \begin{gathered}  F_k ( z ) :=  e^{ \frac i 2 ( z - \bar z ) k } \frac{\theta ( z - z ( k ) ) }{\theta ( z ) }, \ \ \ 
 z ( k ) = \frac{\sqrt 3}{ 4 \pi i} k , \ \ \ a ( k ) := \frac{ 2 \pi i \theta ( z ( k) )} {\theta' ( 0 )}, \\
\theta ( z )
:= \theta_1 ( z | \omega) := - \sum_{ n \in \mathbb Z } \exp ( \pi i (n+\tfrac12) ^2 \omega+ 2 \pi i ( n + \tfrac12 ) (z + \tfrac 12 )  ) ,
\end{gathered} \end{equation}
that is $ \theta $ is the first Jacobi theta function and its simple zeros coincide with 
$ \Lambda $ -- see \cite{tata} or \cite{voca}.  Weierstrass $ \sigma $ function was used explicitly in \cite{dun} and the theta function in 
\cite{magic}, but in fact it is only the canonical nature of Green's function and the set 
$ \Lambda^*$ that matter (though of course constructing a function with vanishes precisely 
at $ \Lambda^*$ hides those special functions).

\section{Existence and multiplicities of magic angles}
\label{s:exi}

So far we have not addressed the question of existence of magic $ \alpha $'s, and in 
particular of existence of real simple $ \alpha $'s (see the definition in \S \ref{s:flat}).
It is not clear if there exist more than one physical magic angles and the current 
experimental and theoretical evidence suggests that there may only be one. The work
of Becker--Oltman--Vogel \cite{bovo} on random perturbations of 
TBG provides some mathematical evidence for that.

In the chiral model rigorous existence and simplicity of the
first real magic angle has however been established:

\begin{theo}[\cite{lawa},\cite{bhz1}]
\label{t:exi} 
For the potential \eqref{eq:defUV} and for the (discrete) set of magic $ \alpha$'s, $ \mathcal A $,  
defined in  \S \ref{s:flat}, we have
\begin{equation}
\label{eq:defa1}     \min \mathcal A \cap [ 0 , \infty ) = \alpha_1 \simeq 0.586 . 
\end{equation}
In addition, in the sense of \eqref{eq:defm}, 
\begin{equation}
\label{eq:defa1m}
m ( \alpha_1 ) =1, 
\end{equation}
that is, $ \alpha_1 $ is simple. 
\end{theo}

Watson and Luskin \cite{lawa} followed the approach of \cite{magic} and proved
existence of a zero of $ v_F ( \alpha ) $ given in \eqref{eq:defv} (see Theorem \ref{t:v}). 
That was done by a careful analysis of the Taylor series at $ 0 $, with precise 
estimates of the remainder, and floating point arithmetic.

The approach of \cite{bhz1} was based on the spectral characterization from \cite{beta} (see \S \ref{s:spec}) and the evaluation, theoretical and numerical, of sums of powers of magic
$ \alpha$'s:

\begin{theo}[\cite{beta},\cite{bhz1}] For the potential in \eqref{eq:defUV} we have
\begin{equation}
\label{eq:tr1}
\sum_{ \alpha \in \mathcal A } \alpha^{-4} = \frac{ 8 \pi } {\sqrt 3 }, 
\end{equation}
and more generally, for $ p \in \mathbb N + 2 $, 
\begin{equation}
\label{eq:tr2} 
\sum_{ \alpha \in \mathcal A } \alpha^{-2p} \in \frac{ \pi } {\sqrt 3} \mathbb Q . 
\end{equation}
In the above sums the multiplicity of $ \alpha \in \mathcal A $ is given by the algebraic multiplicity
of $ 1/\alpha $ as an eigenvalues of $ T_k $, $ k \notin \Lambda^* $, where
$ T_k $ is the Birman--Schwinger operator \eqref{eq:defBS}.
\end{theo}
These identities are based on writing $ \sum_{ \alpha \in \mathcal A } \alpha^{-2p} 
= \tr T_k^{2p} $, and \eqref{eq:tr1} was proved in \cite[\S 3.3]{beta} (the sum 
in \eqref{eq:tr2} with $ p = 4 $ was also given as $ 80 \pi/\sqrt 3$; since there we
considered
action on  $ L^2 ( \mathbb C/ 3 \Lambda ) $ rather than on $ L^2_0 $, the 
multiplicities were nine fold higher; we note that for odd powers of $ T_k $ the traces
are $ 0 $ in view of \eqref{eq:symmA}). 
The far reaching generalization in \eqref{eq:tr2} happened thanks to the 
expansion of the collaboration in \cite{bhz1}. It holds for a greater class of potentials.
The existence of algebraic multiplicities
greater than geometric multiplicities (Jordan blocks) is suggested by numerical 
experiments -- see \cite[\S 10.1]{bhz3}. 

The method for proving \eqref{eq:tr1} provides an algorithm for finding the 
rational number $ (\sqrt 3/\pi ) \tr T_k^{2p} $. This allows a precise evaluation 
of regularized determinants of $ I - T_k $ and that lead to an alternative
proof of \eqref{eq:defa1} and a proof of \eqref{eq:defa1m}.

An immediate consequence of \eqref{eq:tr1}, \eqref{eq:tr2}, the transcendental
nature of $ \pi / \sqrt 3 $, and of Newton identities is (see \cite[Theorem  6]{bhz1} for a more general version):
\begin{theo}[\cite{bhz1}] 
For the potential \eqref{eq:defU}, 
\begin{equation}
\label{eq:infA}
| \mathcal A | = \infty. 
\end{equation}
\end{theo}

Before moving to the discussion of higher multiplities we present some open problems
related to the above theorems. They all seem quite hard.

\begin{op}
\label{10}  Show that \eqref{eq:infA} holds for any non zero potential
satisfying \eqref{eq:defU}.
\end{op}

\begin{op}
\label{11} Using Theorem \ref{t:spec} it is not difficult to see that 
$  | \{ \alpha \in \mathcal A : | \alpha | \leq r \} \leq C r^2 $. Do we have lower bounds?
Is there a way to use methods of Christiansen \cite{Ch99} (``plurisubharmonic magic") 
to obtain results for
generic potentials?
\end{op}

\begin{op}
\label{12} Show that for the potential \eqref{eq:defUV} and $ \alpha_1 $ given in 
\eqref{eq:defa1} we have
\[  \tfrac12  \alpha_1^{2p} \sum_{ \alpha \in \mathcal A } \alpha^{ -2 p} \to 1, \ \ \  p \to \infty, \ \ p \in 
\mathbb N .\]
This seems to be the case numerically as, $ \min\{ | \alpha | : \alpha \in \mathcal A \setminus \{\pm \alpha_1\} \} > 1 $. Any type of asymptotic result about $ \tr T_k^{2p} $ would be interesting.
\end{op}

We now turn attention to higher multiplicities. Figure~\ref{fig:degeneracy} showed numerically
computed multiplicities, including $ \alpha \in \mathcal A \cap \mathbb R $ with $ m ( \alpha) > 1 $
(see \eqref{eq:defm} for the definition of multiplicity). For the BM potential \eqref{eq:defUV} 
``half" of the complex $ \alpha$'s have multiplicity two (indicated by circles; we show 
$ \alpha$'s is the first quadrant):
\begin{center}
\includegraphics[width=14cm]{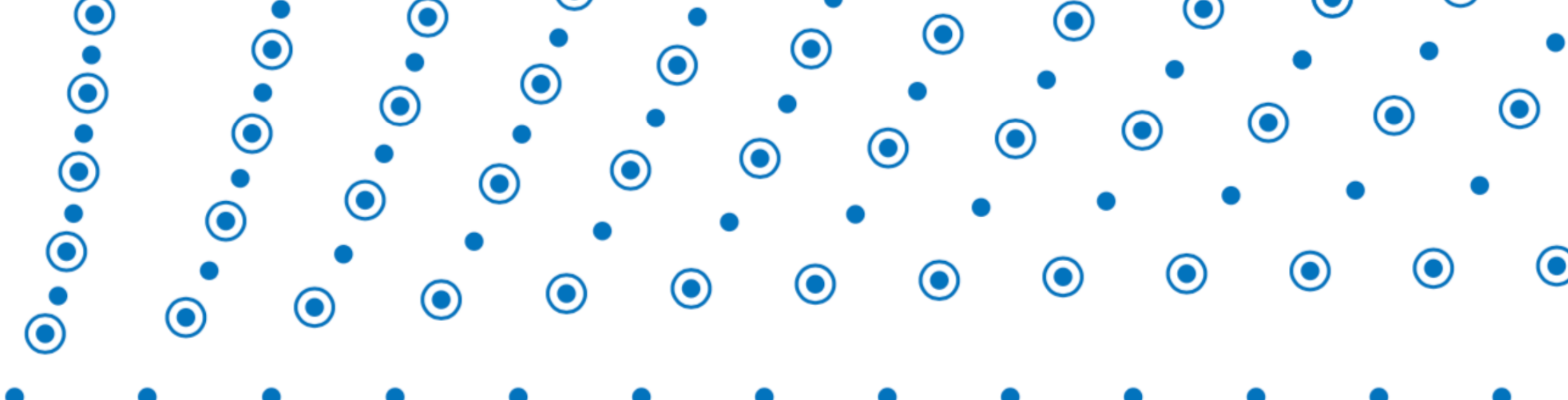}
\end{center}

Using Theorem \ref{t:rigid} below and analysis  of traces of $ T_k^{2p }$ restricted to 
different spaces $ L^2_{p,k} $ we obtain a partial mathematical confirmation of the above
figure:

\begin{theo}[\cite{bhz3}] For the Bistritzer--MacDonald potential \eqref{eq:defUV} 
\[    | \{ \alpha \in \mathcal A : m ( \alpha ) > 1 \} | = \infty, \]
that is, there exist infinitely many (complex) degenerate magic $ \alpha $'s.
\end{theo} 

The double $ \alpha$'s shown in the above figure are protected as we have a surprising rigidity 
result expressed using the spaces defined in \eqref{eq:Lkp}:
  \begin{theo}[\cite{bhz3}]
\label{t:rigid}
For any potential satisfying \eqref{eq:defU} we have,
with the definition of multiplicity \eqref{eq:defm}, 
\begin{equation}
\label{eq:imply} \begin{gathered}
m ( \alpha ) = 1 \ \Longrightarrow \ \dim \ker_{ L^2_{0,2} }  D ( \alpha ) = 1 ,\\
 m (\alpha ) = 2 \ \Longrightarrow \ \dim \ker_{L^2_{0,0} } D ( \alpha ) = 
\dim \ker_{ L^2_{0,1} }  D ( \alpha ) = 1.
\end{gathered} \end{equation}
In particular, a multiplicity two $ \alpha \in \mathcal A $ {\em cannot} be split 
into simple $ \alpha$'s by deforming a potential within the class \eqref{eq:defU}.
\end{theo} 
In \cite[Theorem 4]{bhz3} we also have an analogue of Theorem \ref{t:simple} for the 
case of double $ \alpha$'s. 

\stepcounter{op}

\begin{op} 
\label{14} As suggested by \eqref{eq:z2u} the multiplicity of 
$ \alpha $ is closely related to the number of zeros (counted with multiplicity)
of the eigenstate of $ D( \alpha ) $. For $ 1 \leq m ( \alpha ) \leq 2$, 
Theorem \ref{t:rigid} can be used to obtain the precise description (see 
\S \ref{s:top}). What is the situation for higher multiplicities?
\end{op}

It is natural to ask if generically we only have simple or double magic $ \alpha$'s. 
We have established it by expanding the class of allowed potentials:
\begin{equation}
\label{eq:Hamiltonian1}
   D(\alpha) :=
  2 D_{\bar z } \otimes I_{\mathbb C^2 } + W ( z ) , \ \ \ 
   W ( z ) :=  \begin{pmatrix} 0& \alpha U_+ (z ) \\ \alpha U_-(z) & 0 \end{pmatrix} ,
\end{equation}
where the potentials satisfy
 \begin{equation}
 \label{eq:news}
U_\pm (z+\gamma) = e^{\pm i  \langle \gamma,K \rangle} U_\pm (z), \ \ \gamma \in \Lambda, 
\quad U_\pm 
 (\omega z) = \omega U_\pm (z).
 \end{equation}
 The self-adjoint Hamiltonian $ H ( \alpha ) $ is defined by \eqref{eq:chiH} and commutation
 relations  \eqref{eq:transl},\eqref{eq:rota} still hold. We then have the same Bloch--Floquet
 theory as in \S \ref{s:blfl} and the same definitions of $ \mathcal A $ and $ m ( \alpha ) $
 (see \S \ref{s:flat}). 
 
As the space of allowed potentials $ W $ we use a Hilbert space of 
{\em real analytic} functions equipped with the following norm:
for a fixed $ \delta > 0 $, 
\begin{equation}
\label{eq:norm}
\| W \|_\delta^2 := 
\sum_{\pm} \sum_{ k \in \Lambda^*/3 } |a^\pm_{k } |^2 e^{ 2 | k| \delta}, \   \ \ \ 
U_{\pm} ( z ) = \sum_{  k \in K + \Lambda^* } a^\pm_{k } e^{ \pm i \langle z ,k \rangle } . 
\end{equation}
Then we define $ \mathscr V = \mathscr V_\delta $ by 
\begin{equation}
\label{eq:Vscr}
W \in \mathscr V \ \Longleftrightarrow \ 
\text{ $ W$ satisfies \eqref{eq:news}, } \ 
\| W \|_{\delta } < \infty . 
\end{equation}
With this in place we can state

\setcounter{theo}{13}

\begin{theo}[\cite{bhz3}]
\label{t:gen}
There exists a generic subset (an intersection of open dense sets),
$ \mathscr V_0 \subset \mathscr V $, such that if 
$ W \in \mathscr V_0 $ then for all  $ \alpha \in \mathcal A $ (defined
using \eqref{eq:Hamiltonian1})
\[      1 \leq m ( \alpha ) \leq 2 . \]
\end{theo} 

A more precise formulation related to Theorem \ref{t:rigid} is given in \cite[Theorem 3]{bhz3}.

\begin{op}
\label{15}   Does Theorem \ref{t:gen} hold for a generic set of potentials satisfying
 \eqref{eq:defU}?
 \end{op}

\section{Topology of flat bands}
\label{s:top} 

Topology of flat bands refers to the topology of vector bundles over the $ k$-space
torus $ \mathbb C/\Lambda^* $ obtained by considering eigenfunctions of $ H_k  ( \alpha ) 
= H_k ( \alpha , 0 ) $
(see \eqref{eq:defHk}) for $ \alpha \in \mathcal A $, that is for $ \alpha$'s at which 
we have perfectly flat bands. The eigenfunctions are given by 
\begin{equation}
\label{eq:rank2}  \Phi := \begin{pmatrix} u \\ v \end{pmatrix} , \ \  
u \in \ker_{ H^1_0 } ( D ( \alpha ) + k ) , \ \  v \in \ker_{H^1_0 } ( D( \alpha)^* + \bar k ) , \ \
H_k  ( \alpha ) \Phi = 0 . \end{equation} 
The two components $ u $, $ v $, are completely decoupled and hence we can consider
them separately.  Symmetries of $ D ( \alpha ) $ (see \cite[\S 2.2]{bz2} for 
a quick review) show that we only need to consider $ \ker_{ H^1_0 } ( D ( \alpha ) + k ) $. 
As we already mentioned, the nontrivial topology implies blow up of moments of 
Wannier functions corresponding to lack of localization -- see  \cite[Theorem 9, \S 8.5]{notes} 
and references given there. 

We now assume that $ \alpha \in \mathcal A $ and that 
\begin{equation}
\label{eq:multiass}     1 \leq m ( \alpha ) \leq 2 ,
\end{equation}
that is the band has multiplicity one or two in the sense of \S \ref{s:flat}. 
In view of Theorem \ref{t:simple} and \cite[Theorem 4]{bhz3} we have 
\begin{equation}
\label{eq:defVk}  V ( k ) := \ker_{ H^1_0 }  ( D ( \alpha ) + k ) \subset L^2_0 , \ \ 
\dim V ( k ) = m ( \alpha ) , \ \ k \in \mathbb C ,
\end{equation}
and we can define a trivial vector bundle $ \widetilde E \to \mathbb C $ of rank $ m ( \alpha ) $:
\[    \widetilde E := \{ ( k , v ) : v \in V ( k ) \} \subset \mathbb C \times L^2_0 ( \mathbb C/\Lambda; \mathbb C^2 )  . \]
To define a vector bundle over the torus $ \mathbb C /\Lambda^* $ we need an equivalence relation
on $ \mathbb C \times L^2_0 ( \mathbb C/ \Lambda ; \mathbb C^2 ) $ based on 
\begin{equation}
\label{eq:deftau} 
\begin{gathered} 
 \tau(p)^* H_k ( \alpha ) \tau ( p ) = H_{k+p} ( \alpha ) , \ \ \ 
\tau ( p )^* ( D ( \alpha ) + k ) \tau ( p ) = D ( \alpha ) + k + p , \\
 \tau ( p )^{-1}  V ( k ) = V ( k + p ) , \ \ \ 
  [ \tau ( p ) u ] ( z ) := e^{ i \langle z, p \rangle } v ( z ) , \ \ p \in \Lambda^* . 
  \end{gathered}
  \end{equation}
It is given as follows:
\begin{equation}
\label{eq:deftau1} 
\begin{gathered} \exists \, p \in\Lambda^* \   ( k, u ) \sim_\tau ( k + p , \tau ( p )^{-1} u ) \end{gathered}
\end{equation}
 Using this (see \cite[Lemma 8.4]{notes} or
\cite[Lemma 5.1]{bhz2}), 
\begin{equation}
\label{eq:defE}     E := \widetilde E\, /\sim_\tau  \, \to \mathbb C /\Lambda^* . 
\end{equation}
is a holomorphic vector bundle over $ \mathbb C/\Lambda^* $. In the case of $ m ( \alpha ) = 1$
(and up to precise definitions) this observation was made by Ledwith et al \cite{led}. In view
of \eqref{eq:ukz} the line bundle can be identified with a theta bundle over the torus -- 
see \cite[\S 5.3]{bhz2}.

A natural connection on this vector bundle can be defined either as the Chern connection
or the Berry connection, as they are equal in the holomorphic case -- see \cite[\S 9, Proposition 9.1]{bhz3} 
for a detailed presentation and definitions. The scalar curvature of this connection 
is a two form on $ \mathbb C/\Lambda^* $,
\begin{equation}
\label{eq:defH}  \tr \Theta = H ( k )  d \bar k \wedge d  k , \end{equation}
see \cite[\S 9]{bhz3}. Here $ \Theta $ is the curvature form taking values in $ \Hom ( E, E ) $.
The following observations were made in \cite[\S 5.2]{bhz2} and \cite[\S 9.3]{bhz3}:
\begin{equation}
\label{eq:propH}
H ( k ) \geq 0 , \ \  \  H ( \omega k ) = H ( k ) , \ \ \  H ( k ) = H ( -k ) . 
\end{equation}
In particular, $ \mathcal K = \{ 0 , K , - K \}  $ (see \eqref{eq:defK}) is contained in the set of
critical points of $ H $. 

\begin{op}
\label{16}  Show that for the potential \eqref{eq:defU} (or for
a more general class of potentials?) and $ \alpha \in \mathcal A \cap \mathbb R $
(or simply for $ \alpha_1 $ in \eqref{eq:defa1}), 
$ \mathcal K $ is the set of all critical points
of $ H ( k ) $, and that the maximum is attained at $ 0$ (the $ \Gamma $ point) and
the minimum at $ \pm K $ (the $ K $-points):
\begin{center}
\includegraphics[width=10cm]{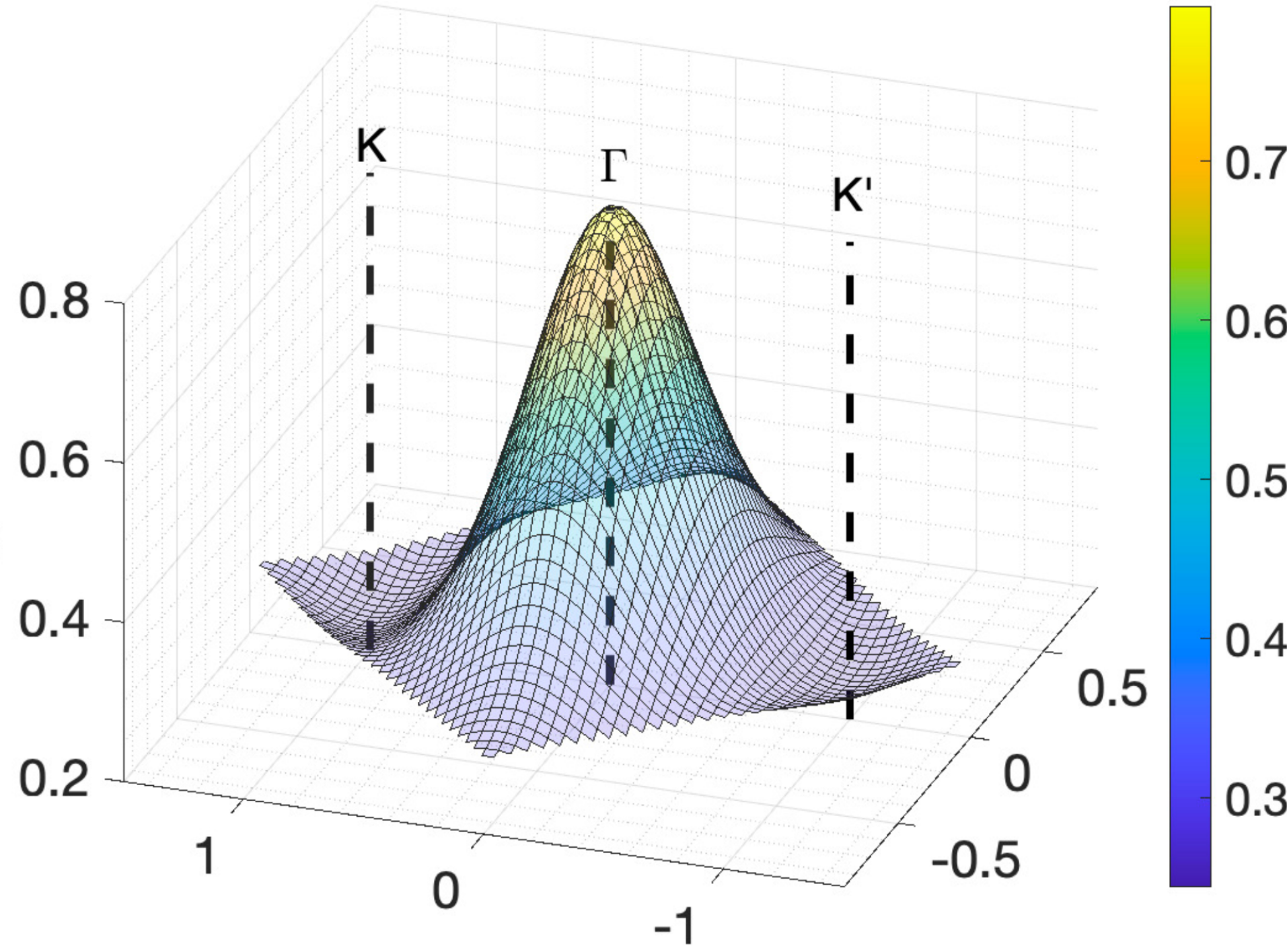}
\end{center}
For a discussion of analogous issues when multiplicity is equal to $ 2 $, see
\cite[\S 10.2]{bhz3}. 
\end{op}

The Chern number for complex vector bundles over a torus is defined using \eqref{eq:defH}:
\begin{equation}
\label{eq:chern}  c_1 ( E ) := \frac{i}{ 2 \pi } \int_{\mathbb C /\Lambda^* }  \tr \Theta = 
 - \frac{1}{  \pi } \int_F H ( k ) d m( k) , \end{equation}
where $ F $ is a fundamental domain of $ \Lambda^* $ and $ d m ( k ) = dx dy $, $ k = x+iy$,
the Lebesgue measure. We have $ c_1 ( E ) \in \mathbb Z $ (see \cite[Theorem 6]{notes} and references
given there) 
and if $ c_1 ( E ) \neq 0$ then the vector bundle is {\em non-trivial}, that is it is not
homeomorphic to $ \mathbb C /\Lambda^* \times \mathbb C^n $. For complex vector bundles 
over tori $ c_1 ( E ) $
is the only topological invariant. (For instance, 
for a simple $ \alpha $ we could consider the complex vector bundle defined
using $ \ker_{ H^1_0 ( \mathbb C, \mathbb C^4 ) } H_k ( \alpha ) $, see \eqref{eq:rank2}.
Its Chern number vanishes and the bundle is trivial.)

For simple $ \alpha $'s a evaluation of $ c_1 ( E) $ follows easily from \eqref{eq:ukz} 
-- see \cite{led} for a direct calculation and \cite[(5.9),(B.8)]{bhz2} for an argument based
on general principles. It turns out \cite[Theorem 5]{bhz3} that the Chern number does not change
if $ \alpha $ is double:

\begin{theo}[\cite{bhz2},\cite{bhz3}]
Suppose that \eqref{eq:multiass} holds and that the complex vector bundle $ E $ is defined
by \eqref{eq:defE}. Then the Chern number defined in \eqref{eq:chern} is given by 
\begin{equation}
\label{eq:chern1}  c_1 ( E ) = -1 .
\end{equation}
\end{theo}

Yang \cite{yang} provided a mathematical justification of the Chern number calculation
in \cite{yphys1},\cite{yphys2} (and of other issues related to flat bands in their setting) 
for two twisted $ n$-layer wafers of graphene. In that
case, the analogue of the line bundle \eqref{eq:defE}  satisfies $ c_1 ( E ) =  -n $. 

\begin{op}
\label{17}  Does \eqref{eq:chern1} hold without the assumption 
\eqref{eq:multiass}? 
\end{op}

\section{Dynamics of Dirac points for in-plane magnetic field}
\label{s:dynip}

Interesting mathematical phenomena arise when a constant magnetic field in the
direction parallel to the two twisted layers of graphene are added. Following
Kwan--Parameswaran--Sondhi \cite{kps} and Qin--MacDonald \cite{wema} the new 
Hamiltonian for the chiral model is given by 
\begin{equation}
\label{eq:defHB}  H_B ( \alpha ) := \begin{pmatrix} 0 & D_B ( \alpha )^* \\
D_B ( \alpha ) & 0 \end{pmatrix},  \ \ \  D_B ( \alpha ) := D ( \alpha ) + \mathcal B, \ \ \
\mathcal B := \begin{pmatrix} B &\ \  0 \\  0 & - B \end{pmatrix} , \end{equation}
where $ B = |B| e^{ 2 \pi i \theta } $ with $| B| $ corresponding to the strength
of the magnetic field and $  2 \pi \theta $ is its, in-plane direction; $ D ( \alpha ) $
is the same as in \eqref{eq:Hamiltonian}. 

\begin{figure}
{\begin{tikzpicture}
\node at (-2,0) {\includegraphics[width=7.6cm]{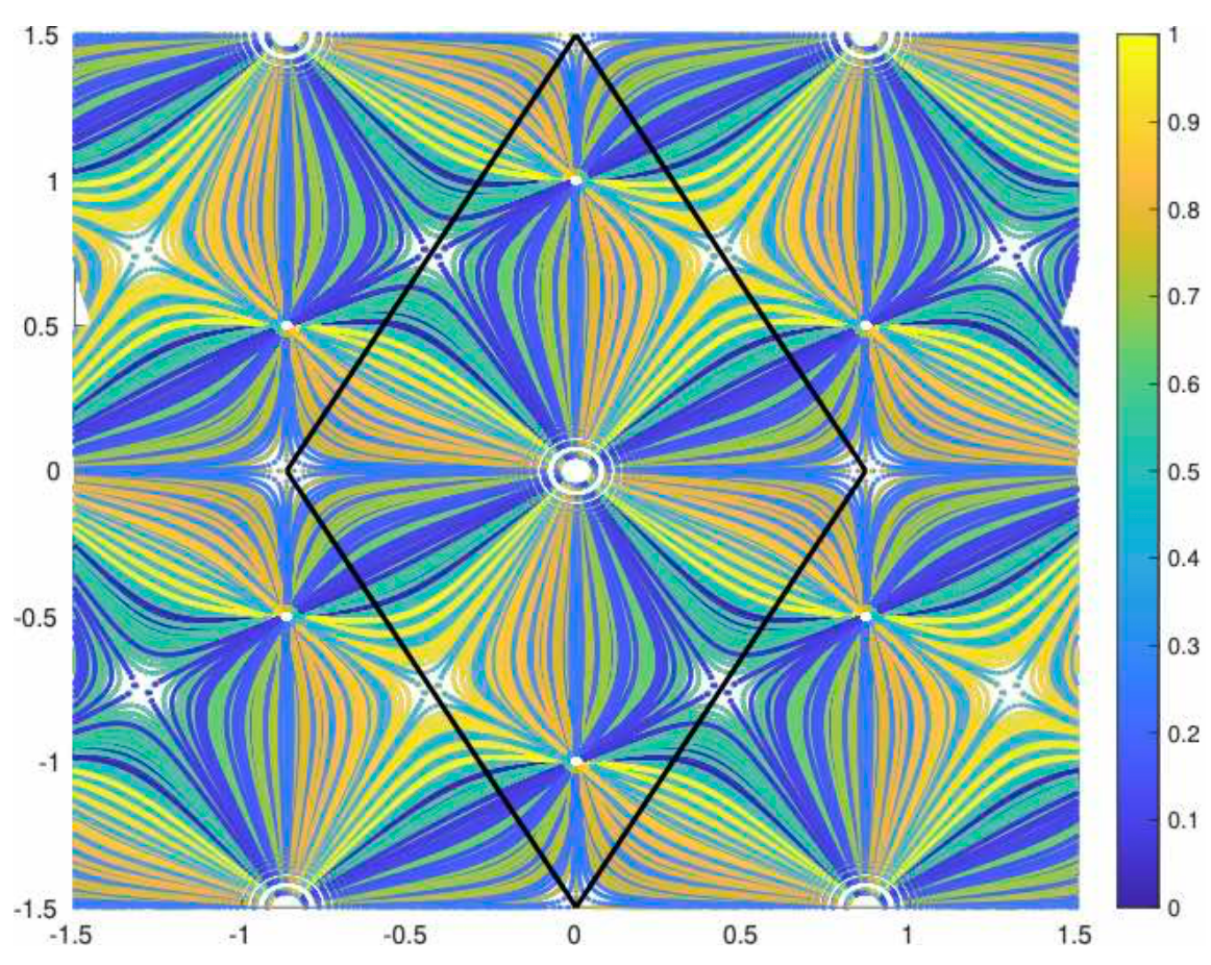}};
\node at (7.6-2,0) {\includegraphics[width=7.6cm]{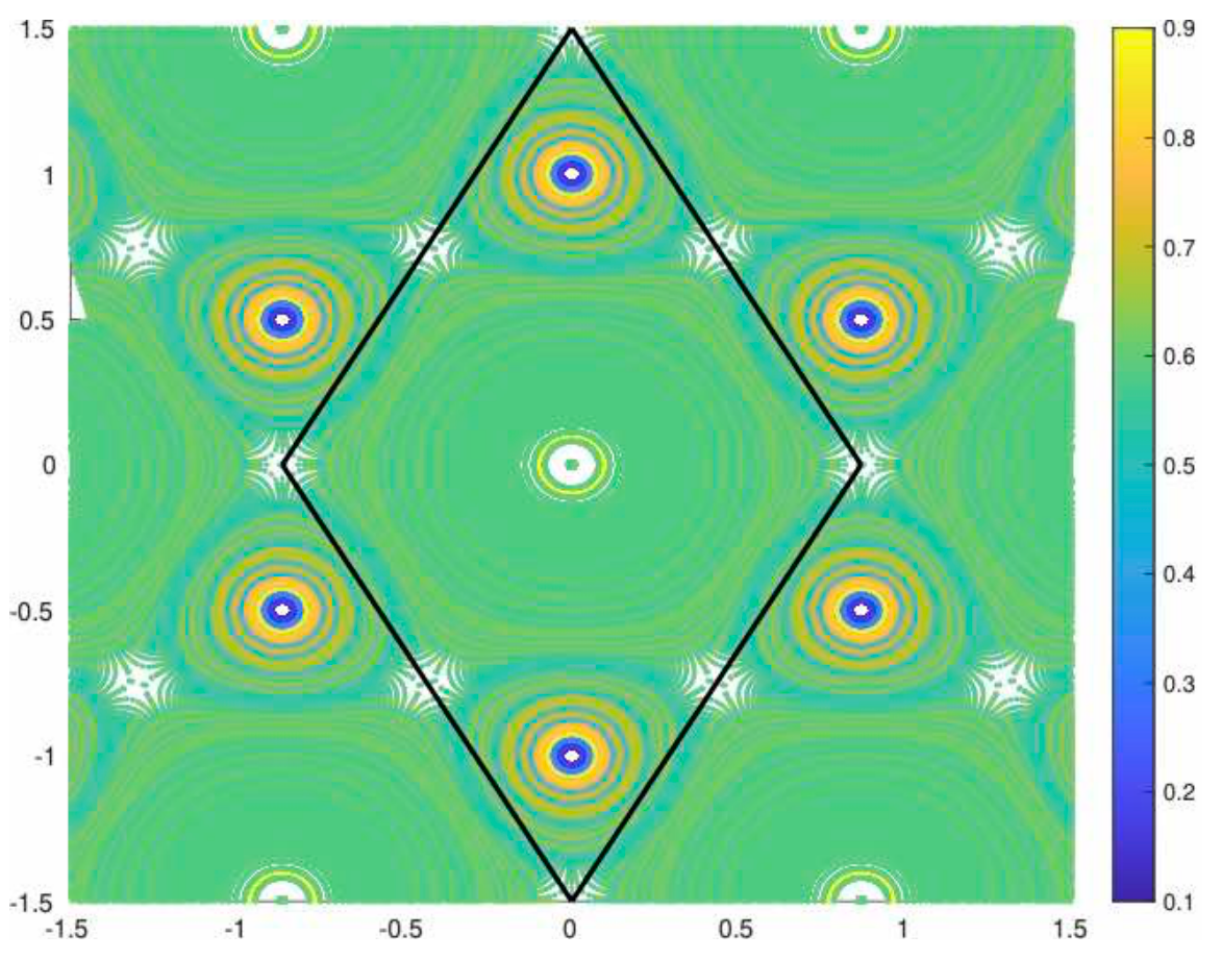}};
\node at (1.3,-3) {$\theta$}; 
\node at (7.6+3.5-2.25,-3.0) {$\alpha$};
\end{tikzpicture}}
\caption{\label{f:HB} The dynamics of Dirac points 
for $ H_B $ in  \eqref{eq:defHB} with the BM potential \eqref{eq:defUV}.  The magnetic field  given by 
$B=B_0 e^{2\pi i \theta}$ with $ B_0 = 0.1 $ 
Colour coding (shown in colour bars) corresponds to different values of $\theta $ on the left, 
and different values of $ \alpha $ on the right. In the left figure $ \alpha $ varies between $ 0.1 $ and $ 0.9 $ and curves of different colour trace the corresponding Dirac points -- see 
\url{https://math.berkeley.edu/~zworski/B01.mp4} for an animated version. When 
$ 3 \theta \in \mathbb N $ we showed in \cite[Theorem 3]{bz1} that the Dirac points
move along straight lines -- see \url{https://math.berkeley.edu/~zworski/Rectangle_1.mp4}
where $ \theta = \frac13 $.
In the right figure, $ \theta $ varies and curves of different colour trace the corresponding
Dirac points. The predominance of green 
(corresponding to the range between 0.5 and 0.6) means that most of the motion happens
near the (first) magic alpha -- for the dance of Dirac points  for fixed $ B $  and as $ \alpha $ varies, 
see 
\url{https://math.berkeley.edu/~zworski/first_band.mp4} 
which shows $ E_1 ( \alpha, k ) /\max_k E_1( \alpha, k )$. (The boundary Brillouin zone is also 
shows; we take the image of the $k$ plane by the map $ k \mapsto z ( k ) $, see \eqref{eq:defz}
so that $ \Lambda^* $ is mapped to $ \mathbb Z + \omega \mathbb Z$.)
}
\end{figure}

For the BMH and the chiral model, the bands close to zero touch at $ 0$ at $ \pm K $ (see
\eqref{eq:defK} -- this are the $ K$-points in our coordinates) and the intersection is expected
to be conic (except for the perfectly flat bands), that is we see two Dirac points  -- 
see Open Problem \ref{2} and the figure there.
Theorem \ref{t:bands} shows that for the chiral model, $ H ( \alpha ) = H ( \alpha, 0 ) $ 
in the notation of \eqref{eq:defBM}, once the bands touch $ 0 $ away from $ \pm K$, 
the bands are perfectly flat.

\begin{figure}
{\begin{tikzpicture}
   \node at (-11,0) {\includegraphics[trim={2.7cm 0 0 0},width=6.2cm]{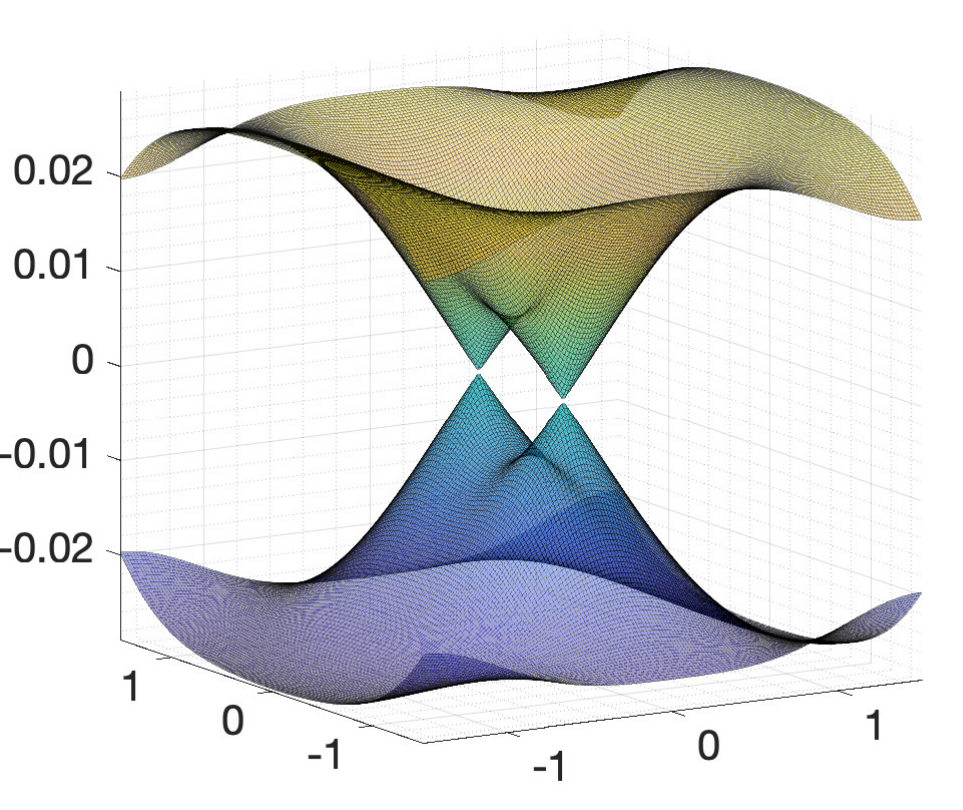}};  
    \node at (-4,0){\includegraphics[trim={1.7cm 0 0 0},width=6.5cm]{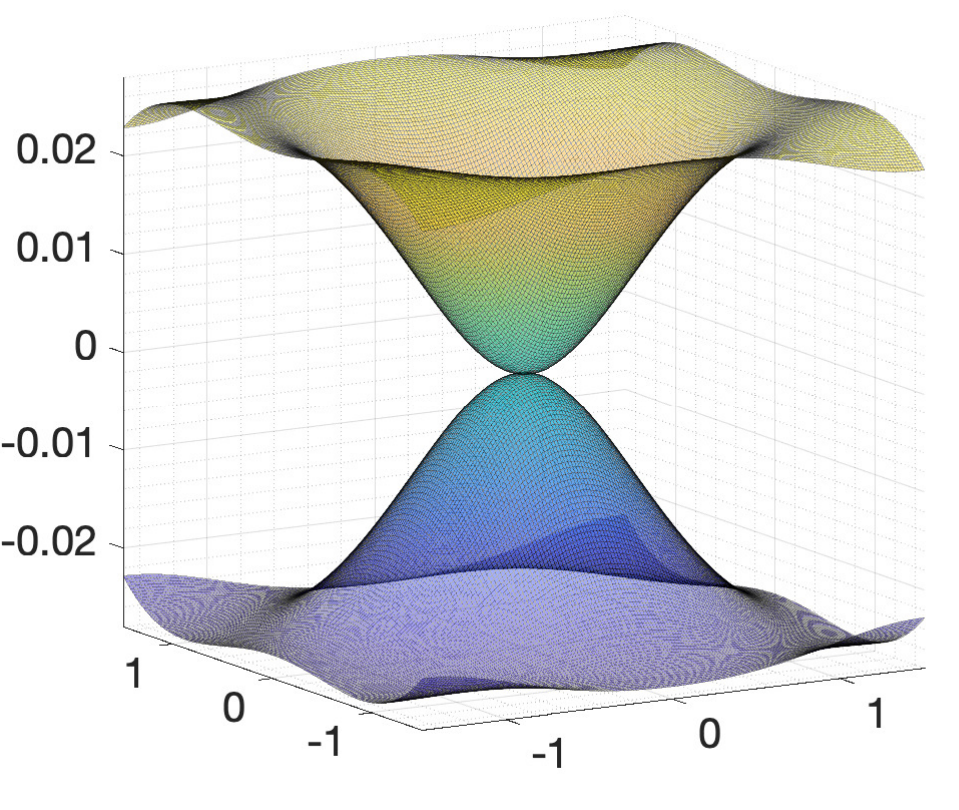}};
     \node at (-3.5+1,-3){$\Im k$};
     \node at (-8+0.8,-2.7){$\Re k$};
     \node at (-9+1.4,0.2){$E$};
       \node at (-3.5-6,-3){$\Im k$};
     \node at (-8-6.3,-2.7){$\Re k$};
     \node at (-16+1,0.2){$E$};
  \end{tikzpicture}}
\caption{\label{f:qbcp} 
When $ B $ is real
in \eqref{eq:defHB} the two Dirac cones approach $ \Gamma $ point as $ \alpha \to \alpha^* = \underline \alpha + \mathcal O ( B^3 ) $ ($ \underline \alpha $ a simple
real magic parameter) on the line $ \Im k= 0 $ (left). For 
$ \alpha = \alpha^* $, the quasi-momentum $ k $ at which the bifurcation happens are 
the boundary of the Brillouin zone and the $ \Gamma $-point which is shown in the figure (right).
The animation \url{https://math.berkeley.edu/~zworski/Rectangle_1.mp4} shows the motion of 
Dirac points in this case.}
\end{figure}
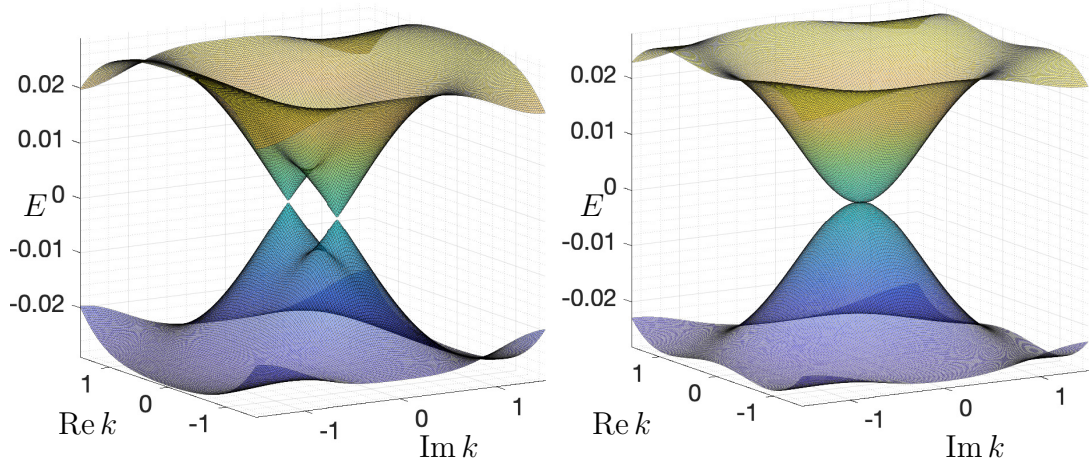

It was observed numerically in \cite{kps} that for the chiral model with in-plane magnetic
field \eqref{eq:defHB} flat bands disappear when $ B \neq 0 $ and the two Dirac points
 move.  Moreover for $ \alpha \in \mathcal A $ the Dirac points seem to coalesce at 
 the $ \Gamma $ point forming a quadratic band crossing point (QBCP) -- see Figure \ref{f:qbcp}.
 In \cite{bz1} we provided a more precise description of the dynamics of Dirac points for 
 small magnetic fields.  In particular finer analysis and numerical evidence suggest that exact
 QBCP appear only when Dirac points move along straight lines which happens when 
 $ 3 \theta \in \mathbb N $ (the direction of the magnetic field is given by $ 2 \pi \theta $)
 -- see \cite[Theorem 3 and Figure 5]{bz1}. 
 
The reason for the Dirac points appearing close to $ \Gamma $ when $ \alpha $ is close
to (simple) elements of $ \mathcal A $, can be elegantly described using properties
of theta functions. Since it is a simple consequence of \eqref{eq:ukz} and \eqref{eq:u*kz}
we recall it, referring to \cite[\S 4]{bz1} for additional details. This also allows to 
present an approach to perturbation theory based on Schur's complement formula
(via Grushin problems in the terminology of Sj\"ostrand who turned Schur's complement
formula into a systematic tool) -- see \cite[\S 2.6]{notes}. Same
approach is used to obtain Theorem \ref{t:pert}.

Suppose that $ \alpha \in \mathcal A $ is simple, and in the notation of \eqref{eq:ukz} and
\eqref{eq:u*kz} the operator (see \S \ref{s:intr} for the review of notation)
\[  \mathscr D ( \alpha, k  ) := \begin{pmatrix} D ( \alpha ) + k & | u^* ( k ) \rangle \\
\langle u ( k ) | & 0 \end{pmatrix} : H^1_0 \times \mathbb C \to L^2_0 \times \mathbb C , \]
is invertible with the inverse given by 
\[  \mathcal E ( \alpha, k ) = \begin{pmatrix} E ( k ) & | u ( k  ) \rangle \\
\langle u^* ( k ) | & E_{-+} ( k )   \end{pmatrix} :  L^2_0 \times \mathbb C \to H^1_0 \times \mathbb C , \]
where $ E_{-+} ( k ) \equiv 0 $ is the effective Hamiltonian: from Schur's complement formula 
\cite[(2.15)]{notes} we see that $ D ( \alpha ) + k $ is invertible iff and only iff $ E_{-+} ( k ) = 0 $. 
Since $ \alpha \in \mathcal A $, $ \Spec_{ L^2_0 }D( \alpha ) = \mathbb C $ this is consistent
with $ E_{-+} ( k ) \equiv 0 $. For $ |B| \ll 1 $, we can consider $ D_B ( \alpha ) $ as a perturbation 
of $ D ( \alpha ) $ and we still have invertibility 
\[ \begin{gathered}  \begin{pmatrix} D_B  ( \alpha ) + k & | u^* ( k ) \rangle \\
\langle u ( k ) | & 0 \end{pmatrix}^{-1} = \begin{pmatrix} E^B ( k ) & E_+^B ( k ) \\
E_-^B ( k ) & E_{-+}^B (k ) \end{pmatrix}, \\
E_{-+}^{B} ( k ) = - \langle u^* ( k ) | \mathcal B | u ( k ) \rangle + \mathcal O ( B^2 ) ,
\end{gathered}  \]
see \cite[Proposition 2.12]{notes}. From \eqref{eq:ukz} and \eqref{eq:u*kz} we then
obtain that
\[  E_{-+}^B ( k ) = - c(k)^{-2} B ( G ( k ) + \mathcal O ( B ) ) , \ \ \
G ( k ) = 2 \int_{\mathbb C/\Lambda } F_k ( z ) F_{-k} ( z ) \varphi_0 ( z ) \psi_0 ( z ) dm ( z) , \]
where $ F_k $ is defined in \eqref{eq:defFk}. This definition combined with a theta function 
identity
\[ \theta ( z + u ) \theta ( z - u ) \theta ( \tfrac12 )^2= \theta^2 ( z ) \theta^2 ( u + \tfrac12 ) - \theta^2 ( z +
\tfrac12 ) \theta^2 ( u) , \]
and symmetries of $ \psi_0 $ and $ \varphi_0 $ (see \S \cite[\S 4.1]{bz1}) gives 
\begin{equation}
\label{eq:G2g}  G( k ) = g_0 \, \frac{ \theta ( z ( k ) )^2}{ \theta ( \tfrac12 ) ^2 } , \ \ \ 
g_0  =  g_0 ( \underline \alpha )  := 
2 \int_{\mathbb C/\Lambda }   \ \theta ( z + \tfrac12 )^2  \frac{ \varphi_0 ( z ) \psi_0 ( z )}{ \theta ( z ) ^2 }    dm ( z).
\end{equation}
For the Bistritzer--MacDonald potential and the first magic angle, $ \alpha_1 $ 
(see Theorem \ref{t:exi})  $ |g_0| \simeq 0.07 \neq 0 $. We now see that $ k $ is a Dirac point for \eqref{eq:defHB} with $ \alpha = \alpha_1 $  if and only if $ E_{-+}^B ( k ) = 0 $, and in particular
\begin{equation}  
\label{eq:B2th}   
k \in \Spec_{ L^2_0 } D_B ( \alpha_1 ) \ \Longrightarrow \ \theta( z ( k ) )^2 + \mathcal O ( B ) = 0  . \end{equation}
(For $ g_0 ( \alpha ) $ for other real magic $ \alpha$'s see \cite[Table 1]{bz1}.)

Since $ \theta ( z ( k ) )^2 $ vanishes quadratically at $ 0 $ (the $ \Gamma $ point), 
equation  \eqref{eq:B2th} shows that at $ \alpha = \alpha_1 $ and for $ B$ small 
the Dirac points are near the $ \Gamma $ point.  It also suggests QBCP -- see Figure~\ref{f:qbcp} and
\cite[\S 5]{bz1} for a discussion of the bifurcation at $ \Gamma $ and other points. 

The study of the effective Hamiltonian $ E_{-+}^B ( k ) $ (a scalar function in our case) and some additional
arguments give the following result (see \cite[\S 2]{bz1} for more detailed statements):

\begin{theo}
\label{th:Gamma}
Suppose $ \underline \alpha \in \mathcal A $ is simple and $ g_0 ( \underline{\alpha} ) \neq 0 $
where $ g_0 $ is defined in \eqref{eq:G2g}. 
Then there
exists $ \delta_0 > 0 $  such that for $ 0 < |B | < \delta_0 $ and  $  | \alpha - \underline{\alpha } | <  \delta_0  $,
the spectrum of $ D_B ( \alpha )$ on $ L^2_0 $ is discrete (that is the set of Dirac points) and 
\begin{equation}
\label{eq:count}  \vert \Spec_{L^2_0}(D_B(\alpha)) \cap \CC/\Gamma^* \vert =2  ,
\end{equation} 
where the elements of the spectrum are included according to their (algebraic) multiplicity. 
In addition, for a fixed constant $ a_0 > 0 $ and for every $ \epsilon $ there exists $ \delta $ such that
for $ 0 < |B | < \delta $, $ |\alpha - \underline \alpha | < a_0 \delta |B| $, 
\begin{equation}
\label{eq:nearG}  \Spec_{ L^2_0 } ( D_B ( \alpha ) ) \subset \Lambda^* + D (0 , \epsilon ) ,
\end{equation}
where we recall that elements of $ \Lambda^* $, in particular $ 0 $, correspond to the $ \Gamma $ point.
\end{theo}

A more detailed description would be very desirable. Among things which were left open in \cite{bz1} 
is the behaviour near $ K$ points when $ 3 \theta\in \mathbb N $ -- see \cite[Figure 5]{bz1}. 
We state one, somewhat vaguely formulated, problem:

\begin{op}
\label{18} Is there a dynamical system which fully explains Figure~\ref{f:HB}?
Basic symmetries of Dirac points are described in \cite[(2.10)]{bz1} but the clean 
structure may be due to the special BM potential \eqref{eq:defUV}. It becomes more complicated
for other potentials -- see \cite[Figure 1]{bz1}.
\end{op}

The quantitative behaviour of Dirac points seems to remain similar for BMH and clarifying that 
would also be nice. The agreement is particularly striking for $ 3 \theta \in \mathbb N $ -- 
see \url{https://math.berkeley.edu/~zworski/Dirac_BMH.mp4} ($ \alpha_0 = \lambda $, $ \alpha_1 = \alpha$) 
where a
comparison of the movement of 
Dirac points for chiral, weakly interacting, and BMH ($ \lambda = 0.7 \alpha$) is animated. It is harder to catch Dirac points when $ \lambda \neq 0 $ as we do not have a simple characterization as
spectrum of $ D_B ( \alpha ) $ on $ L^2_0 $. Hence the neighbourhoods  of the Dirac points are shown.

\section{Small angle limit as a semiclassical limit}
\label{s:semi}

The small angle limit corresponds to letting $ \alpha \to \infty $. In that case it is
natural to write $ \alpha =  \lambda/ h $, $ h \in (0 , 1] $, $ \lambda \in K \Subset \mathbb C \setminus 0$
and consider asymptotic behaviour as $ h \to 0 $. When consider real and positive alpha we can simply take $ \lambda = 1 $. 

The operator $ D ( \alpha ) $ in \eqref{eq:Hamiltonian} then becomes (up to an irrelevant factor of $ h^{-2} $)
\begin{equation}
\label{eq:defP}   P ( x, h D ) := \begin{pmatrix} 2 h D_{\bar z } & \lambda U ( z ) \\
\lambda U ( - z ) & 2 h D_{\bar z } \end{pmatrix} , \ \ \ D_{\bar z } = (1/2i) ( \partial_{x_1} + i \partial_{x_2 }) ,  \end{equation}
which is a semiclassical differential system in the sense of \cite[Appendix E.1.1]{res}. Its matrix 
valued symbol is  given by 
\begin{equation}
\label{eq:symbp}  p ( x, \xi ) = \begin{pmatrix} 2 \bar \zeta & U ( z ) \\
U (-z ) & 2 \bar \zeta \end{pmatrix}, \ \ \ z = x_1 + i x_2, \ \ \zeta = \tfrac12 ( \xi_1 - i \xi_2 ) . 
\end{equation}

Theorem \ref{t:bands} shows that (with $ H^1_0 = H^1_{\rm{loc}} \cap L^2_0 $ defined in 
\eqref{eq:Lk}) 
\begin{equation}
\label{eq:semA}   \begin{split}  h \lambda \in \mathcal A  \ & \  \Longleftrightarrow \ \ \Spec_{L^2_0 }  P ( x, h D ) = \mathbb C   \\ \ & \
\Longleftrightarrow \ \ \exists \, u \in H^1_0, \ u \neq 0 , \ \ P ( x, h D ) u = 0. 
\end{split} \end{equation} 
We note that $ E_\ell ( \lambda/h , k )^2 $, defined in \eqref{eq:specHk} (essentially 
the bands of $H ( \alpha ) $),   are the eigenvalues of the self-adjoint operator
\begin{equation}
\label{eq:defP2} 
 P_2 ( x, h D, hk ) :=  ( P ( x , h D ) + h k )^* ( P ( x, h D ) + h k ) . 
 \end{equation} 
 Since we only need to consider $ k $
in a fundamental domain of $ \Lambda^* $, $ hk $ is a lower order terms when $ h \to 0 $.

In \S \ref{s:squeeze} we will see one reason for the difficulty of finding $ \lambda$'s 
with  exactly $3 \Lambda$--periodic solutions to $ P ( x, h D ) u = 0 $ (or $ u \in L^2_0 $)
when $ h $ is small, that is, the difficulty of using \eqref{eq:semA}
 to characterize magic $ \alpha =  \lambda/h $. 

Instead of \eqref{eq:semA} one could attempt to analyse semiclassical the spectral characterization 
of Theorem \ref{t:spec}: for $ k \notin \mathcal K  $ (see \eqref{eq:defK}; we could take $ k = 0 $), 
\[  \lambda h \in \mathcal A \ \Longleftrightarrow \ \lambda^{-1} \in \Spec_{ L^2_0 } 
\left( ( 2h D_{\bar z } - h k )^{-1} W ( z ) \right) , \ \ \ W ( z ) := \begin{pmatrix} 0 & U ( z ) \\
U ( - z) & 0 \end{pmatrix} , \]
which of course seems like a tautology. The problem here lies in the fact that 
$ (2 h D_{\bar z } - h k )^{-1} $, with the Schwartz kernel explicitely given in \eqref{eq:defFk}, 
is essentially independent of $ h $ and is {\em not} a semiclassical pseudodifferential 
operator: $ hk $ is a lower order term and the symbol of $ 2 h D_{\bar z }$, $ 2 \bar \zeta $
has all of $ \mathbb C $ as its range. 
\begin{figure}
\includegraphics[width=7.15cm]{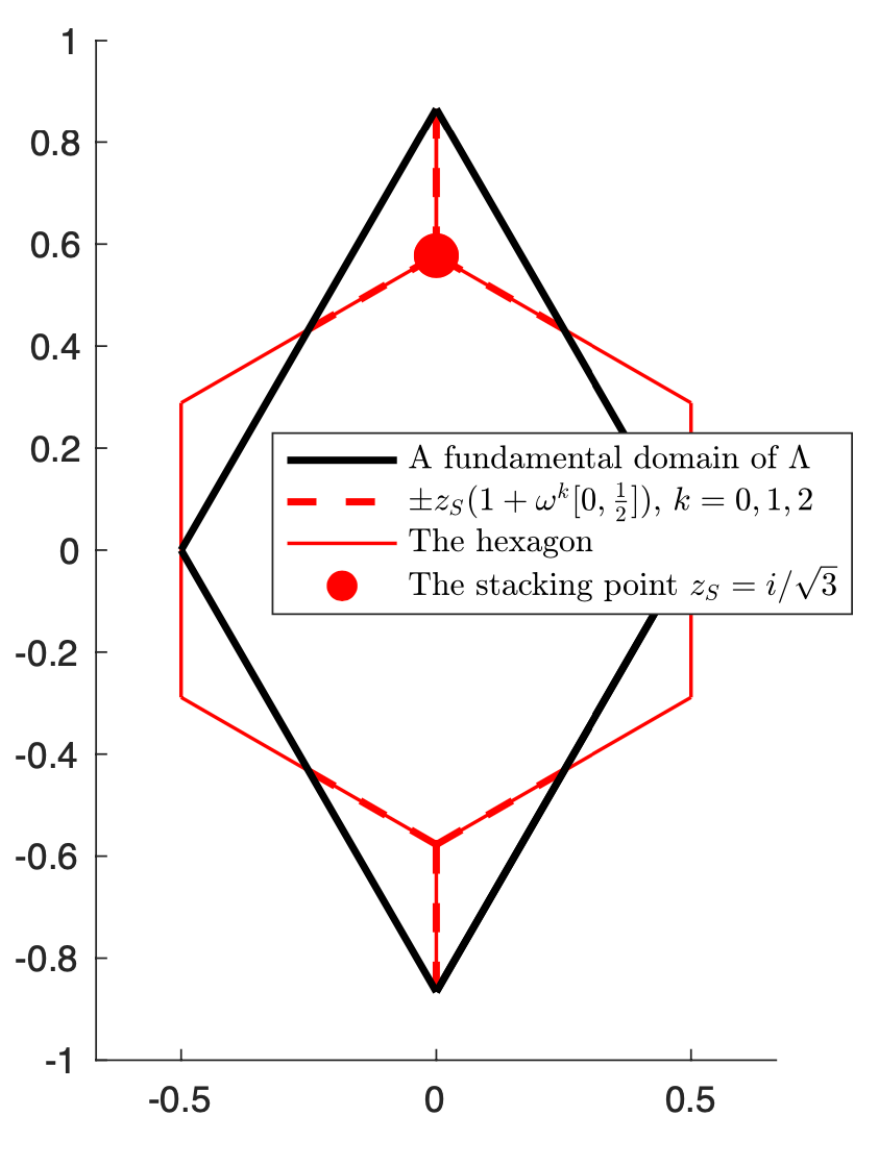} \includegraphics[width=7.425cm]{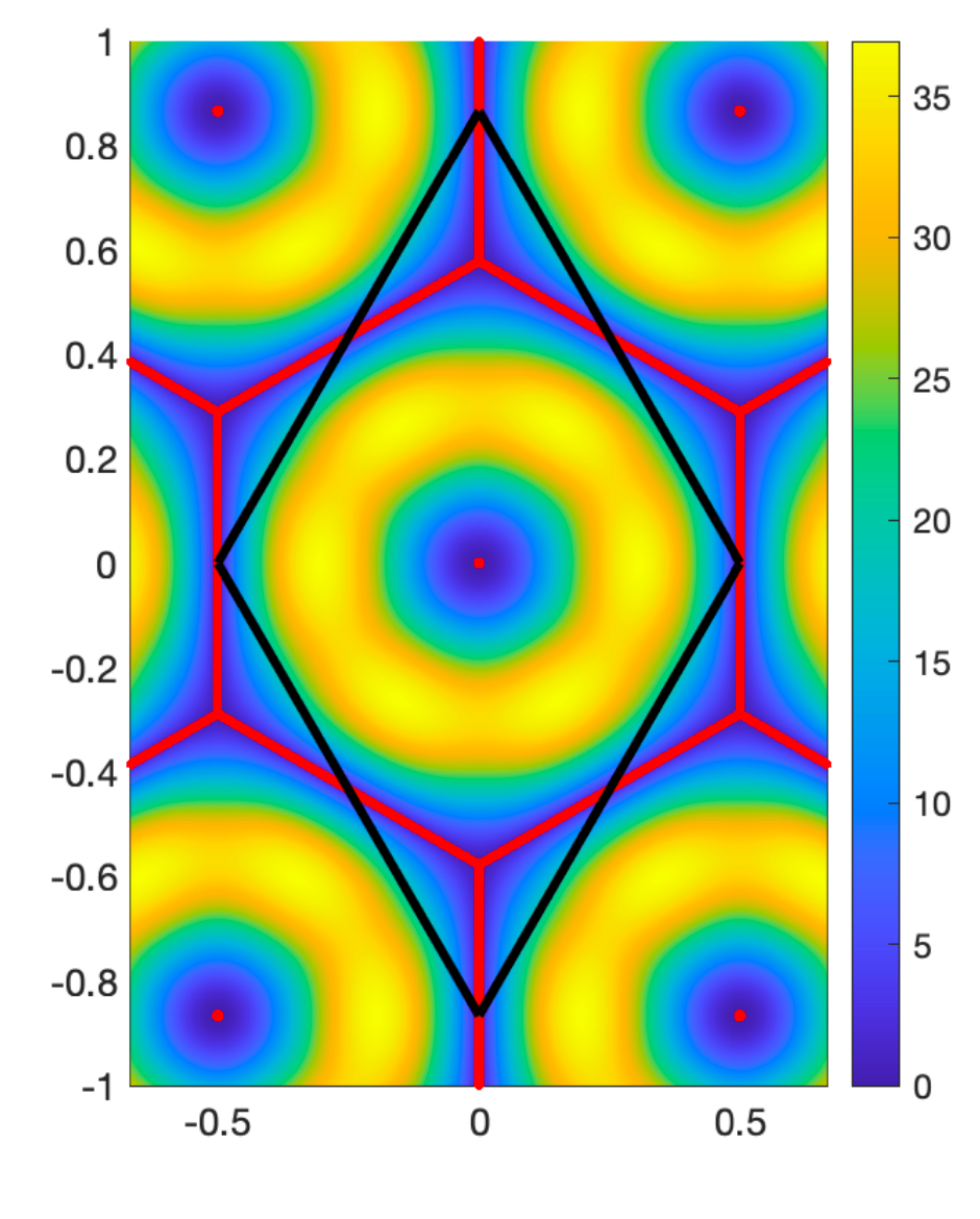}
\caption{Left: the vertices of the hexagon in a fundamental domain of $ \Lambda $
are given by the
{\em stacking points} $ \pm z_S $, $ z_S = i/\sqrt 3 $ (we use the coordinates of
\S \ref{s:stan}). They are non-zero
points of high symmetry in the sense that $ \pm \omega z_S \equiv \pm z_S \!\! \mod  \Lambda$.
Right: the contour plot of $ |\{ q, \bar q \}_{q^{-1} ( 0 ) } |$ for $ q $ given by the determinant of the 
semiclassical symbol of $ D ( \alpha ) $ (see \eqref{eq:defq}), $ \alpha = 1/h $; the set
where $ \{ q, \bar q \}_{q^{-1} ( 0 ) } = 0 $ is in \red{red}. We should stress that the structure of that
set becomes more complicated for other potentials $ U $ satisfying the required symmetries -- see
 \cite[Figure 6]{beta}.
\label{f:hiz}}
\end{figure}

We finally remark that Open Problem \ref{1} (and also \ref{8}) is semiclassical in nature: it states
a quantization rule 
\[  \lambda_{n+1} - \lambda_n  =  \gamma h + \mathcal O ( h^2 ) , \ \ \ \gamma \simeq \tfrac12. \]

\subsection{Exponential squeezing of bands}
\label{s:squeeze}
In \cite{beta} we observed that the results on the existence of localized quasimodes 
for non-normal semiclassical differential operators with analytic coefficients implies
existence of many exponentially small (as $ \alpha \to \infty $) Bloch eigenvalues 
for the chiral model. That means that as $ \alpha $ gets large it is hard to distinguish
an exactly flat band from many bands that seem flat. Since the phenomenon is 
semiclassical we use the notation of this section:
\begin{theo}[\cite{beta}] 
\label{t:sq}
Suppose that $ U $ is given by \eqref{eq:defU} and $ E_\ell ( 1/h, k ) $ are defined
in \eqref{eq:specHk}. Then, there exist constants $c_0, c_1, c_2  > 0 $ such that
\begin{equation}
\label{eq:sq} 
| E_\ell ( 1/h, k )  | \leq c_1 e^{ - c_0 / h } , \ \ \ | \ell | \leq c_2/h . \end{equation}
\end{theo}

The proof is based on a result of Dencker--Sj\"ostrand--Zworski \cite[Theorem 1.2${}'$]{dsz}
(see also \cite[\S II.2.8]{HiS}) which in turn was based on works of
H\"ormander and of Kashiwara, Kawai, and Sato. Roughly,  it states the following
fact: suppose that $ Q ( x, hD , h ) $ is a (scalar) semiclassical differential operator 
with analytic coefficients and $ q ( x, \xi ) $ is its principal symbol. Then 
\begin{equation*}
q ( x_0, \xi_0 ) = 0 , \  \{ \Re q, \Im q \} ( x_0 , \xi_0 ) < 0  \Longrightarrow 
\left\{ \begin{array}{l}  \exists \, \ u ( h ) \in C^\infty,  \| u ( h) \|_{L^2 } = 1,  \\ \| Q ( x, h D , h ) u ( h ) \|_{L^2 } \leq
C e^{-C/h} , \\
\text{$ u ( h ) $ is {\em microlocalized} at $ ( x_0, \xi_0 ) $,} \end{array} \right. 
\end{equation*}
see \cite{dsz} and references given there. Here $ \{ a, b \} $ denotes the {\em Poisson bracket} 
which in our $2D$ case and using the notation $ z$ and $ \zeta $ in \eqref{eq:symbp} is
given by
\[ \{ a, b \} = \partial_\zeta a\, \partial_z b  - \partial_\zeta b\, \partial_z a + \partial_{\bar \zeta } a\,
\partial_{\bar z } b - \partial_{\bar \zeta } b\, \partial_{
\bar z } a , \]
see \cite[\S 2.4]{semi} for an introduction to its geometric significance.

The type microlocalization we have for 
$ u( h)$, implies in particular that  that $ |u ( h, x ) | \leq e^{ - |x-x_0|^2/Ch } $, which
means that $ u (h )$ ``lives" in $ B( x_0, h^{\frac12 - \varepsilon}  ) $, for any $ \epsilon> 0 $. 
From such local approximate solutions we can built many approximate solutions with 
any periodicity properties. (A model to keep in mind is the annihilation operator 
$ Q ( x, h D ) = h D_{x_1 } - i x_1 $ with $ ( x_0, \xi_0 ) = ( 0 , 0 )\in \mathbb R^2 \times \mathbb R^2  $; 
we can then 
take $ u ( h , x ) = c ( h ) e^{ -x_1^2/h - x_2^2/h } $.) 

At points $z_0 $ with $ U ( z_0 )  \neq 0 $ an easy reduction (see \cite[Proof of 
Proposition 4.1]{beta}) shows that to construct $ u ( h ) \in C^\infty ( \mathbb C ; \mathbb C^2 ) $
localized at $ z_0 $ and satisfying 
\begin{equation}
\label{eq:u2v}  \| ( P ( x, h D , h ) + h k ) u ( h ) \| \leq C e^{ -1/Ch } \| u ( h ) \|, 
\end{equation}
it is enough to find  $ v( h) \in C^\infty ( \mathbb C;\mathbb C ) $, 
localized to $ z_0 $ such that $\| Q ( x, h D_x , h ) v (h ) \|_{L^2}  \leq C e^{-1/Ch} \| v ( h ) \|_{L^2} $
where $ Q $ is a {\em scalar} operator with the principal symbol given by the determinant
of $ p $ in \eqref{eq:symbp}:
\begin{equation}
\label{eq:defq}  q ( x, \xi ) =  ( 2 \bar \zeta )^2 - \lambda^2 U ( z ) U ( -z )  , \ \ \
z = x_1 + i x_2, \ \ \zeta = \tfrac12 ( \xi_1 - i \xi_2 ) . \end{equation}
Hence, in view of the discussion above, we need to look for $ ( x_0 , \xi_0 )$ 
such that $ q (x_0, \xi_0 )  = 0 $ and $ \{ \Re q , \Im q \} ( x_0, \xi_0 ) < 0 $. Such points
are indeed plentiful -- see Figure \ref{f:hiz} for the case of $ \lambda = 1$ and 
\cite[\S 4]{beta} for more examples. 

Once we have \eqref{eq:u2v} we obtain an en exponentially accurate approximate solution to 
$ P_2 ( x, h D , hk ) u( h ) = 0 $ where $ P_2 $ was defined in 
\eqref{eq:defP2}. Self-adjointness of $ P_2 $ then implies existence of exponentially 
small eigenvalues. Using many localized approximate solutions we can bound their number
from below by $ 1/h $, see \cite[\S 4]{beta}. 

\begin{op}
\label{19}  Relate the geometry of level sets of
$ z \mapsto | \{ q,  \bar q \}|_{ q^{-1} ( z ) = 0 } | $ (see Figure \ref{f:hiz}) to the concentration 
of mass of the protected states $ u_K ( \lambda/ h ) $ (see Theorem \ref{t:prot})
 as $ \lambda $ varies in a compact set and $ h \to 0 $. For an animated example 
 see  {\rm \url{https://math.berkeley.edu/~zworski/bracket_dynamics.mp4}} where $ h = 1/8 $ and $ \lambda
 $ varies on a circle of radius $ 1 $. This problem is related to the issues discussed in 
 \S \ref{s:forb} below.
 \end{op}

\subsection{Classically forbidden regions}
\label{s:forb}
The contour plot of $ z \mapsto \log | u_K ( \alpha , z ) |$ as $ \alpha $ changes (and 
$ U $ is given in \eqref{eq:defUV}) as well as the link in Open Problem \ref{19}, suggest that
solutions to $  D ( \alpha ) + k ) u = 0$, $ u \in H^1_0 $ (nontrivial only for $ \alpha \in 
\mathcal A $ if $ k \not =  \pm K $) {\em decay exponentially} in $ \alpha $ near the hexagon spanned
by the stacking points (see Figure~\ref{f:hiz}) and near the center of the hexagon.
From the semiclassical point of view presented in this section this means decay
$ e^{-c/h }$ which typically corresponds to classically forbidden regions. 

The standard notion of classically forbidden regions is based on ellipticity: if 
$ Q  $ is a principally scalar semiclassical differential operator, elliptic in the classical sense (that is, 
for fixed $ h $), with analytic coefficients and a scalar principal symbol $ q ( x, \xi ) $ then 
(with $  {\rm{neigh}} (x_0 ) $ denoting {\em some} neighbourhood of $ x_0 $)
\begin{equation} 
\label{eq:defQ}  q|_{ \pi^{-1} ( x_0 ) } \neq 0, \ \ 
Q  u = 0 \text{ in $  {\rm{neigh}} (x_0 ) $},  \ \| u \|_{ L^2} = 1 \   \Longrightarrow  \ 
\| u \|_{ L^2 ( {\rm{neigh}} (x_0 ) )} \leq C e^{ -c/h} , 
\end{equation}
see \cite[Theorem 4.1.5]{mart}, \cite[Proposition 6.4]{hiz}. (A typical example is
given by $ Q = - h^2 \Delta + V ( x ) $ where $ V \in C^\infty $ is real valued -- there
is no need for analyticity in that case -- see \cite[Theorem 7.3]{semi}; in that
case the condition is simply that $ V ( x_0 ) > 0 $ as then for all $ \xi$, $q( x_0, \xi ) 
= \xi^2 + V ( x_0 ) > 0 $.) 

\begin{figure}
\includegraphics[width=14cm]{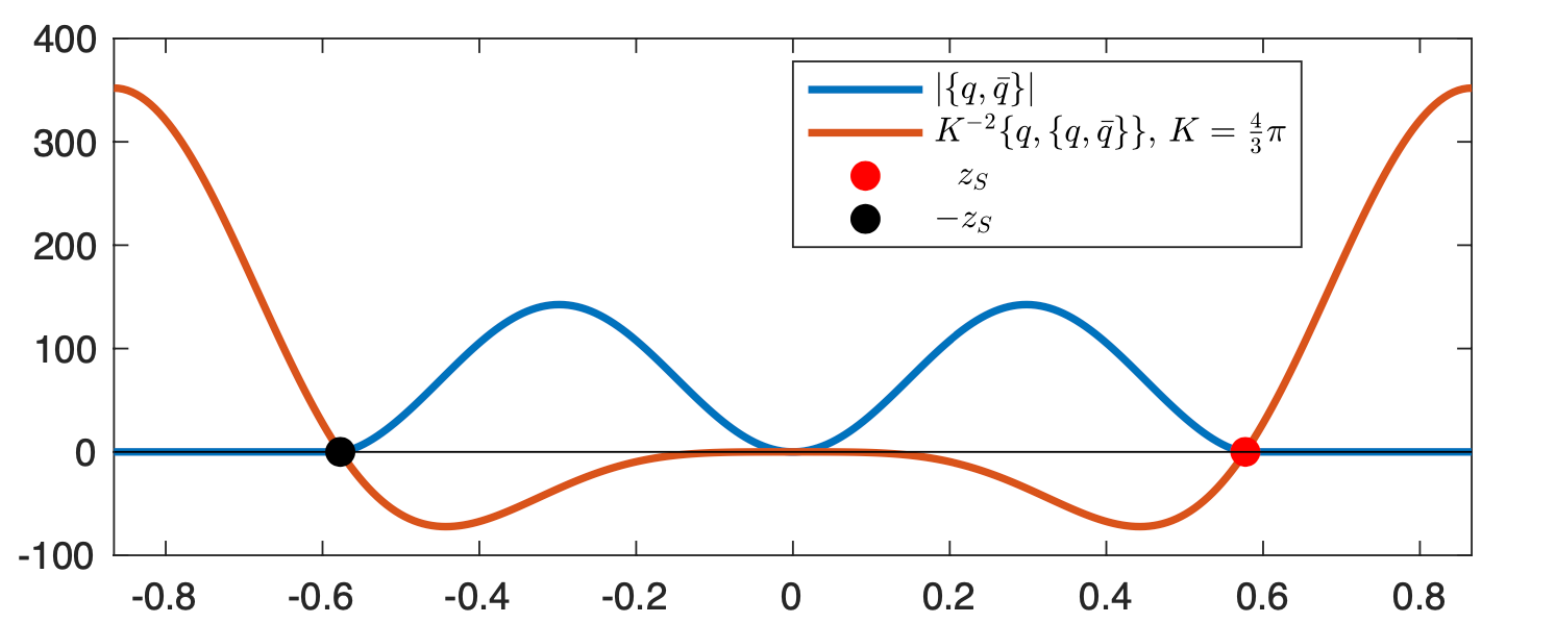}
\caption{Plots of $ |\{ q, \bar q\} | $ and of (rescaled) $ \{ q, \{ q , \bar q \} \} $ above the
intersection of the imaginary axis and the fundamental domain in Figure~\ref{f:hiz}. The edges of
the hexagon emanate right of $ z_S $ and left of $ - z_S$.
\label{f:hiz1}}
\end{figure}

In the case of the operator $ P ( x, h D ) $ given in \eqref{eq:defP} there are no 
classically forbidden region: for every $ x \in \mathbb R^2 $ there exists
$ \xi \in \mathbb R^2 $ at which the determinant of the principal symbol (see
\eqref{eq:defq}) vanishes, $ q ( x, \xi )= 0 $. 

The remedy for this is to use analogues of results on (analytic) hypoellipticity
due to Tr\'epreau (with different proofs, following an approach
 due to Sj\"ostrand and reviewed in \cite{HiS},  provided by Himonas), which followed ideas of
Egorov, H\"ormander, and Kashiwara (we defer to \cite{hiz} for pointers to 
the literature). Hypoellipticity here refers to having the same conclusion 
$ \| u \|_{ L^2 ( {\rm{neigh}} (x_0 ) )} \leq C e^{ -c/h}  $ as in \eqref{eq:defQ} but without
the assumption that $q|_{ q^{-1} ( x_0 ) } \neq 0$. 

 A  semiclassical version of a general hypoelliptic result we need is given as follows:
let $ Q $ satisfy the same general assumptions as before \eqref{eq:defQ};
\begin{equation}
\label{eq:3br} \left. \begin{array}{l} \{ q , \bar q\}|_{ \pi^{-1 } ( x_0 ) \cap q^{-1} ( 0 )  } =  0  , \\ 
\{ q, \{ q, \bar q \} \}|_{ \pi^{-1 } ( x_0 ) \cap q^{-1} ( 0 )  } \neq 0, \\ 
 Q  u = 0 \text{ in $  {\rm{neigh}} (x_0 ) $},  \ \| u \|_{ L^2} = 1 \end
{array} \right\}   \Longrightarrow  \ 
\| u \|_{ L^2 ( {\rm{neigh}} (x_0 ) )} \leq C e^{ -c/h} ,  \end{equation}
see \cite[Theorem 2]{hiz}. 

To see why such a result could be true 
consider a simple one dimensional example:  $ q ( x, \xi ) = \xi + i x^2 $, $ ( x, \xi ) \in 
\mathbb R \times \mathbb R $,  $ x_0 = 0 $. 
Then $ \{ q, \bar q \} ( x_0, \xi )  = - 4 i x_0   = 0 $, $\{ q , \{ q , \bar q \} \} ( x_0, \xi ) = - 4 i$, so the 
condition holds. 
If $ 0 = q ( x, h D ) u = ( h/ i) ( \partial_x - x^2/h ) u $, then $ u ( x , h )  = u(0, h  ) e^{ \frac13 x^3/ h } $.  For this to be uniformly bounded near $ 0 $, we need $ u ( 0, h ) = e^{ - c /h } $, $c> 0 $.   So $ | u ( x, h ) | \leq e^{ -c/2h}  $ for $ |x| $ small. We remark that similar bracket conditions in the 
semiclassical setting appeared recently in the work of Sj\"ostrand--Vogel \cite{SV} who provided
fine tunneling estimates for a model operator. Any extension of their results to more general operators
should have consequences in our setting as well.

As in \eqref{eq:scalar} we can reduce the problem of looking at solutions to 
$ h( D( \alpha ) + k ) = P ( x, h D ) + hk $ to a principally scalar problem, 
with the principal symbol given by $ q ( x, \xi ) $ in \eqref{eq:defq}. It then turns out that 
the condition in \eqref{eq:3br} holds at any $ x_0 $ on an {\em open} edge of the hexagon 
spanned by the stacking points -- see Figure~\ref{f:hiz1} for the case of $ \lambda = 1 $
and $ U $ given in \eqref{eq:defUV}. Remarkably, due to the special properties of the
BM potential, the sign properties can be established analytically -- see \cite[\S 3]{hiz}.

At $ \pm z_S $ the 
condition in \eqref{eq:3br} does not hold. However,  $ \pi^{-1} (\pm z_S ) \cap 
q^{-1} ( 0 ) = \{ ( \pm z_S , 0 ) \} $ and 
\begin{equation}
\label{eq:4br}  \{ q, \bar q \} ( \pm z_S , 0 ) = 0 , \ \ \
 \{ q , \{ q , \{ q , \{ q , \bar q \}\}\}\} ( \pm z_S , 0 ) \neq 0. \end{equation}
  General hypoellipticity 
results of Tr\'epreau do not apply to this case but a detailed analysis of our 
specific principal symbol \cite{tza} 
allows an application of the same strategy as in 
the proof of \eqref{eq:3br} to obtain exponential decay near the stacking points.

Since the conditions in \eqref{eq:3br} and \eqref{eq:4br}  classical in the sense of involving the 
symbol (that is, the ``classical observable",  $ q ( x, \xi ) $) and Poisson brackets
(objects underlying classical dynamics), we obtain the following result about
classically forbidden regions 
\begin{theo}[\cite{hiz},\cite{tza}] 
\label{t:hiz}
There exists a fixed open neighbourhood, $ \Omega $, of the
hexagon spanned by the stacking points (see Figure \ref{f:hiz})
and $ c > 0$  
such that if $ u ( h ) \in H^1_0 $ satisfies $ ( P ( x, h D ) + hk ) u = 0 $ and
$ \|u ( h ) \|_{L^2_0} = 1 $, then 
\begin{equation}
\label{eq:hiz}     \| u ( h ) \|_{ L^2(\Omega ) } \leq c^{-1}  e^{ - c/h } . 
\end{equation} 
\end{theo}

The situation is more complicated at the center of the hexagon, $ z_0 = 0 $. In that case, the operator
is not of principal type, that is, $ q (0, 0 ) = 0 $ ($ \pi^{-1} ( 0 ) \cap q^{-1} ( 0 ) = \{ ( 0, 0 ) \}$)
and $ d q ( 0 , 0 ) = 0 $. This means that lower order terms should matter. That is 
confirmed by comparing \eqref{eq:scalar} with the scalar model $ Q ( \alpha ) $ (with no
lower order terms). For $ Q ( \alpha ) $, unlike for the chiral model, we do not see
exponential decay near $ 0 $ (the decay near the hexagon based on the properties of pricipal symbol
$ q $ persists)
: on the left $\log |u | $ for $ u $ a protected state for $ D ( \alpha ) $ and
on the right same for  $ Q ( \alpha )$:
\begin{center}
\includegraphics[width=7.5cm]{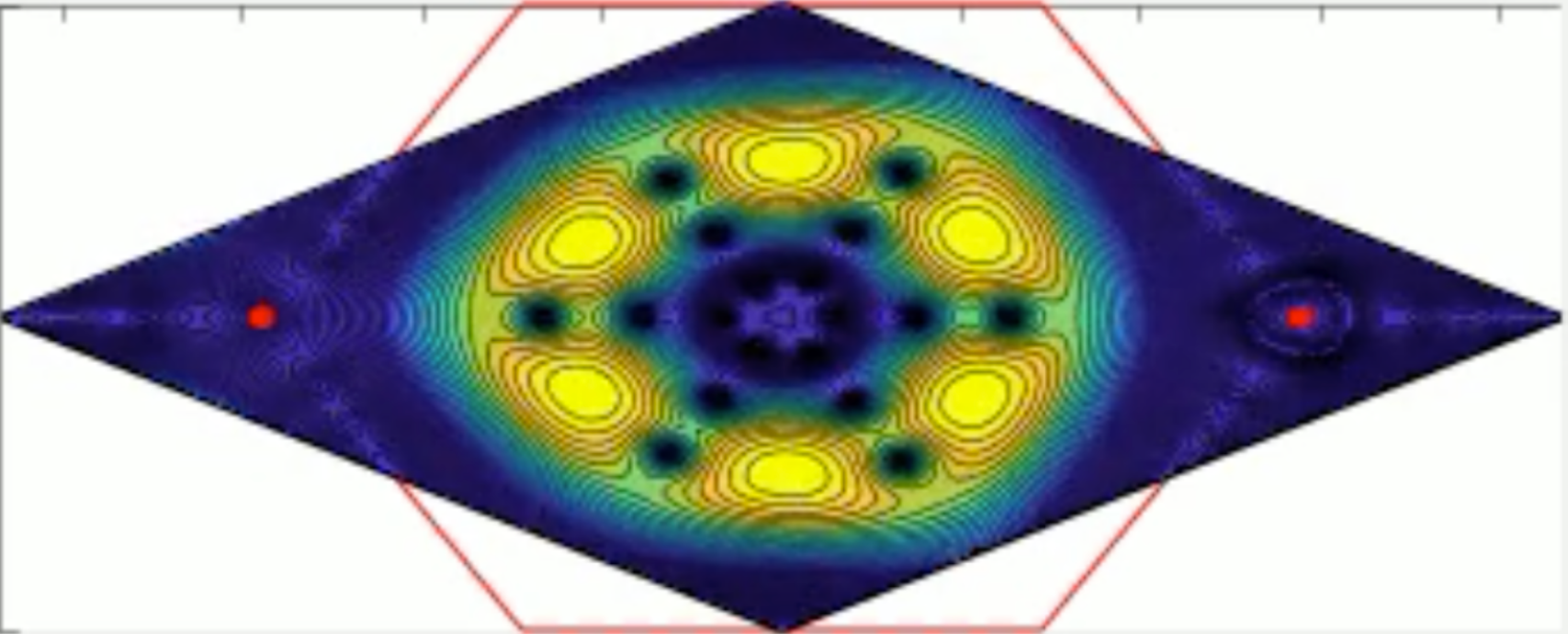}   \includegraphics[width=7.5cm]{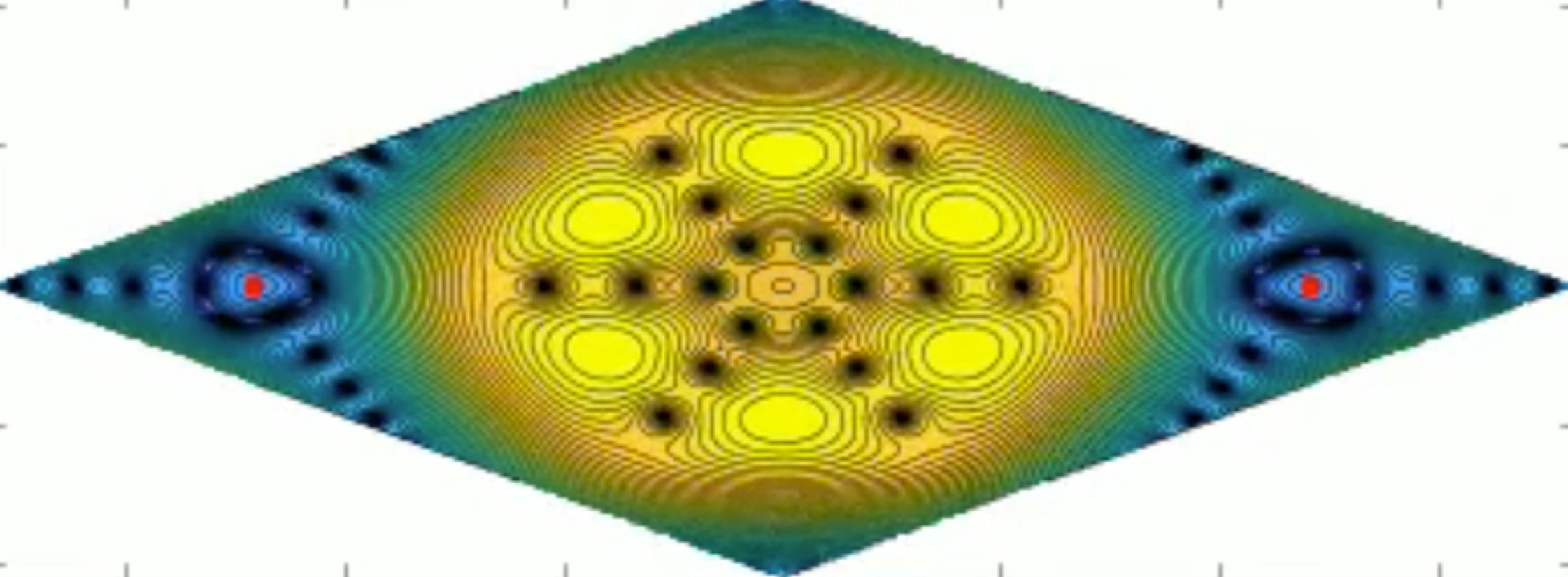}
\end{center}

\begin{op}
\label{20}  Show that  
there exist a fixed neighbourhood $ \Omega $ of $ 0 $ (see Figure~\ref{f:hiz}) 
and $ c > 0 $ 
 such that if $ ( P ( x, h D ) + hk ) u = 0 $, where $ P $ is given in 
 \eqref{eq:defP}, and $ \|u ( h ) \|_{L^2_0} = 1 $, then  
$   \| u ( h ) \|_{ L^2(\Omega ) } \leq c^{-1}   e^{ - c/h } $. 
\end{op}

\section*{Appendix by Mengxuan Yang and Zhongkai Tao}

\renewcommand{\theequation}{A.\arabic{equation}}
\refstepcounter{section}
\renewcommand{\thesection}{A}
\setcounter{equation}{0}

We prove the existence of conic singularities in the first band of the chiral limit \cite{magic} of the Bistritzer--MacDonald 
Hamiltonian \cite{BM11} of twisted bilayer graphene when $\alpha\notin\mathcal{A}$.

\begin{theo}
\label{thm:cone}
    Near $\pm K$ points, the first band $E_1(\alpha,k)$ is given by
    \begin{equation}
        \label{eq:cone}
        E_1(\alpha, k) = c(\alpha) \cdot |k\pm K| + \mathcal{O}(|k\pm K|^{2}),
    \end{equation}
    where $c(\alpha) \geq 0$ with the equality holds if and only if $\alpha\in \mathcal{A}$.
\end{theo}

\begin{figure}
\centering
\includegraphics[width=8cm]{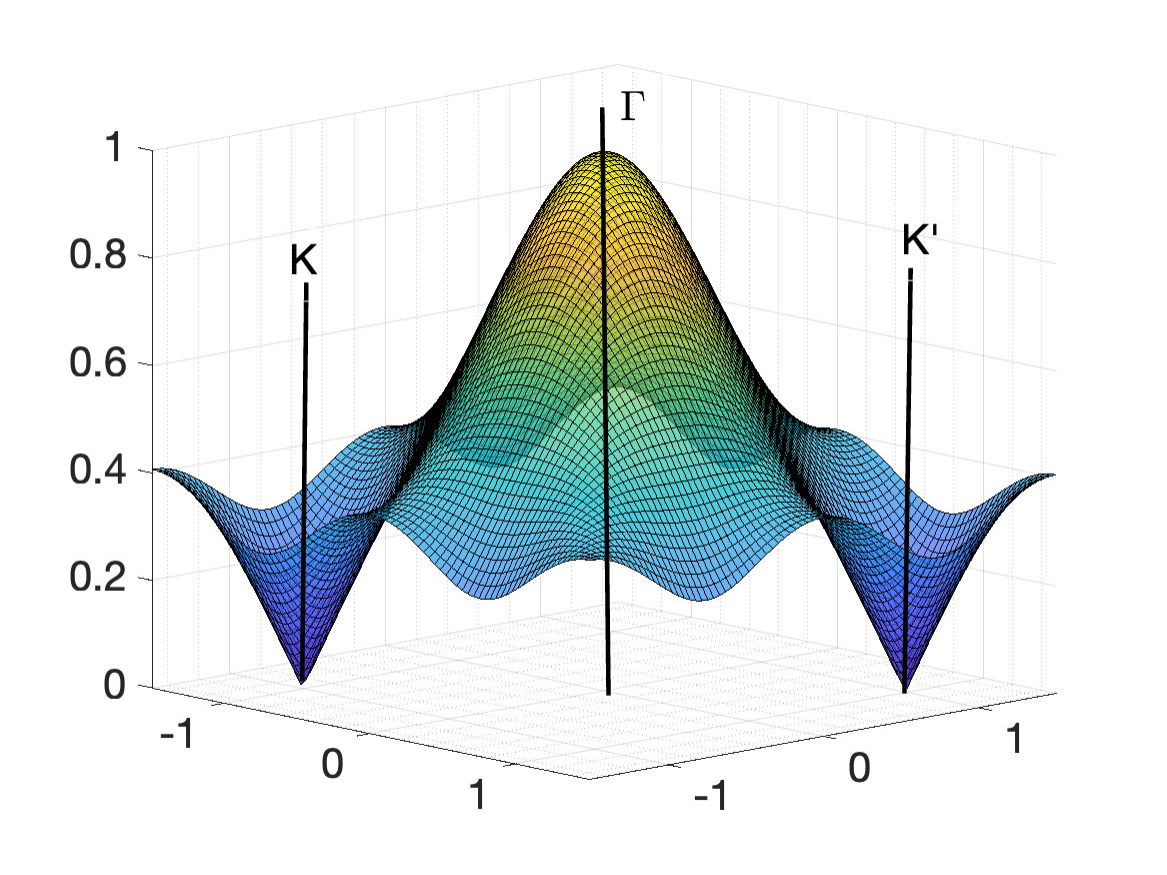}
\caption{\label{fig:cone} Two Dirac cones at $K$ and $K'$ points.}
\end{figure}

A key fact used in the proof is the \emph{existence of protected eigenstates} \cite{magic,beta}
described in Theorem \ref{t:prot}. We also remark that a dual result is the existence of protected states for the operator $D(\alpha)^*$: there exists $ v_{\pm K} ( \alpha ) \in  H^1_{0} ( \mathbb C; \mathbb C^2 ) $ such that
$ \tau(K) v_{K} ( 0 ) = (1,0)^T $, $ \tau(-K) v_{-K} (0) = (0,1)^T $, 
\begin{equation} 
\begin{gathered}
 v_{\pm K}(\alpha)\in \ker_{L^2_0( \mathbb C ; \mathbb C^2 )} ( D( \alpha )^* \pm \bar{K} ). 
\end{gathered} 
\end{equation}
It also follows from the proof that the generalized eigenspace also has dimension $1$, i.e., the spectrum is simple.

Now we prove Theorem \ref{thm:cone} by setting up a Grushin problem to compute the first band $E_1(\alpha, k)$ near $k=\pm K$ for $\alpha\notin\mathcal{A}$. We refer to \cite[Appendix C]{res} for a 
presentation of this method. The proof of Theorem \ref{thm:cone}  is based on the following general fact:
suppose that 
$X_1\subset X_2$ are two Banach spaces and $P:X_1\to X_2$ be a Fredholm operator of index $0$
such that 
    \begin{align*}
        \ker P={\rm span}\{\varphi\}, \quad \ker P^*={\rm span} \{\varphi_*\}.
    \end{align*}
    Then there is a dichotomy: 
    \begin{equation}
    \label{eq:pmt} 
        \begin{gathered}
       \begin{split} & \text{$P-z$ is invertible in a punctured neighbourhood of $z=0$;}\\
       & \text{ if moreover the eigenvalue $z=0$ is simple, then $\langle \varphi,\varphi_*\rangle\neq 0$}
       \end{split} \\
        \text{or}  \\
        \text{$P-z$ is not invertible for all $z$, and $\langle \varphi,\varphi_*\rangle=0$.}
    \end{gathered} 
\end{equation}
\begin{proof}[Proof of \eqref{eq:pmt}]
    The first part of the dichotomy follows from the analytic Fredholm theory (see \cite[Theorem C.8]{res}), which says if $P-z$ is invertible at one point then $(P-z)^{-1}$ is a meromorphic family.

    Now suppose $P-z$ is invertible in a neighbourhood of $z=0$ and $0$ is a simple eigenvalue, then $(P-z)^{-1}$ has the following expansion near $z=0$:
    \begin{align*}
        (z-P)^{-1}=A_0(z)+\frac{\Pi}{z}
    \end{align*}
    where $A_0(z)$ is holomorphic and $\Pi$ is a rank one projector. From the expansion we see
    $P\Pi=\Pi P=0$. So 
    \begin{align*}
        {\rm im}\,\Pi\subset \ker P={\rm span}\{\varphi\},\quad {\rm im}\,P\subset \ker \Pi.
    \end{align*}
    Thus $\Pi$ is of the form $\Pi(y)=\langle y,v_*\rangle \varphi$ for some $v_*\in X_2^*$. Moreover, $\langle Px, v_*\rangle=0$ for any $x\in X_1$, which implies $P^*v_*=0$. Thus $v_*=c\varphi_*$ for some $c\in\mathbb{C}\setminus\{0\}$. Since $\Pi^2=\Pi$, we conclude $\langle \varphi,\varphi_*\rangle\neq 0$.

    Suppose $P-z$ is not invertible for any $z$, then we consider the following Grushin problem:
    \begin{align*}
        \begin{pmatrix}
            P-z &R_-\\
            R_+ &0
        \end{pmatrix}: X_1\times \mathbb{C}\to X_2\times\mathbb{C}
    \end{align*}
    where $\varphi_*(R_- 1)=1$ and $R_+ \varphi =1$. One can compute from \cite[Proposition C.3]{res} that $E_{-+}(z)=z\langle \varphi,\varphi_*\rangle +\mathcal{O}(|z|^2)$. By assumption $E_{-+}(z)=0$, so we conclude $\langle \varphi,\varphi_*\rangle=0$.
\end{proof}

We can now give 
\begin{proof} [Proof of Theorem \ref{thm:cone}]
For the chiral Hamiltonian
\begin{equation}
H_k ( \alpha) : H^1_0 ( \mathbb C ; \mathbb C^4 ) \to 
L^2_0 ( \mathbb C ; \mathbb C^4 ),  \ \ \alpha \in \mathbb C,  \end{equation}
we consider the existence of a Dirac cone at $K$ point, as the point $-K$ is similar. By the existence of protected states, there exist two normalized protected states $\varphi(\alpha;z),\psi (\alpha;z)\in \ker_{L^2_0(\CC;\CC^4)} H_K(\alpha)$ such that 
\begin{equation}
    \varphi(\alpha;z) = (u_K(\alpha),0_{\CC^2})^T,\ \ \psi (\alpha;z) = (0_{\CC^2}, v_K(\alpha))^T.
\end{equation}
We consider the Grushin problem for the operator $H_k(\alpha)-z$ near $k=K$: 
\begin{equation}
  \label{eq:Grushin-BM}
  \mathcal{H}_k =
    \begin{pmatrix}
    H_k(\alpha) - z & R_-\\
    R_+ & 0
    \end{pmatrix}: H^1_{0} ( \mathbb C ; \mathbb C^4 ) \oplus \CC^2 \longrightarrow L^2_{{0}} ( \mathbb C ; \mathbb C^4 ) \oplus \CC^2
\end{equation}
with 
\begin{gather*}
  R_-: (u_-^{(1)}, u_-^{(2)})^T \mapsto u_-^{(1)} \varphi + u_-^{(2)} \psi,\ 
  R_+: u \mapsto (\langle u, \varphi\rangle, \langle u, \psi\rangle)^T . 
\end{gather*}
For $k=K$, the Grushin problem \eqref{eq:Grushin-BM} is invertible with the inverse given by 
\begin{equation}
  \label{eq:Grushin-BM-inverse}
  \mathcal{E} = 
  \begin{pmatrix}
    E & E_+\\
    E_- & E_{-+}
    \end{pmatrix}: L_{0}^2 ( \mathbb C ; \mathbb C^4 ) \oplus \CC^2 \longrightarrow H_{0}^1 ( \mathbb C ; \mathbb C^4 ) \oplus \CC^2
\end{equation}
with 
\begin{gather*}
    Ev= \sum_{j\neq \pm1}\frac{1}{E_j-z}\langle v, \varphi_j\rangle \varphi_j,\ \ E_+v_+= R_-v_+, \ \ 
    E_-v= R_+v,\ \ 
    E_{-+}= \begin{pmatrix}
      z & \\
       & z
      \end{pmatrix}, 
\end{gather*}
where $\{\varphi_j\}$ is an orthonormal basis of eigenfunctions of $H_K(\alpha)$ with eigenvalue $E_j$ such that $\varphi_1 = \varphi$ and $\varphi_{-1} = \psi$. By \cite[Proposition C.3]{res}, the perturbed Grushin problem \eqref{eq:Grushin-BM} is well-posed for $|k-K|$ sufficiently small and the eigenvalues of $H_k(\alpha)$ are given by zeros of the determinant of
\begin{equation}
   F_{-+}=E_{-+}+\sum_{k=1}^\infty(-1)^kE_-A(EA)^{k-1}E_+,\ \ A = \begin{pmatrix}
       & \overline{k-K} \\
       k-K & 
      \end{pmatrix}. 
\end{equation}
In particular, the leading order term is given by
\begin{equation}
\label{eq:1st_order}
    E_-AE_+ = \begin{pmatrix}
       & \overline{(k-K)} \langle v_K, u_K \rangle \\
        (k-K) \langle u_K, v_K \rangle & 
      \end{pmatrix}
\end{equation}
This yields that $E_{\pm 1}(\alpha, k) = \pm|\langle v_K, u_K \rangle| \cdot |k-K| + \mathcal{O}(|k-K|^2)$ near $k=0$, where $\langle v_K, u_K \rangle =0$ if and only if $\alpha\in\mathcal{A}$ by 
\eqref{eq:pmt}.
\end{proof}

\end{document}